\documentclass[a4paper, 12 pt, twoside, reqno]{amsart}
     \usepackage[utf8]{inputenc}
    \usepackage{slashed}
    \usepackage[english]{babel}
    \usepackage[all]{xy}
    \usepackage{enumerate}
    \usepackage{setspace}
    \usepackage{hyperref}
    \usepackage{tikz}
\usetikzlibrary{calc}
\usepackage{tikz-cd}
    
    \usepackage[euler-digits,euler-hat-accent]{eulervm}
    \usepackage{amsthm}
    
    \usepackage{times}
    \usepackage{amsmath}
    \usepackage{amsfonts}
    \usepackage{amssymb}
    \usepackage{amsthm}
    \usepackage{graphicx}
    \usepackage{array}
   \usepackage[dvipsnames]{xcolor}
    \usepackage{mathrsfs}
    
    \usepackage{eucal}
    \usepackage{esint} %%% for \fint
    \usepackage{tikz}
    \usepackage{upgreek}
    
    \usepackage{mathtools}

    \usepackage{bm}
    
    \allowdisplaybreaks
    
    \newcommand{\LM}{{\bm{\mathrm L}}}
    
    \theoremstyle{plain}
    \newtheorem{theorem}{Theorem}[section]
    
    \newtheorem{corollary}[theorem]{Corollary}
    \newtheorem{proposition}[theorem]{Proposition}
    \newtheorem{lemma}[theorem]{Lemma}
    \newtheorem{conjecture}[theorem]{Conjecture}
    
    \theoremstyle{definition}
    \newtheorem{definition}[theorem]{Definition}

    \theoremstyle{remark}
    \newtheorem{remark}[theorem]{Remark} 
    \newtheorem{example}[theorem]{Example} 
    \newtheorem{question}[theorem]{Question} 
    \newcommand{\SM}{\mathtt s}

    \numberwithin{equation}{section}
    \numberwithin{figure}{section}
    \numberwithin{table}{section}
    
    \newcommand{\R}{\mathbb{R}}
    \newcommand{\N}{\mathbb{N}}
    \newcommand{\C}{\mathbb{C}} 
    
    \newcommand{\Z}{\mathbb{Z}}
    \newcommand{\T}{\mathbb{T}}

    \newcommand{\G}{\mathcal{G}}
                     
    \newcommand{ \ii}{\,\mathrm{i}\,}
    
    \newcommand{\Bb}{\mathcal{B}}
                    \newcommand{\ie}{\textsl{i.\,e.\,}}

    \newcommand{\Oo}{\mathscr O}

    \newcommand{\Hh}{\mathcal{H}}

    \newcommand{\A}{\mathfrak{A}}

    \newcommand{\pr}{\mathbb{P}}

    \hypersetup{
    pdftoolbar=true,        % show Acrobat’s toolbar?
    pdfmenubar=true,        % show Acrobat’s menu?
    pdffitwindow=true,     % window fit to page when opened
    pdfstartview=true,    % fits the width of the page to the window
    pdftitle={Abelian code models},    % title
    pdfauthor={Danilo Polo Ojito},     % author
    breaklinks=true, %permet le retour a la ligne dans les liens trop longs
    colorlinks=true,       % false: boxed links; true: colored links
    linkcolor=purple,         % color of internal links (change box color 
    citecolor=teal, % color of links to bibliography
    urlcolor=blue, %couleur des hyperliens
    bookmarksopen=true, %si les signets Acrobat sont crees,
    filecolor=magenta,      % color of file links
    }

\begin{document}

    \title[ground-state spaces of generalized stabilizer codes]{ground-state spaces of generalized stabilizer codes}

\author[C. Cheng]{Cathy Cheng}
\address{Department of Mathematics, Pennsylvania State University\\ University Park, PA 16802, USA
\\
\href{mailto:ybc5551@psu.edu}{ybc5551@psu.edu}}
\author[N. Higson]{Nigel Higson}
\address{Department of Mathematics, Pennsylvania State University\\ University Park, PA 16802, USA
\\
\href{mailto:higson@psu.edu}{higson@psu.edu}}

\author[D. P. Ojito]{Danilo Polo Ojito}
\address{Department of Mathematics\\ Pontificia Universidad Javeriana, Bogota, Colombia
	\href{danilopolo@javeriana.edu.co}{danilopolo@javeriana.edu.co}}

\author[E. Prodan]{Emil Prodan}

\address{Department of Physics and
 Department of Mathematical Sciences 
\\Yeshiva University, New York, NY 10016, USA \href{mailto:prodan@yu.edu}{prodan@yu.edu}
}

    \vspace{2mm}
    
    \date{\today}
    
    \maketitle
   \begin{abstract}
In this paper, we determine the complete ground-state spaces for 
a broad class of generalized stabilizer quantum models defined over CW-decompositions of 
non-compact manifolds of arbitrary dimensions and topological type. We introduce
a $C^*$-algebraic framework in which the quantum spin algebra
$\A$ is realized as the twisted group $C^*$-algebra of an
infinite discrete Pauli module $P$ endowed with a symplectic
structure. We establish a correspondence between stabilizer
models and isotropic subgroups of $P$: Every Lagrangian
extension of such a subgroup determines a $C^*$-diagonal
of $\A$, whose Weyl groupoid encodes the dynamics through
a continuous 1-cocycle. This realization yields an explicit
family of extremal ground-states for every admissible
Lagrangian model, together with their decomposition into
sectors. When the sector space is $T_0$, this gives a complete characterization of the ground-state spaces. We also prove that every Lagrangian
stabilizer model admits a unique frustration-free ground
state with a spectral gap. Our framework applies in arbitrary dimension, including higher abelian quantum doubles, homological codes, and a broad class of stabilizer models.

 \end{abstract}

    {\scriptsize \tableofcontents}

    \section{Introduction}   
The computation and classification of the spaces of ground-states for dynamics over quantum spin algebras is of critical importance for the programs of topological classification of phases of matter, topological quantum computation and quantum error correction. The common setting involves a UHF algebra $\A = \bigotimes_\Sigma{\rm M}$ of physical observables over an infinite discrete subset $\Sigma\subset \mathbb R^d$, with ${\rm M}$ a matrix algebra, together with a strongly continuous
one-parameter group of automorphisms
$\alpha\colon\mathbb R\to{\rm Aut}(\A)$. Its generator $\bm{\delta}$ is a densely defined $*$-derivation,
typically specified on a dense $*$-subalgebra of local observables by $
\bm{\delta}(A)=\lim_n \ii[H_n,A],$ 
where $\{H_n\}$ is a net of local Hamiltonians and the limit is taken in norm. The ground-state space then consists of all states $\SM$ on $\A$ satisfying
\begin{equation}\label{eq: GS1}
    -\ii \SM(A^\ast \bm{\delta}(A))\; \geq \;0,\qquad \forall  A\in {\rm Dom}(\bm{\delta})
\end{equation}
 When the local Hamiltonians are positive, this space contains the so-called \emph{frustration-free} states, characterized by $\SM(H_n)=0$ for all $n$. These states play a central role
in quantum error correction and topological quantum computation
\cite{Terha2015,Kit0,Kit00}. Ground states that are not
frustration-free can describe anyonic sectors in topologically
ordered models, motivating the study of the full ground-state
space \cite{WangBook,Ogata2022,Perez}. While partial information about the ground-states of various models abounds in the physics literature, computing the \textit{full} space of ground-states, \ie identifying
all states satisfying \eqref{eq: GS1} for a given dynamics
$\bm{\delta}$, has proven to be extremely difficult. This  has so far been accomplished for only a few one-dimensional models \cite{Araki1985,Matsui1996,Koma1998,OgataI,OgataII} and, more recently, for a class of two-dimensional models \cite{CNN}. Furthermore, no unifying framework has emerged so far, since the success of the mentioned computations rested on functional analytic techniques specialized to the structures and particularities of those cases.

\medskip 

On the other hand, the equilibrium states at finite temperatures over groupoid $C^\ast$-algebras afford a characterization in terms
of quasi-invariant measures for dynamics induced by groupoid
$1$-cocycles \cite{RenaultBook}.  See \cite{Nesh1} for the state of the art of this problem. In the same context of dynamics generated by $1$-cocyles, a fine characterization of the ground-states has been achieved relatively recently in \cite{LLN2019} by Laca, Larsen and Neshveyev. This characterization identifies a boundary subset of the groupoid's unit space, determined by the $1$-cocycle, such that
the state space of the restricted groupoid $C^\ast$-algebra
is affinely homeomorphic to the ground-state space. In \cite{LLN2019}, the authors use this formalism to complete the characterization of the equilibrium and ground-states for certain natural time evolutions on Hecke algebras. In \cite{Thomsen2014,Thomsen2020,Thomsen2021}, Thomsen analyzed the ground-state spaces for time-evolutions over UHF algebras generated by generalized gauge potentials. He showed, for example, that the spaces of KMS and ground-states bear no relation to each other, as the latter can be affinely homeomorphic to the state space of any AF-algebra. As Thomsen himself acknowledges in \cite{Thomsen2021}, his characterization of the ground-states can be translated in the groupoid language of \cite{LLN2019}. These examples demonstrate the effectiveness of the groupoid
formalism solving \eqref{eq: GS1} once the dynamics can be ``diagonalized,'' in the
sense that the algebra $\A$ admits a groupoid presentation in
which $\bm{\delta}$ is induced by a $1$-cocycle. We follow this strategy to compute and classify all the solutions of \eqref{eq: GS1} for a broad class of stabilizer code models.

\medskip 

Many examples in the physics literature involve dynamics
$\bm{\delta}$ on $\A$ whose local Hamiltonians $H_n$ are sums
of commuting projections. This class includes models arising
from actions of (quasi-)Hopf algebras
\cite{Kit1,BuerschaperJMP2013,LevinPRB2005}, which, at least
in two dimensions, are conjectured to provide representatives
of every non-chiral topological phase of matter \cite{Perez}.
The connection between a subclass of these models and the
groupoid framework described above was established in
\cite{DE1,PoloProdan2026}, building on a crucial observation
in \cite{AlickiJPA2007}: the local interaction terms of the
toric code generate a maximal abelian subalgebra (masa) of
the CAR algebra. This result was strengthened in
\cite{DE1}, where this masa was shown to be a $C^\ast$-diagonal and generalized for any abelian quantum double model in \cite{PoloProdan2026}. Consequently, the theory of $C^\ast$-inclusions developed by
Kumjian and Renault \cite{Kum,Ren1} provides a presentation
of the observable algebra through the associated twisted
Weyl groupoid. In this presentation, the dynamics of the
models considered in \cite{PoloProdan2026} are induced by
$1$-cocycles, allowing a complete characterization of their
KMS states by the methods described above. These observations,
together with the results of \cite{LLN2019}, suggest that
a unified framework for the complete characterization of
ground-state spaces is also within reach.

\medskip
% Our models are
%  In a finite spin system,
% an abelian subgroup of Pauli operators defines a code space
% as their common eigenspace with eigenvalue $1$. Operators
% on this space describe the encoded logical degrees of
% freedom \cite{Ball2023,Terha2015}. Equivalently, the code
% states are the frustration-free ground states of
% $H_{\mathscr S}=\sum_s(1-P_s)$, where $s$ runs over a
% generating set of the stabilizer subgroup and $P_s$ is
% the spectral projection of $s$ onto eigenvalue $1$.
% We extend this perspective to infinite systems, using
% $C^\ast$-algebras to describe both the logical observables
% and the full ground-state space. We extend the stabilizer framework to infinite systems built
% from a finite abelian group $G$. Let $\Sigma$ be a
% CW-decomposition of a noncompact $n$-dimensional manifold,
% with dual CW-structure $\widetilde\Sigma$. The Pauli module
% $P_k$ consists of finitely supported configurations valued
% in $G$ on the $k$-cells of $\Sigma$ and in its Pontryagin
% dual $\widehat G$ on the $(n-k)$-cells of
% $\widetilde\Sigma$. Central extensions of $P_k$ define
% generalized Pauli groups $\mathcal P_k$, and the associated
% twisted group $C^\ast$-algebras
% $\A_{\sigma_k}:=C^\ast(P_k,\sigma_k)$ serve as algebras of
% physical observables. We assume that the commutator pairing
% $\omega_k$ is non-degenerate.

Our focus is on stabilizer code models, whose local Hamiltonians are sums of mutually commuting projections constructed from Pauli operators. These models play a central role in quantum error correction and fault-tolerant quantum computation \cite{Ball2023, Gottesman1997,CRSS1997,GottesmanFT1998, Kit0, Kit00, Kit1,Dennis2002}. Motivated by their symplectic description \cite{Haah2013,RubaYang2024,RubaYang2025}, we develop a unified
$C^*$-algebraic framework for stabilizer codes on infinite
discrete sets in terms of isotropic subgroups of a generalized
Pauli module. Within this framework, we describe the
ground-state spaces of a broad class of these generalized stabilizer code models. We now outline our framework, the models it encompasses, and the resulting description of their ground-state spaces. We start by considering a locally finite CW-decomposition $\Sigma$ of a
noncompact $n$-dimensional manifold $M$, equipped with a dual
CW-structure $\widetilde\Sigma$.  We work in the noncompact case, motivated by codes on infinite
discrete sets, although the same formalism can be adapted easily to the
compact case. The second ingredient is a finite abelian group $G$ with Pontryagin dual $\widehat G$, which generates a Pauli module
$$
P_k
\;=\;
C_c(\Sigma_k,G)
\oplus
C_c(\widetilde\Sigma_{n-k},\widehat G),$$
of finitely supported continuous functions. This construction places a $|G|$-dimensional qudit on each
$k$-cell of $\Sigma$, with elements of $P_k$ labeling generalized Pauli
operators acting on these qudits.  Although $P_k$ is abelian, the associated Pauli operators need not commute. Their multiplication is encoded by a normalized $2$-cocycle $\sigma_k$ on $P_k$, which defines a generalized Pauli group $\mathcal P_{\sigma_k}$ as a central extension of $P_k$ by $\mathbb T$. The commutator pairing $\omega_k$ of $\sigma_k$ (see \eqref{Eq:SymplecticForm}) is assumed to be non-degenerate and defines a symplectic structure on $P_k$. The corresponding algebra
of physical observables is the \emph{twisted group} $C^*$-algebra
$$
\A_{\sigma_k}\;:=\;C^*(P_k,\sigma_k),$$
generated by formal unitaries $\{U_p\}_{p\in P_k}$ satisfying the relation \eqref{Eq:Relations}. These unitaries with their phases form a group isomorphic to $\mathcal{P}_{\sigma_k}$, and we identify the two throughout this work. Our first structural result identifies the algebra $\A_{\sigma_k}$ with the quantum spin UHF algebra.

\begin{proposition}
For every normalized $2$-cocycle $\sigma_k$ with non-degenerate $\omega_k$
$$
\A_{\sigma_k}\;\cong\; {\rm M}_{|G|}^{\otimes \Sigma_k}\;= \;\bigotimes_{\Sigma_k}{\rm M}_{|G|}(\C).$$
\end{proposition}
Although this isomorphism is independent of the cocycle, in most parts of this work we use the standard cocycle \ref{sec: pauli} to retain the natural local structure induced by finite
subsets of $\Sigma_k$ used in physics.

 In this setting, a code is a nonempty weak-$*$ closed convex subset $\mathtt C$ of the state space of $\A_{\sigma_k}$. Its stabilizer $\mathscr S_{\mathtt C}$ consists of all Pauli unitaries $U\in\mathcal P_{\sigma_k}$ satisfying $\SM(U)=1$ for every $\SM\in\mathtt C$, and the pair $(\mathtt C,\mathscr S_{\mathtt C})$ is called a stabilizer code. Every stabilizer $\mathscr S_{\mathtt C}$ projects onto an isotropic subgroup $L\subseteq(P_k,\omega_k)$. Conversely, every isotropic subgroup $L$ admits compatible lifts $l\mapsto U_l$, satisfying $U_lU_{l'}=U_{ll'}$, which define a code $\mathtt C_L$. Thus, isotropic subgroups, together with compatible lifts, encode the stabilizer constraints, and $\mathtt C_L$ is the largest code satisfying them. These constraints also give rise to models whose local Hamiltonians are sums of mutually commuting projections. Given a generating set $\LM$ of $L$, the formal
Hamiltonian
$$
H_{\LM}\;=\;\sum_{l\in\LM}(1-P_l),$$
where $P_l$ is the spectral projection of $U_l$ onto
eigenvalue $1$, has $\mathtt C_L$ as its frustration-free
state space. Under the admissibility condition of
Definition~\ref{Def:AdmissibleL}, this interaction defines
a dynamics $\alpha_{\LM}$ on $\A_{\sigma_k}$ with generator
$\bm{\delta}_{\LM}$ according to \ref{propo: dynamics}.  We refer to the dynamics $\alpha_{\LM}$ as the stabilizer code model. Once compatible lifts have been fixed, the isotropic subgroup $L$ determines the stabilizer constraints, while $\LM$ specifies the dynamics $\alpha_{\LM}$. Our main goal is to characterize the full ground-state
space $\mathfrak G_{\LM}$ of $\bm{\delta}_{\LM}$. As a first step toward understanding $\mathfrak G_{\LM}$,
we characterize its subset
$\mathfrak G_{\LM}^F$ of frustration-free ground states.

For an isotropic subgroup $L$,
let $L^\perp$ denote its symplectic orthogonal \eqref{eq: isotropic complement}. The \emph{logical
algebra}
$\A^{\rm log}_L\simeq C^*(L^\perp/L,\widetilde\sigma_k)$ is a twisted group algebra that describes the degrees of freedom that remain after imposing
the stabilizer constraints. Our first main result identifies
$\mathfrak G_{L}^F$ with the state space of this algebra
and establishes uniqueness of the frustration-free ground
state in the Lagrangian case, \ie $L=L^\bot.$
\begin{theorem}\label{teo: GSFF1}
Let $L\subset P_k$ be isotropic. The frustration-free
ground-state space $\mathfrak G_L^F$ of $\bm{\delta}_{\LM}$ is affinely
homeomorphic to the state space of $\A^{\rm log}_L$. In particular, if $L$ is Lagrangian, then
$\A^{\rm log}_L\cong\C$, and $\bm{\delta}_{\LM}$ has
a unique frustration-free gapped ground state.
\end{theorem}
The proof of Theorem~\ref{teo: GSFF1} and the underlying constructions are presented in Section~\ref{sec: logical algebra} and Theorem~\ref{theo: gapped}. Thus, the logical algebra $\A^{\rm log}_L$ completely determines the space $\mathfrak G_L^F$. In particular, the Lagrangian condition provides an algebraic criterion for the existence of a unique frustration-free ground state with a spectral gap. In Section~\ref{Sec:Applications}, we illustrate these results through explicit descriptions of abelian quantum double and color code models on the plane, whose stabilizer subgroups are Lagrangian. We also treat the Ising model, whose stabilizer subgroup is isotropic but not Lagrangian, and compute $\A^{\rm log}_L$.

\medskip
 
 Theorem~\ref{teo: GSFF1} does not, however, describe
the full ground-state space, since $\mathfrak G_{\LM}$
may also contain states that are not frustration-free. To go beyond the frustration-free setting, our second main result provides the announced  groupoid diagonalization of $\bm{\delta}_{\LM}$. This brings stabilizer code models into the framework of $C^*$-inclusions developed by Kumjian and Renault, allowing us to study their equilibrium states through groupoid methods and, in particular, to apply the ground-state characterization of Laca, Larsen and Neshveyev.
\begin{theorem}\label{teo: diagonal1}
Let $L\subset P_k$ be an admissible isotropic subgroup. If $L$ is Lagrangian, then:
\begin{enumerate}[i]
    \item The commutative $C^*$-algebra
    $\A_L=C^*\{U_l\mid l\in L\}$
    is a $C^*$-diagonal of $\A_{\sigma_k}$.
    \item Let $\G_L$ be the Weyl groupoid of 
    $(\A_{\sigma_k},\A_L)$. It is AF and yields an isomorphism
    $$
    \A_{\sigma_k}\;\simeq\; C^*\G_L.
    $$

    \item The dynamics
    $\alpha_{\LM}$ is represented by a $1$-cocycle
    $c_{\LM}$ on $\G_L$:
    $$
    \bigl(\alpha_{\LM}^t(f)\bigr)(\zeta)
    \;=\;
    e^{\ii t c_{\LM}(\zeta)}f(\zeta),
    \qquad
    f\in C_c(\G_L),\;\zeta\in\G_L,\;t\in\R.
    $$
\end{enumerate}
More generally, if $L$ is not Lagrangian, then for any
Lagrangian extension $\widetilde L$ of $L$, the dynamics
$\alpha_{\LM}$ is represented by a $1$-cocycle
$c_{\LM}$ on $\G_{\widetilde L}$.
\end{theorem}
The proof and further details are given in Sections~\ref{sec: weyl groupoid}, \ref{sec: iso class}, and \ref{sec: dynamics by group}. Moreover, explicit presentations for $\G_L$ and $c_{\LM}$ are also provided in these Sections.  The strength of this result lies in the simplification it provides: in the groupoid picture, the dynamics $\alpha_{\LM}$ acts simply by multiplication by the phase determined by $c_{\LM}$. This turns the study of the dynamics into the analysis of $c_{\LM}$, making both ground states and KMS states accessible through the methods described above. Here we focus on the ground-state space. The study of KMS states and the use of this framework to investigate superselection sectors, braiding, and Haag duality are left to future work. 

\medskip

Now we have the ingredients to describe $\mathfrak G_{\LM}$. We first give a concrete realization of the Weyl groupoid $\G_{L}$ and the boundary set of $c_{\LM}$. For a Lagrangian subgroup $L$, the Weyl groupoid admits the explicit description as the transformation groupoid  $
\G_L\cong\widehat L\rtimes(P_k/L)$, where $\widehat L$ is the dual group of $L$ and $P_k/L$ acts by multiplication on $\widehat L$ via 
$[p]\mapsto [p]\triangleright\chi=\omega_k(p,\text{--})\chi$.
In this presentation, the dynamics is induced by the
cocycle $c_{\LM}$ given explicitly in \eqref{Eq:GoodCL}.
Following \cite{LLN2019}, we define the boundary set of $c_{\LM}$
$$
Z_{\LM}
\;=\;
\big\{\chi\in\widehat L\mid
c_{\LM}([p],\chi)\geq0,\;\forall [p]\in P_k/L\big\}$$
and the boundary groupoid as the restriction
$\G_{\LM}^b:=\G_L|_{Z_{\LM}}$.
The ground-state characterization of \cite{LLN2019}
then gives the following counterpart of
Theorem~\ref{teo: GSFF1}.

\begin{proposition}
    The space $\mathfrak G_{\LM}$ is affinely homeomorphic to the state space of the groupoid $C^*$-algebra $C^*\G^b_{\LM}$.
\end{proposition}
This Proposition turns the description of ground states into
the study of the boundary algebra $C^*\G^b_{\LM}$. Under this correspondence, pure states of the boundary
algebra give rise to pure ground states, and their weak-$*$
closed convex hull recovers $\mathfrak G_{\LM}$.
We therefore focus on $\mathtt P(C^*\G^b_{\LM})$. A natural family of such states arises from the orbit space
$\Gamma_{\LM}:=Z_{\LM}/\G_{\LM}^b$, equipped with the
quotient topology. Since $\G_\LM^b$ turns out to be principal, each orbit $\Oo_\gamma$ determines
a regular representation $\iota_\gamma$ on
$\ell^2(\Oo_\gamma)$, whose projective space parametrizes
a single class of pure ground states under unitary
equivalence of their GNS representations. Distinct orbits
determine inequivalent classes. More explicitly, let $\mathbb P(\ell^2(\Oo_\gamma))$
denote the projective space of $\ell^2(\Oo_\gamma)$.
Under the identification $\A_{\sigma_k}\cong C^*\G_L$,
each ray $[\xi]$, represented by a unit vector
$\xi\in\ell^2(\Oo_\gamma)$, defines a pure ground state
of $\bm{\delta}_{\LM}$ via
$$
\SM_{\gamma,[\xi]}(f)
\;=\;
\sum_{\chi\in\Oo_\gamma}
\sum_{\substack{[p]\in P_k/L\\
c_{\LM}([p],\chi)=0}}
\overline{\xi(\chi)}\,
f\big([p]^{-1},[p]\triangleright\chi\big)\,
\xi\big([p]\triangleright\chi\big),
\qquad f\in C_c(\G_L).
$$
This expression is independent of the chosen unit
representative of $[\xi]$. Our third main result shows
that this construction is injective and, when
$\Gamma_{\LM}$ is $T_0$, exhausts all pure ground states.
In that case, it also determines $\mathfrak G_{\LM}$
space by taking the weak-$*$ closed convex hull.
\begin{theorem}\label{Theo: GSintiial}
Let $L$ be an admissible Lagrangian subgroup. There is a well-defined injective map
$$
\bigsqcup_{\gamma\in\Gamma_{\LM}}
\mathbb P\bigl(\ell^2(\Oo_\gamma)\bigr)
\longrightarrow
{\rm Ext}(\mathfrak G_{\LM}),
\qquad
[\xi]\longmapsto\SM_{\gamma,[\xi]},
$$
where ${\rm Ext}(\mathfrak G_{\LM})$ denotes the set
of pure ground states of $\bm{\delta}_{\LM}$.
If $\Gamma_{\LM}$ is $T_0$, this map is bijective, and
$$
\mathfrak G_{\LM}
\;=\;
\overline{\operatorname{co}}^{\,w^*}
\big\{
\SM_{\gamma,[\xi]}
\mid
\gamma\in\Gamma_{\LM},\;
[\xi]\in\mathbb P\bigl(\ell^2(\Oo_\gamma)\bigr)
\big\}.
$$
In this case, every ground state is a weak-$*$ limit
of finite convex combinations of the pure ground states
$\SM_{\gamma,[\xi]}$.
\end{theorem}
The proof of this result is given in Section~\ref{sec: computation GS}.    In the examples discussed in
Section~\ref{Sec:Applications}, $\Gamma_{\LM}$ agree with
the superselection sectors of the models.  This extends to the present
framework the results obtained in \cite{CNN}
for abelian quantum double models, which we recover
in Section~\ref{Sec:Applications}. From this
perspective, Theorem~\ref{Theo: GSintiial} constructs
an explicit family of pure ground states organized
by sector. When $\Gamma_{\LM}$ is $T_0$, this family
exhausts all pure ground states, and every ground state
is a weak-$*$ limit of finite convex combinations of
states from these sectors.
Whether $\Gamma_{\LM}$ is $T_0$ for every
admissible Lagrangian model remains open; we conjecture
that this is the case in~\ref{conj: T0}.  We conclude with a concrete sufficient condition for
$\Gamma_{\LM}$ to be $T_0$. For $\chi\in\widehat L$, define its support $
{\rm supp}\,\chi:=
\{l\in\LM:\chi(l)\neq 1\}
$. Let $\Xi_{\LM}\subset\widehat L$ be the set of
characters with finite support. On this set, define
$T\colon\Xi_{\LM}\to\N_0$ by
$T(\chi)=|{\rm supp}\,\chi|$.

\begin{theorem}
Let $L$ be an admissible Lagrangian subgroup.
If $Z_{\LM}\subseteq\Xi_{\LM}$, then $\Gamma_{\LM}$
is $T_0$. If, in addition, $T$ is uniformly bounded
on $Z_{\LM}$, then the boundary algebra
$C^*\G_{\LM}^b$ is solvable in the sense of \cite{Dynin}.
\end{theorem}
The proof is given in Section~\ref{Sec:SpaceGS}.
The finite-support condition therefore ensures that the states in Theorem \ref{Theo: GSintiial} completely determine $\mathfrak G_{\LM}$. The additional boundedness condition yields a finite
filtration of the boundary algebra by ideals, providing
a stratified description of its structure. Since
${\rm supp}\,\chi$ records the violated stabilizer
constraints, this description connects the algebraic
structure with the excitation configurations compatible
with the ground-state condition. In Section~\ref{Sec:Applications}, we apply these results
to concrete physical models and verify explicitly that
$Z_{\LM}\subseteq\Xi_{\LM}$ and $T$ is uniformly bounded. For abelian Kitaev models,
we recover the complete ground-state classification
of \cite{CNN}, illustrating how the groupoid description
encodes the familiar physical picture. We then obtain
ground-state classifications for color code models
and for the biased and unbiased Ising models in
dimension $1$.

\medskip

\noindent{\bf Acknowledgments:}
Danilo Polo Ojito and Emil Prodan were supported by the
U.S. National Science Foundation through grant CMMI-2131760
and by the U.S. Army Research Office through contract
W911NF-23-1-0127. Nigel Higson was supported by the
U.S. National Science Foundation through grant DMS-1952669.

\section{Stabilizer Codes and $C^\ast$-Algebras}
The purpose of this section is to give a $C^*$-algebraic description of stabilizer codes on finite and infinite lattices. The basic idea is to place the qudits on the cells of a CW-decomposition $\Sigma$ of a manifold and encode the corresponding Pauli operators into a discrete group $P$. We work throughout with infinite lattices, although all our results apply equally to the finite case.

    \subsection{Generalized Pauli groups and their algebras of observables}

Let $M$ be a smooth non-compact $n$-manifold. Then $M$ always admits a locally finite CW-decomposition, {\it i.e.} a locally finite cellular complex whose geometric realizations are homeomorphic to the manifold. Furthermore, any such CW-structure $\Sigma$ accepts a dual $\widetilde \Sigma$, which can be generated via the barycentric construction \cite{RoeBook}. In the following, we will fix a pair $(\Sigma,\widetilde{\Sigma})$ of such CW-structure and its dual, together with a non-trivial finite abelian group $G$, whose Pontryagin dual will be denoted by $\widehat G$. In this context, for a fixed but arbitrary $k$ from $\{0,1,\ldots,n\}$, we consider the discrete abelian group
\begin{equation}\label{Eq:Pk}
     P_k
    \;:=\; 
    C_c(\Sigma_k,G)
    \oplus
    C_c(\widetilde \Sigma_{n-k},\widehat G),
\end{equation}
of compactly supported functions from the $k$-cells and their duals to the respective groups.  The group $P_k$ encodes the finitely supported Pauli labels associated with the $k$-cells of $\Sigma$, while the corresponding Pauli operators will arise from a central extension of $P_k$.

\begin{remark}
The geometric and topological properties of the manifold $M$
serve primarily as motivation: our groupoid diagonalization
and ground-state results rely on the algebraic framework
developed below, without further
assumptions. These properties enter in
Section~\ref{Sec:Applications}, in particular to verify the
Lagrangian property of specific models. We retain this setting
with a view toward future work on the localization and
transport of excitations.  \hfill $\blacktriangleleft$
\end{remark}

\iffalse 
we will consider the compactly supported cellular cycles and their complexes with coefficients in $G$ and $
    \widehat G$, respectively,
    \[
    C_n(\Sigma,G) \stackrel \partial \longrightarrow 
    C_{n-1}(\Sigma,G) \stackrel \partial \longrightarrow \cdots\stackrel \partial \longrightarrow 
    C_0(\Sigma,G) 
    \]
    and 
    \[
    C_n(\widetilde{\Sigma},\widehat G) \stackrel \partial \longrightarrow 
    C_{n-1}(\widetilde{\Sigma},\widehat G) \stackrel \partial \longrightarrow \cdots\stackrel \partial \longrightarrow 
    C_0(\widetilde{\Sigma}, \widehat G).
    \]
\fi

To any finite abelian group $A$, one can associate a generalized Pauli group defined as the nontrivial central extension of $A \times \widehat A$ under the standard pairing $\langle (a,\chi),(b,\psi)\rangle=\chi(b) \psi(a)^{-1}$. We will consider here generalizations of this notion:

    \begin{definition}
    A $k$-generalized Pauli group $\mathcal{P}_k$ is a central extension
    $$
    1\longrightarrow \mathbb T
    \longrightarrow \mathcal P_k
    \longrightarrow P_k
    \longrightarrow 1.
  $$
    \end{definition}
The central extensions of $P_k$ by $\mathbb T$ are classified,
    up to equivalence, by the cohomology group $
    H^2(P_k,\mathbb T).$
    More precisely, every class
    $
    [\sigma_k]\in H^2(P_k,\mathbb T)
    $
    determines a central extension
    \[
    1\longrightarrow \mathbb T
    \longrightarrow \mathcal P_{\sigma_k}
    \longrightarrow P_k
    \longrightarrow 1,
    \]
    and two extensions are equivalent if and only if they define the same cohomology class. Consequently, in order to keep track of the corresponding cohomology class,
we shall denote the associated Pauli group by $\mathcal P_{\sigma_k}$ rather
than simply $\mathcal P_k$. 
\begin{definition}
    The algebra of observables associated to a $k$-generalized Pauli group determined by $[\sigma_k]\in H^2(P_k, \mathbb T)$ is the twisted group $C^\ast$-algebra
    $$\A_{\sigma_k}\;=\;C^*(P_k, \sigma_k)$$
\end{definition}
\begin{remark}
Since $P_k$ is amenable, here the full and reduced twisted group $C^*$-algebras coincide $
C^*(P_k,\sigma_k)\simeq C_r^*(P_k,\sigma_k).
$     \hfill $\blacktriangleleft$

\end{remark}

Let us recall that $\A_{\sigma_k}$ is generated as a $C^\ast$-algebra by a set of unitaries $\{U_p\}_{p\in P_k}$, which satisfy the relations    
    \begin{equation}\label{Eq:Relations}
    U_pU_q
    \;=\;
    \sigma_k(p,q)\,U_{pq},\qquad 
    U_qU_p
    \;=\;
    \omega_k(p,q)\,U_pU_q.
    \end{equation}
where 
\begin{equation}\label{Eq:SymplecticForm}
    \omega_k(p,q)\;=\;\overline{\sigma_k(p,q)}\, \sigma_k(q,p).
\end{equation} 
The elements of $\mathcal P_{\sigma_k}$ can be presented as tuples $(z,p) \in \mathbb T \times P_k$ with multiplication rule
$$
(z,p)(z',q)\;=\;(\sigma_k(p,q)zz',pq).
$$
Then the assignment $(z,p) \mapsto zU_p \in \A_{\sigma_k}$ embeds $\mathcal P_{\sigma_k}$ as a subgroup of $\mathscr U(\A_{\sigma_k})$, the group of unitary elements of $\A_{\sigma_k}$. We will not make any distinction between this image and the group $\mathcal{P}_{\sigma_k}$. Note that we can always choose a representative of the cohomology class of the 2-cocycle such that $\sigma_k(e,p)=1=\sigma_k(p,e)$ for all $p \in P_k$, which we will do throughout.

\begin{remark}
    The map $\omega_k: P_k \times P_k \to \mathbb T$ defined in \eqref{Eq:SymplecticForm} is called the commutator pairing of the extension. It is an alternating bicharacter that depends only on the class of $\sigma_k$ and, in fact, $H^2(P_k,\mathbb T) \simeq {\rm Hom}(\Lambda^2 P_k,\mathbb T)$ \cite[Th.~6.4]{BrownBook}. Any such bicharacter is determined by its restrictions $\eta(c,\tilde c;c',\tilde c')$ to the elementary subgroups of $P_k$ supported by pairs of cells and, by using its bicharacteristic, we must conclude that $\omega_k$ always takes the form 
\begin{equation}\label{Eq:PairingForm}
\omega_k(p,q)\;=\;\prod \eta(c,\tilde c;c',\tilde c')\big (p(c,\tilde c),q(c',\tilde c')\big ),
\end{equation}
where the product is over the cells $c,c' \in \Sigma_k$ and $\tilde c,\tilde c' \in \widetilde{\Sigma}_{n-k}$. Furthermore, $\eta(c,\tilde c;c'\tilde c')$ are bicharacters of $G \times \widehat G$ fulfilling the usual constraints. Note that there are no constraints on the support of $\eta$, hence we can have long-range pairings of cells.    \hfill $\blacktriangleleft$
\end{remark}

As seen above, the relevant cohomology is highly nontrivial but note that, while two cohomologous cocycles $\sigma_k$ induce isomorphic twisted algebras, there can be isomorphic twisted $C^\ast$-algebras that are induced from non-cohomologous cocycles. Thus, the main question about the structures we just introduced is the following:

\begin{question}\label{Q:IsoClasses}
How many non-isomorphic twisted group $C^*$-algebras $\A_{\sigma_k}=C^*(P_k,\sigma_k)$ arise from the cohomology classes of $\sigma_k$?
\end{question}

This question will be fully settled in Theorem \ref{theo: groupoid realization}.

\subsection{The standard $k$-generalized Pauli group}\label{sec: pauli} This is the example we understand the most. For example, we can explicitly characterize the isomorphism class of its algebra of observables. For $k=0,1,\dots, n$, let $
    \beta_k \colon \Sigma_k \times \widetilde \Sigma_{n-k}\to \{0,1\}$
   be the dual cellular incidence pairing defined by
\begin{equation}\label{eq: pairing}
    \beta_k(c,\tilde c)
\;=\;
\begin{cases}
1,& \text{if $c$ and $\tilde c$ are dual cells},\\[0.2cm]
0,& \text{otherwise},
\end{cases}
\end{equation} 
and consider the 2-cocycle $\gamma_k\colon P_k\times P_k\to\mathbb T$
    defined by
    \begin{equation}\label{eq: pauli cocyclo}
            \gamma_k\big((f,\eta),(g,\zeta)\big)
    \;=\;
    \prod_{c,\tilde c}
    \langle f(c),\zeta(\tilde c)\rangle^{\beta_k(c,\tilde c)},
    \end{equation}
    where $
    \langle\cdot,\cdot\rangle:
    G\times\widehat G
    \to
    \mathbb T$
    is the canonical Pontryagin pairing. Recall that the elements $(f,\eta)\in P_k$ are compactly supported functions $f\colon \Sigma_k\to G$ and $\eta\colon \widetilde{ \Sigma}_{n-k}\to \widehat{G}$. Therefore, the product is finite and $\gamma_k$ is a well-defined 2-cocycle that determines a cohomology class $
    [\gamma_k]\in H^2(P_k,\mathbb T),$
    and a central extension
    \[
    1\longrightarrow \mathbb T
    \longrightarrow \mathcal P_{\gamma_k}
    \longrightarrow P_k
    \longrightarrow 1.
    \]
    We call $\mathcal P_{\gamma_k}$ the standard $k$-generalized Pauli group. 

    \begin{proposition}
        $\A_{\gamma_k}$ is naturally isomorphic to the UHF algebra $ {\rm M}_{|G|}^{\otimes \Sigma_k}$.
    \end{proposition}
\begin{proof}
Let $\Sigma_k^N$ be a filtration of $\Sigma_k$ by finite subsets, and let $\widetilde \Sigma_{n-k}^N$ be the corresponding  dual filtration of $\widetilde{\Sigma}_{n-k}$ such that $|\Sigma_k^N|=|\widetilde \Sigma_{n-k}^N|$. Consider the finite subgroup of $P_k$
$$
P_k^N
\;=\;
\bigoplus_{\Sigma_k^N}G
\oplus
\bigoplus_{\widetilde\Sigma_{n-k}^N}\widehat G
\;\simeq\;
\bigoplus_{\Sigma_k^N}(G\oplus\widehat G).
$$
consisting of functions supported on $\Sigma_k^N$ and $\widetilde \Sigma_{n-k}^N$. By the definition of $\gamma_k$, its restriction to $P_k^N$ is non-degenerate. Hence $
C^*(P_k^N,\gamma_k|_{P_k^N\times P_k^N})
\simeq
{\rm M}_{|G|}^{\otimes \Sigma_k^N}.$ Since $P_k=\bigcup_N P_k^N$, taking the inductive limit gives $
\A_{\gamma_k}
\simeq
{\rm M}_{|G|}^{\otimes\Sigma_k}.
$
\end{proof}

\begin{remark}
    The above proof relies heavily on the strict local character of the pairing $\gamma_k$. It may {\it fail} for the generic pairings \eqref{Eq:PairingForm} because, due to the non-local character of $\omega_k$, non-degeneracy will not be preserved when restricting to finite subgroups supported by finite subsets of the cellular complexes.  Thus, different cocycles may induce different notions of locality and different filtrations of the observable algebra, even when the underlying $C^*$-algebras are isomorphic, as we shall see in Theorem~\ref{theo: groupoid realization}.   \hfill $\blacktriangleleft$
\end{remark}

    It should be mentioned that when $G=\mathbb Z_2$, this construction recovers the basic Pauli group and the algebra of observables is the quantum spin-$\frac{1}{2}$ algebra over the $k$-cells of $\Sigma$.

\subsection{Stabilizer codes by $C^\ast$-methods} We denote by $\mathtt S(\A_{\sigma_k})$ the state space of the algebra of observables, endowed with its natural weak-$\ast$ topology, and introduce the $C^\ast$-version of stabilizer codes:

\begin{definition}
    A {\it code} is a non-empty weak-$*$ closed convex subset $\mathtt C \subseteq \mathtt S(\A_{\sigma_k}).$ The {\it stabilizer} of a code $\mathtt C$ is the following subset of $\mathcal P_{\sigma_k}$:
    $$
    \mathscr S_\mathtt C
    \;:=\;
    \big\{
    U\in \mathcal P_{\sigma_k}
    \mid
    \mathtt s(U)=1
    \text{ for all }
    \mathtt s\in \mathtt C
    \big\}.$$
    A stabilizer code is the pair $(\mathtt C, \mathscr S_\mathtt C)$.
\end{definition}

Stabilizer codes can be effectively characterized using the observation that a central extension endows $P_k$ with the symplectic structure induced by the alternating bicharacter $\omega_k$. Throughout, we assume that this commutator pairing is non-degenerated. A subgroup $L\subseteq P_k$ is called \emph{isotropic} with respect to $\omega_k$ if
    $$
    \omega_k(l,l')
    \;=\;
    1, 
    \qquad
    \forall \, l,l'\in L.$$
    The symplectic complement of a subset $S\subseteq P_k$ is  the set defined as
  \begin{equation}\label{eq: isotropic complement}
        S^\perp
    \;:=\;
    \big\{
    p\in P_k
    \mid 
    \omega_k(p,s)=1
    \text{ for all } s \in S
    \big\}
  \end{equation}
  An isotropic subgroup $L\subseteq P_k$ is called \emph{Lagrangian} if $
    L=L^\perp.$ 
The following well-known Lemma will be used throughout this work.
      \begin{lemma}\label{Lem:SwithU}
        If $U$ is a unitary element in a unital $C^*$-algebra $\mathcal A$ and $\SM$ is a state of $\mathcal A$ such that $|\SM(U)|=1$, then
        $$\SM(UA)\;=\; \SM(U)\SM(A)\;=\;\SM(AU),\qquad \forall\; A\in \mathcal A.$$
    \end{lemma}

The above fact allows us to clarify the structure of a stabilizer code:
    
    \begin{proposition}\label{prop: abelian}
    The stabilizer  $\mathscr S_\mathtt C$ is an abelian subgroup of $\mathcal P_{\sigma_k}$. Moreover, its projection $L$ in $P_k$ is an isotropic subgroup and $L \simeq \mathscr S_\mathtt C$.
    \end{proposition}

    \begin{proof}
    Let $z_p U_p$ and $z_qU_q$ be from $\mathscr S_\mathtt C$. Our first task is to prove that $z_p z_q U_{p}U_{q}$ also belongs to $\mathscr S_\mathtt C$. From definition of $\mathscr S_\mathtt C$, $\mathtt s(z_pU_p)=\mathtt s(z_qU_q)=1$ for $\mathtt{s}\in \mathtt{C}.$ Since $z_pU_p$ is unitary, Lemma \ref{Lem:SwithU} assures us that
    $$
    \mathtt s\big((z_pU_p)(z_qU_q) \big)\;=\;\mathtt s\big((z_qU_q)(z_pU_p) \big)\;=\;1,$$
    and the first task is completed. Furthermore, since $U_qU_p=\omega_k(p,q) U_pU_q$, it must be that $\omega(p,q)=1$ and, as such, the elements $z_pU_p$ and $z_q U_q$ of $\mathcal P_{\sigma_k}$ commute. Lastly,
    $$
    \mathtt s((z_pU_p)^{-1}) \;= \;\mathtt s(\bar z_pU_p^*)\; =\; \overline{\mathtt s(z_p U_p)} \;=\; 1.$$
    Therefore $\mathscr S_\mathtt C$ is indeed an abelian subgroup of $\mathcal P_{\sigma_k}$. It is also clear now that its projection $L$ is an isotropic subgroup of $P_k$. For the last statement, we observe that the lift of any $l \in L$ is unique in $\mathscr S_\mathtt C$. Indeed, if $z_l U_l$ and $z'_l U_l$ both sit in $\mathscr S_\mathtt C$, then, by taking a state from $\mathtt C$, we have
    \[
    z_l \mathtt s(U_l) \;=\; z'_l \mathtt s(U_l) \;=\; 1
    \]
    which proves the point. Thus, the projection $\mathscr S_\mathtt C \to L$ is a bijective group morphism.
    \end{proof}

 \begin{remark}
        The equation $\SM(U)=1$ is a generalization of the standard stabilizer constraint briefly discussed in our introductory remarks. Indeed, in the GNS representation $(\pi,\mathcal{H},\psi_\SM)$ of $\SM$, one obtains the eigenvalue equation
\begin{equation}\label{eq: violations}
   \pi_\SM(U)\psi_\SM\;=\; \psi_\SM,\qquad \forall \,U\in \mathscr{S}_\mathtt C, \; \SM\in \mathtt C. 
\end{equation}
 \end{remark}

In the opposite direction, we can associate a code $\mathtt C_{L}$ to each isotropic subgroup $L \subset P_k$. Indeed, observe the following:

\begin{proposition}\label{Prop:SvsL}
Any abelian subgroup $\mathscr S \subset \mathcal P_{\sigma_k}$ projects onto an isotropic subgroup $L \subset P_k$. Reciprocally, any isotropic subgroup $L \subset P_k$ accepts an embedding as an abelian subgroup $\mathscr S_L$ of $\mathcal P_{\sigma_k}$.
\end{proposition}

\begin{proof}
The direct statement is obvious. For the reciprocal one, since $U_l U_{l'} = U_{l'}U_l$ and $U_l U_{l'} =\sigma_k(l,l')U_{ll'}$ for any $l,l' \in L$, the 2-cocycle must be symmetric $\sigma_k(l,l')=\sigma_k(l',l)$. This assures us that $\sigma_k$ restricted to $L$ is trivial and, as such, there is a coboundary $\lambda: L \to \mathbb T$ such that $\sigma_k(l,l') =\lambda(l)\lambda(l')/\lambda(ll')$. Then the assignment $l \mapsto (\overline{\lambda(l)},l)$ supplies the desired embedding.
\end{proof}

Now, if $L\subset P_k$ is isotropic, then one can define a code $\mathtt C_{L}$ by setting
\begin{equation}\label{eq: code lagrangian}
    \mathtt C_{L}\;=\;\big\{ \SM\in \mathtt S(\A_{\sigma_k})\mid \SM(U)= 1,\;\forall\,U\in \mathscr S_L\big\}
\end{equation}
Note that $\mathtt C_{L}$ is always a non-empty closed convex subset of $\mathtt S(\A_{\sigma_k})$.

\begin{proposition}
    The relation between $L$ and $\mathtt C_{L}$ is one-on-one if we restrict to Lagrangian subgroups $L \subseteq P_k$.
\end{proposition}

\begin{proof}
    The stabilizer of $\mathtt C_{L}$ contains $\mathscr S_L$. Thus, its projection is an isotropic subgroup of $P_k$ which must include $L$ and, as such, when $L$ is Lagrangian, the two must coincide.
\end{proof}

\begin{proposition}\label{prop: code generators}
If a subset $\bm{\mathrm L}$ generates $\mathscr S_L$ as a group, then
$$
\mathtt C_{L}
\;=\;
\big\{\SM\in \mathtt S(\A_{\sigma_k})\mid 
\SM(U)=1,\ \forall\, U \in\bm{\mathrm L}\big\}.$$
Consequently, the stabilizer code depends only on $\mathscr S_L$ and not on the choice of generators.
\end{proposition}

\begin{definition}
    The \emph{stabilizer algebra} for a code $\mathtt{C}$ is the sub-$C^*$-algebra $
    \A_{\mathtt C}
    \subseteq
    \A_{\sigma_k}$ generated by  its stabilizer $\mathscr S_\mathtt C$.
\end{definition}

    As an immediate consequence of Proposition \ref{prop: abelian} one  gets the following:
    
\begin{proposition}
    $\A_{\mathtt C}$ coincides with the group $C^\ast$-algebra $C^\ast \mathscr S_{\mathtt C}.$ 
    
%     In particular, if $\mathtt C=\mathtt C_{L}$ for an isotropic subgroup $L\subseteq P_k$, then $
% \A_L\equiv \A_{\mathtt C_{L}}\simeq C^*L.$
\end{proposition}
    
In the following and throughout, we extend the 1-cochain $\lambda$ from the proof of Proposition~\ref{Prop:SvsL} to the whole $P_k$ by assigning trivial values over $P_k \setminus L$, and we replace the generators $U_p$ by $\overline{\lambda(p)} U_p$, for which we use the same symbols. By doing so, the isomorphism $\mathscr S_c \simeq L$ is implemented by $l \mapsto U_l$. The new generators satisfy the same relations \eqref{Eq:Relations}, but with the substitution
\[
\sigma_k \;\mapsto\; \sigma_k(p,q)\frac{\lambda(pq)}{\lambda(p)\lambda(q)}.
\]
Note that this new 2-cocycle has the property that $\sigma_k =1$ when restricted to $L \times L$. This, of course, does not change the isomorphism class of the algebra of physical observables. It just produces a new presentation of it that is conveniently adapted to a given isotropic subgroup $L \subseteq P_k$. Consequently, the unitaries $\{U_l\}_{l\in L}$ form an ordinary
unitary representation of $L$, and
$$
\A_L
\;=\;
C^*\{U_l\mid l\in L\}
\;\simeq\;
C^*L
\;\simeq\;
C(\widehat L).
$$
where $\widehat L$ is the Pontryagin dual of $L.$ In particular, for a code $\mathtt C$ one has that
 $$\A_{\mathtt C}\;\simeq\; C^*\mathscr{S}_\mathtt C\;\simeq\;C(\widehat{\mathscr{S}}_\mathtt C)$$  We shall refer to $\widehat{\mathscr{S}}_\mathtt C$ as the configuration space of the code $\mathtt C,$ since its
elements encode all possible syndromes relative to $\mathtt C$, namely,
all possible violations of the stabilizer constraints.  Using the Gelfand identification $
    \A_{\mathtt C}
    \simeq
    C(\widehat{\mathscr{S}}_\mathtt C),$
    each generator $U\in \A_{\mathtt C}$ defines a function
    $
    U\colon \widehat{\mathscr{S}}_\mathtt C\to \mathbb C.$ This gives rise to the syndrome map
    \begin{equation}
          \mathfrak s
    \colon
    \mathscr S_\mathtt C\times \widehat{\mathscr S}_\mathtt C
    \to
    \T,
    \qquad
    \mathfrak s(U,\chi)
    \;=\;
  U(\chi).
    \end{equation}
\section{Isomorphism class of the observable algebra}\label{Sec:IsoClass}

This section gives a definitive answer to the fundamental question \ref{Q:IsoClasses} for Lagrangian codes. In the process, it develops a natural framework for dealing with stabilizer codes, as well as many useful tools of analysis.

    \subsection{The $C^*$-diagonal of a Lagrangian}
We start by recalling the general definition of a $C^*$-diagonal based on \cite{Kum, Ren1}
\begin{definition}
Let $\A$ be an abstract separable $C^*$-algebra.    A sub-$C^\ast$-algebra $\Bb \subset \A$ is said to have the unique extension property (UEP) if any pure state over $\Bb$ uniquely extends over $\A$, in which case the extension is automatically a pure state.
\end{definition} 
The normalizer set of a $C^\ast$-inclusion $\Bb \subset \A$ is defined as
\[
\mathscr{N}(\mathcal{B})
\;:=\;
\big\{
V \in \A
\;\mid\;
V B V^* \in \mathcal{B}
\ \text{and}\
V^* B V \in \mathcal{B}
\ \text{for all } B \in \mathcal{B}
\big\}.
\]
It follows from the very definition that $\mathcal{B}\subset \mathscr{N}(\Bb)$. The sub-$C^\ast$-algebra $\Bb$ is said to be regular if the linear span of $\mathscr{N}(\Bb)$ is dense in $\A$, \ie $C^*( \mathscr{N}(\Bb))=\A.$  Furthermore:

\begin{definition}[\cite{Kum}]\label{def: diagonal} An abelian sub-$C^\ast$-algebra $\Bb$ of $\A$ containing an approximate identity for $\A$ is called a $C^\ast$-diagonal if it is regular and has the UEP.
\end{definition}
\begin{remark}
    Because of the separability of $\A$, any $C^*$-diagonal $\mathcal{B}$ of $\A$ is a maximal abelian subalgebra (masa) \cite[Corollary 2.7]{AB2}. Moreover, it comes with a unique faithful conditional expectation $E\colon \A\to \mathcal{B}$ (see \cite[Theorem 3.4]{And1}). \hfill $\blacktriangleleft$
\end{remark}

We have seen in Proposition \ref{prop: abelian}, for a code $\mathtt C$, the image of $\mathscr S_\mathtt C$ in $P_k$ is always an isotropic subgroup. The situation becomes particularly interesting when this subgroup is Lagrangian. In that case, the associated stabilizer algebra $\A_{\mathtt C}$ gives rise to a $C^*$-diagonal of $\A_{\sigma_k}$, providing a powerful tool for analyzing the structure of error correction code models. This is the content of the following fundamental result.

    \begin{theorem}\label{teo: diagonal}
        Let $L\subset P_k$ be a Lagrangian subgroup and $\mathtt C_{L}$ its associated code according to \eqref{eq: code lagrangian}. Then the commutative $C^*$-algebra $$\A_L\;\equiv\;\A_{\mathtt C_{L}}\;=\;C^*\{ U_l\mid l\in L\}$$ is a $C^*$-diagonal of $\A_{\sigma_k}$.
    \end{theorem}
    
    \begin{proof}
    Clearly $\A_L\simeq C^*L$ is commutative and its Gelfand spectrum coincides with the Pontryagin dual $\widehat{L}$ of $L.$ Thus any pure state $\SM$ of $\A_L$ can be identified with a character $\chi$ of $L.$ Such a state on the generators satisfies
    \begin{equation}\label{eq: normalization}
        \SM(U_l)\;=\; \chi(l) \in \mathbb T\;\implies\; |\SM(U_l)|\;=\;1\quad \forall l\in L
    \end{equation}
    On the other hand, any state on $\A_{\sigma_k}$ is completely determined by its values on the $U_p's$ with $p\in P_k$. Moreover, for $p\in P_k$ one has 
    $$U_l^\ast U_pU_l \;=\; \omega_k(l,p)U_p$$
    The Lagrangian condition implies that if $p\notin L$ there exists $l\in L$ for which $\omega_k(l,p)\neq 1$. Let ${\SM}_\chi$ be any state extending $\SM$ to $\A_{\sigma_k}$. By \eqref{eq: normalization} and Lemma \ref{Lem:SwithU}, one gets
    $${\mathtt{s}}_\chi(U_p)\;=\;{\mathtt{s}}_\chi(U_l^\ast U_pU_l)\;=\;\omega_k(l,p){\mathtt{s}}_\chi(U_p)$$
    The above implies that
    \begin{equation}\label{Eq:Lift}
    {\mathtt{s}}_\chi(U_p)\;=\;
    \begin{cases}
    \chi(p), & p\in L,\\
    0, & p\notin L.
    \end{cases}
    \end{equation}
    Consequently, this determines ${\mathtt{s}}_\chi$ uniquely and, as such, any pure state on $\A_L$ has a unique extension over $\A_{\sigma_k}$, and this unique extension is necessarily pure.
    
   The remaining step is to verify that $\A_L$ is regular. However, this is a direct consequence of the fact that $\{U_p\}_{p\in P_k}$ are normalizers of $\A_L$, \ie 
    $$U_p^\ast U_lU_p \;=\; \omega_k(p,l)U_l$$
     This concludes the proof.
    \end{proof}

\begin{corollary}\label{coro: unique FF}
    If $L$ is a Lagrangian subgroup of $P_k$, then its code $\mathtt C_{L}$ consists of a single pure state, $\mathtt C_{L} =\{\mathtt s_0\}$, where $\SM_0\equiv \SM_{\chi_0}$ with $\chi_0$ the unit element of $\widehat L$.
\end{corollary}

\begin{proof}
    Any state $\mathtt s$ from $\mathtt C_{L}$ has the property that $\mathtt s(U_l) = 1$ for all $l \in L$. Thus, when restricted to $C^\ast L$, it takes the same values as $\chi_0$ on a dense subset, hence $\mathtt s|_{C^\ast L} = \chi_0$. Thus, given the UEP of $\A_L$, there can be one and only one state inside $\mathtt C_{L}$, which is necessarily pure and given by $\mathtt s_{\chi_0}$.
\end{proof}

\subsection{The logical algebra of an isotropic subgroup}\label{sec: logical algebra}

The previous result shows that Lagrangian subgroups determine codes with no non-trivial logical degrees of freedom.  In contrast, a non-Lagrangian isotropic subgroup $L \subset P_k$ has a non-trivial code $\mathtt C_{L}$ and, furthermore, accepts a non-trivial logical algebra $\A_L^{\rm log}$ such that $\mathtt C_{L}$ coincides with the state space of $\A_L^{\rm log}$. As in the standard finite-dimensional version of stabilizer code theory, briefly discussed in the introduction, we now have the means to define the logical degrees of freedom. We prove the statement using guidance from \cite[Section 3.3]{ADL2025}. 

\begin{definition}
   For an isotropic subgroup $L\subset P_k$ consider the unital sub-$C^*$-algebra
\begin{equation}
    \mathcal{B}_L\;=\; C^*\big\{ U_l\mid l\in L^\bot\big\}\;\subset \;\A_{\sigma_k}
\end{equation}
and its closed two-sided ideal $\mathcal J_L$ generated by the elements $U_l-\mathbf1$, with $l\in L\subset L^\perp$. Then the logical algebra associated with the isotropic subgroup $L$ is the quotient $C^\ast$-algebra
\begin{equation}
    \A^{\rm log}_L\;:=\; \mathcal{B}_L/\mathcal{J}_L \;\simeq\; C^\ast (L^\perp/L, \widetilde{\sigma}_k).
\end{equation} 
where $\widetilde{\sigma}_k$ is the normalized $2$-cocycle on $L^\perp/L$ induced by a section
$r\colon L^\perp/L\to L^\perp$, with $r(e)=e$, and explicitly described by
$$
    \widetilde{\sigma}_k(x,y)
    \;=\;
    \frac{\sigma_k(r(x),r(y))}
    {\sigma_k\big(r(x)r(y)r(xy)^{-1},r(xy)\big)},\qquad \forall x,y\in L^\bot/L.
$$
Its cohomology class is independent of the choice of section. 
\end{definition}

The following is a generalization of \cite[Theorem 3.14]{ADL2025}:

\begin{theorem}\label{teo: GSFF}
    Let $L\subset P_k$ be isotropic. Then $\mathtt C_L$ is affinely homeomorphic to the state space $\mathtt S(\A^{\rm log}_L)$ of the logical algebra.
\end{theorem}
\begin{proof}
For each $\varphi \in \widehat{P_k/L^\perp}$, the rule $\varphi \cdot U_p:=\varphi([p])U_p$ defines a point-norm continuous action of the compact abelian group
$\widehat{P_k/L^\perp}$ on $\A_{\sigma_k}$. Averaging this action with
respect to its normalized Haar measure $\mu$ gives a conditional
expectation $E_{L^\bot}\colon \A_{\sigma_k} \to \mathcal{B}_L$,
$$E_{L^\bot}(U_p)\;=\;\int_{\widehat{P_k/L^\bot}}\varphi([p])U_p{\rm d}\mu(\varphi)\;=\;
\begin{cases}
U_p,&p\in L^\perp,\\
0,&p\notin L^\perp.
\end{cases}$$
Now let $\SM\in \mathtt C_{L}$. The same argument used in the proof of
Theorem~\ref{teo: diagonal} shows that $\SM(U_p)=0$ for all $p\notin L^\bot.$ Thus, $\SM$ is completely determined by its restriction to $
\mathcal B_L$ and $\SM=\SM\circ E_{L^\bot}$. On the other hand, since $\SM(U_l)=1$ and the $U_l's$ are central in
$\mathcal B_L$, Lemma \ref{Lem:SwithU} implies that $
\SM\bigl(B(U_l-\mathbf1)C\bigr)=0$
for all $B,C\in\mathcal B_L$ and $l\in L$. Thus,
$\SM|_{\mathcal B_L}$ vanishes on $\mathcal J_L$ and, as such, it induces a unique
state $\rho_\SM$ on $\A^{\mathrm{log}}_L$ satisfying
$$
\SM\;=\;\rho_\SM\circ q\circ E_{L^\perp},$$
where
$q\colon\mathcal B_L\to\A^{\mathrm{log}}_L$
is the quotient map.
Conversely, let $\rho$ be a state on $\A^{\mathrm{log}}_L$. Then $\SM_\rho=\rho\circ q\circ  E_{L^\bot}\in \mathtt C_{L}$ and, furthermore, the assignments
$\SM\mapsto\rho_\SM$ and $\rho\mapsto\SM_\rho$ are mutually inverse, affine, and weak$^*$ continuous since they are obtained
by composing restriction and pullback maps induced by bounded linear maps between $C^*$-algebras. 
\end{proof}

   \subsection{The groupoid of a Lagrangian}\label{sec: weyl groupoid}
    Since $\omega_k$ is a bicharacter, there is a natural left action of the abelian group $P_k/L$ on $\A_L$ determined by
    \begin{equation}\label{Eq:Action}
        [p] \triangleright U_l\;=\;U_p^\ast U_l U_p \;=\;\omega_k(p,l)U_l, \quad l \in L,
    \end{equation}
    where we note that the evaluation of $\omega_k$ depends only on the class of $p$ in $P_k/L$. By Gelfand duality, the action of $P_k/L$ on $\A_L$ induces a continuous action on its Gelfand spectrum $\widehat{L}$, which, by an abuse of notation, we shall denote by the same symbol. More specifically, under the identification $\A_L \simeq C(\widehat L)$, the unitary generators $U_l$ become the functions $U_l(\chi) = \chi(l)$ for $\chi \in \widehat L$, and \eqref{Eq:Action} reads
    \begin{equation}\label{Eq:ChiP}
    (U_p^\ast U_l U_p)(\chi) \;= \;U_l(\chi_{[p]} \chi), \quad \chi_{[p]}\; :=\; \omega_k(p,\text{--}) \in \widehat L.
    \end{equation}
    As such, the induced action of $P_k/L$ on $\widehat L$ is
    $$
    [p] \triangleright \chi \;:=\; \chi_{[p]} \chi. $$
    
\begin{definition} The groupoid associated to the Lagrangian $L\subset P_k$ is the transformation groupoid $$\G_L\;:=\;\widehat{L}\rtimes (P_k/L).$$
\end{definition}

\begin{remark}\label{Re:LGroupoid}
   It is worth pointing out that the elements of this groupoid are triples 
    \begin{equation}\label{Eq:Triples}
    ([p]\triangleright \chi,[p],\chi) \in \widehat L \times P_k/L \times \widehat L
    \end{equation}
    and $([p]\triangleright \chi,[p],\chi)$ is composable with $([p']\triangleright \chi',[p'],\chi')$ iff $\chi'=[p] \triangleright \chi$, in which case
    $$
    \big([p']\triangleright ([p]\triangleright \chi),\,[p'],\, [p] \triangleright \chi\big)\big([p]\triangleright \chi,\,[p],\,\chi\big) \;=\; \big([p'p]\triangleright \chi,[p'p],\chi\big).$$
    The topology of $\G_L$ is inherited from $\widehat L \times P_k/L \times \widehat L$. Note that we could have written the triples in the opposite order but, when the (abelian) group action is written as from the left, there is only one natural order of the triples, the one in \eqref{Eq:Triples} \cite{WilliamsBook}. Also, we can drop $[p]\triangleright \chi$ in \eqref{Eq:Triples} and work with tuples $([p],\chi)$, although the calculations are always easier with triples. Nevertheless, we will do so to simplify the notation. \hfill $\blacktriangleleft$
\end{remark}

An important property of this groupoid is that it is AF. AF-groupoids were introduced in \cite{RenaultBook}, but the concept evolved over time. In particular, two additional definitions were introduced in \cite{FarsiMJM2019} and \cite{GTS}, which eventually were all proven in \cite{ClarkPAMS2025} to describe the same class of groupoids. For us, it will be convenient to subscribe to the definition given in \cite{GTS}:

\begin{definition}
   Let $R$ be an equivalence relation on $X$ equipped with a topology that makes it an \'etale groupoid. One says  that $R$ is a compact \'etale equivalence
relation (CEER) if $R \setminus X$ is a compact subset of $R$. One says that $R$ is an AF-equivalence
relation if $X=R^{(0)}$ is locally compact, Hausdorff, second countable, totally disconnected, and if
$R$ is the inductive limit of a sequence of CEERs.
\end{definition}
    
    \begin{lemma}\label{lem: AF}
        $\G_L$ is an AF-groupoid.
    \end{lemma}
   \begin{proof}
   Notice that every finitely generated subgroup of $P_k$ is finite. Hence $P_k$ is locally finite, and consequently so is its quotient $P_k/L$. On the other hand, $L$ is a countable infinite locally finite discrete abelian group, so its Pontryagin dual $\widehat L$ is a compact metrizable totally disconnected abelian group. Moreover, $\widehat L$ has no isolated points, for otherwise it would be a compact discrete group and therefore finite, contradicting the infinitude of $L$. As a conclusion, $\widehat L$ is a Cantor space.

    Since $P_k/L$ is a locally finite group and $\widehat{L}$ is Cantor space, by \cite[Theorem~3.8]{GTS} is enough to verify that the action of $P_k/L$ on $\widehat L$ is free and minimal. Indeed, the map
$\Theta\colon P_k/L\to \widehat L,$ $[p] \mapsto \chi_{[p]}$ is a well-defined injective group homomorphism, because $\omega_k$ is a bicharacter and $L$ is Lagrangian, and the action of $P_k/L$ on $\widehat L$ is precisely the action by translations of the subgroup $\Theta(P_k/L)\subseteq \widehat L$. Consequently, freeness is a direct consequence of injectivity of $\Theta$. Showing minimality is equivalent to proving that the image of $\Theta$ is dense in $\widehat L$. Assume, toward a contradiction, that $\overline{\Theta(P_k/L)}\neq \widehat L$.
Then,  there exists a nontrivial element
$l_0\in L$ such that $
\eta(l_0)=1,$ for all  $\eta\in \overline{\Theta(P_k/L)}.$ In particular,
$$
\omega_k(p,l_0)\;=\;\Theta([p])(l_0)\;=\;1,
\qquad
\forall p\in P_k.$$
The non-degeneracy of $\omega_k$ implies that
$l_0$ is the identity of $P_k$
contradicting the choice of $l_0$. Hence, $
\overline{\Theta(P_k/L)}=\widehat L.$ This completes the proof.
\end{proof}

% We can actually give an explicit filtration of $\G_L$ by CEERs, which we will use later for certain purposes.

% \begin{proposition}\label{Prop:CEER_Filtration}
%     Let $(F^n)_{n\in \N}$ be an increasing sequence of finite subgroups of $P_k/L$ such that $P_k/L=\bigcup F^n$. Then $\G_L^n:=\widehat L \rtimes F^n$ arrange in an inductive tower of CEERs whose limit is $\G_L$.
% \end{proposition}

We briefly recall the Krieger invariant of an AF-groupoid $(\G, \G^0,\mathfrak s,\mathfrak r)$ based on \cite{Krieger, Mat}.  Let $C_c(\G^0,\Z)$ be the space of compactly supported functions and let $\partial\colon C_c(\G,\Z)\to C_c(\G^0,\Z)$ defined as
$$(\partial f)(u)\;=\;\sum_{\mathfrak s(g)=u}f(g)-\sum_{\mathfrak r(g)=u}f(g).
$$
The zeroth homology group  of $\G$ is the abelian group
$
H_0(\G)
=
C_c(\G^{0},\mathbb Z)\big/\operatorname{Im}\partial
$. It carries the natural positive cone $H_0(\G)^+$ induced by the non-negative functions in $C_c(\G^{0},\mathbb Z)$. When $\G^{0}$ is compact, the constant function $1_{\G^{0}}$ determines a distinguished order unit. The ordered triple $
\big(H_0(\G),H_0(\G)^+,[1_{\G^{0}}]\big)
$
is called the Krieger invariant of $\G$.
\begin{theorem}[\cite{Krieger, Mat}]
    Two AF-groupoids are isomorphic if and only if their Krieger invariants are isomorphic as scaled ordered abelian groups.
\end{theorem}

Now we are in a position to compute the Krieger invariant for $\G_L:$
\begin{proposition}\label{prop: Krieger}
    The Krieger's invariant $
\big(H_0(\G_L),H_0(\G_L)^+,[1_{\widehat L}]\big)$ of $\G_L$ is given by
$$\big(\mathbb Z[1/n],\mathbb Z_+[1/n],1\big),
\qquad n=|G|.$$
\end{proposition}
\begin{proof}
Let $\nu$ be the normalized Haar measure on $\widehat L$. Since the action
of $P_k/L$ on $\widehat L$ is by translations, $\nu$ is invariant. Hence
integration defines a well-defined positive homomorphism
$$
\Phi\colon H_0(\G_L)\to \mathbb R,
\qquad
\Phi([f])\;=\;\int_{\widehat L} f\,d\nu .$$
We first show that $
\operatorname{Im}(\Phi)\subseteq \mathbb Z[1/n].$
Indeed, the inclusion $L\subseteq P_k$ induces, by restriction of characters,
a continuous surjective homomorphism $
\rho\colon\widehat P_k\to \widehat L.$
By uniqueness of normalized Haar measure, 
$\rho_*(\mu)=\nu,$ where $\mu$ denotes the normalized Haar measure on $\widehat P_k$. One can explicitly describe $\widehat{P}_k$ as the Cantor set given by $$\widehat{P}_k\;\simeq\;\widehat{G}^{\Sigma_{k}}\times {G}^{\widetilde \Sigma_{n-k}}\;\simeq \; G^{\N}$$
By using the last identification, any $\chi\in \widehat P_k$ is a function $\chi\colon {\N}\to G$. Define a $m$-cylinder of $\widehat P_k$ by setting $$C(m,\chi)\;=\;\big\{ \chi'\in \widehat P_k\mid \chi|_{\{1,\dots,m\}}=\chi'|_{\{1,\dots,m\}}\big\}$$
Notice that $$\mu\big(C(m,\chi)\big)\;=\;\frac{1}{n^m}\in \Z[1/n],\qquad \forall m\in \N$$
and any clopen subset of $\widehat P$ has a measure contained in $\Z[1/n].$ 
In particular for each $C\subseteq \widehat L$ clopen, 
$\rho^{-1}(C)$ is also clopen in $\widehat P_k$ and one gets
$$
\nu(C)
\;=\;
\mu(\rho^{-1}(C))
\in \mathbb Z[1/n].
$$
Since $C(\widehat L,\mathbb Z)$ is generated by characteristic functions of
clopen subsets, it follows that
\[
\operatorname{Im}(\Phi)\;\subseteq \;\mathbb Z[1/n].
\]
To obtain the reverse inclusion, let $m\geq 1$. By Lemma \ref{lem: finite subgroup}
there exists a finite subgroup $F_m$ of $L$ such that
$|F_m|=|G|^m=n^m.$ It yields a continuous
surjective homomorphism $
r_m\colon\widehat L\to \widehat{F}_m$ by restriction.
Notice that
$
|\widehat{F}_m|=|F_m|=n^m.$ Therefore, for every $\chi\in \widehat{F}_m$, the fiber
$r_m^{-1}(\{\chi\})$ is a clopen subset of $\widehat L$. Since again the push-forward of Haar measure
under $r_m$ is the Haar measure on $\widehat{F}_m$, it follows that
$$
\nu\!\left(r_m^{-1}(\{\chi\})\right)
\;=\;
\frac{1}{|\widehat{F}_m|}
\;=\;
\frac1{n^m}.$$
Hence
$\frac1{n^m}\in \operatorname{Im}(\Phi)$
for every $m\geq 0$. Consequently, $
\mathbb Z[1/n]\;\subseteq\; \operatorname{Im}(\Phi).$
Since the reverse inclusion was established above, we conclude that $
\operatorname{Im}(\Phi)=\mathbb Z[1/n].$

Finally, by the same argument as in \cite[Lemma~4.4]{DE1}, the map
$\Phi$ is injective and order preserving. Therefore,
$$
(H_0(\G_L),H_0(\G_L)^+,[1_{\widehat L}])
\;\cong\;
(\mathbb Z[1/n],\mathbb Z_+[1/n],1)$$
and the statement follows.
\end{proof}

\subsection{Resolving the isomorphism class}\label{sec: iso class}

The computation of the Krieger invariant yields our first classification result:
\begin{corollary}
All Lagrangian subgroups of $P_k$ induce isomorphic AF-groupoids.
\end{corollary}

As immediate consequence, one gets the following model for the groupoid algebra.
\begin{proposition}\label{prop: trivial twist}
    Every twist on $\G_L$ is trivial and 
    $$C^*\G_L\;\simeq \;{\rm M}_{|G|}^{\otimes {\Sigma_k}}$$
\end{proposition}
\begin{proof}
    By Lemma \ref{lem: AF}, the groupoid $\G_L$ is AF. Thus, by
\cite[Proposition~4.6]{DE1}, the twist is trivial on AF-relations.  Finally, the isomorphism is a consequence of the Krieger's invariant in \ref{prop: Krieger}.
\end{proof}

\begin{theorem}\label{theo: groupoid realization}
$\G_L$ is the Weyl groupoid of the $C^*$-diagonal pair $(\A_L,\A_{\sigma_k})$. Consequently, 
$$\A_{\sigma_k}\;\simeq\; C^*\G_L\;\simeq\;{\rm M}_{|G|}^{\otimes {\Sigma_k}},$$
for any cocycle $\sigma_k$ such that $\omega_k$ is non-degenerate.
\end{theorem}

\begin{proof}
Choose a normalized set-theoretic section $
s\colon P_k/L\to P_k$ such that $s([1])=1.$ For $x\in P_k/L $ define
$V_x:=U_{s(x)}.$ Conjugations by $V_x^*$ define automorphisms of $\A_L$
$$
V_x^*U_lV_x
\;=\;
{\omega_k(s(x),l)}U_l,
$$
which induce an action $\alpha\colon P_k/L\to {\rm Aut}(\A_L)$ that coincides with \eqref{Eq:Action}. The extension induced by the quotient map $P_k \to P_k/L$ may not be trivial, so let $\kappa$ be its 2-cocycle, \ie 
\[
\kappa(x,y)
\;:=\;
s(x)s(y)s(xy)^{-1}\in L, \quad x,y\in P_k/L.
\]
Using the twisted multiplication relations in $\A_{\sigma_k}$, we
obtain
$$
V_x^*V_y^*
\;=\;
\nu(x,y)V_{xy}^*,$$
where
\begin{equation}\label{eq: twisted multiplier}
\nu(x,y)\;=\;
\overline{\sigma_k\bigl(s(y),s(x)\bigr)}
\,\sigma_k\bigl(s(xy),\kappa(x,y)\bigr)
\,U_{\kappa(x,y)}^* \in\mathscr U(\A_L),
\end{equation}
where $\mathscr U(\A_L)$ stands for the set of unitaries of $\A_L.$ The associativity of the multiplication in $\A_{\sigma_k}$
implies that $(\alpha,\nu)$ is a twisted action of $P_k/L$ on $\A_L$. The inclusions
$$
\A_L\;\subseteq\;\A_{\sigma_k},
\qquad
x\longmapsto V_x^*,$$
therefore form a twisted covariant surjective representation of
$(\A_L,P_k/L,\alpha,\nu)$ which generates the entire $\A_{\sigma_k}$. Consequently,
there is a surjective $*$-homomorphism
$$
\Phi\colon \A_L\rtimes _{\alpha,\nu}(P_k/L)\to \A_{\sigma_k}, \quad \Phi(U_l)\;=\;U_l,
\quad
\Phi(v_x)\;=\;V_x^*.$$
By Gelfand duality, $\nu$ induces a twist $\tau_\nu$ on $\G_L$ for which
$$\A_L\rtimes _{\alpha,\nu}(P_k/L)\;\simeq\; C^*(\G_L,\tau_\nu)\;\simeq\;C^*\G_L$$
where $C^*(\G_L,\tau_\nu)$ stands for the twisted groupoid algebra and the last isomorphism is a consequence of that any twist on $\G_L$ is trivial by Proposition \ref{prop: trivial twist}. Thus, $\Phi$ is an isomorphism since $C^*\G_L$ is simple by Proposition \ref{prop: trivial twist}, and this completes the proof .
\end{proof}
\begin{remark}
Theorem~\ref{theo: groupoid realization} completely answers Question \ref{Q:IsoClasses}: all cocycles $\sigma_k$  with non-degenerate commutator pairing give rise to the same algebra of observables, up to isomorphism. However, different cocycles may induce different notions of locality and hence different filtrations of this algebra. \hfill $\blacktriangleleft$
\end{remark}   

\section{Dynamics induced from isotropic subgroups}

In this section, we develop a general framework for constructing
stabilizer models from isotropic subgroups of $P$. Since we want to exploit the locality and the natural filtration of the observable algebra induced by finite subsets of $\Sigma$, we specialize to the standard Pauli extension described in Section~\ref{sec: pauli}. We will also fix parameter $k$ once and for all and, as such, we will drop $\sigma$ and $k$ from our notation. 

\medskip 

Throughout, we will work exclusively with the following class of isotropic subspaces:

\begin{definition}\label{Def:AdmissibleL}
    An admissible isotropic subgroup $L\subset P$ is one that admits a set $\bm{\mathrm L}$ of generators such that
    $$
   \big |\{l \in \bm{\mathrm L} \; | \; \omega(p,l) \neq 1 \} \big| \;<\; \infty, \quad \forall\; p \in P.$$
    The set of admissible isotropic spaces will be denoted by $\mathcal L$.
\end{definition}

\begin{remark}
    Recalling the definition \eqref{Eq:Pk} of $P$, we see that, if we can find cellular cycles generating $L$ such that, for any compact subset $K$ of $\Sigma_k \times \widetilde \Sigma_{n-k}$, there is a finite number of those cellular cycles with support intersecting $K$, then $L$ belongs to $\mathcal L$. This is the case if, for example, the generating cellular cycles have compact supports with uniformly bounded diameters and form a locally finite family. This is the situation seen in all concrete models worked out by physicists (see section~\ref{Sec:Applications}).  \hfill $\blacktriangleleft$
\end{remark}

\subsection{Canonical models}
By construction, $U_l^{|G|}=1$ for each $l\in L$ since $l\mapsto U_l$ forms a unitary representation of $L$. As such, the spectrum of $U_l$ is finite and included in the set of $|G|$-th roots of unity. We will make use of the spectral projections onto the eigenvalue $1$,
\begin{equation}
  P_{l}\;=\; \frac{1}{|G|}\sum_{i=1}^{|G|} U_l^i.
\end{equation}
It is clear that these projections commute with each other. We define the formal Hamiltonian
\begin{equation}
    H_{\LM}\;=\;\sum_{l\in \bm{\mathrm L}} ({\bf 1}-P_l)
\end{equation}
where $\bm{\mathrm L}$ is a set of generators of $L$.  Observe that the expression for $H_{\LM}$ is generally formal, since $L$ may be infinitely generated. To make sense of the corresponding dynamics, we choose a
filtration of $\LM$ by increasing finite subsets
$
\LM_1\subseteq \LM_2\subseteq \cdots \subseteq \LM,$ with
$\LM=\bigcup_{n\in\mathbb N}\LM_n,$
and define the local Hamiltonians $$
H_{\LM_n}
\;=\;
\sum_{l\in \LM_n}(\mathbf 1-P_l).$$

\begin{proposition}\label{propo: dynamics}
Let $L \in \mathcal L$ and $\bm{\mathrm L}$ be a generating set as in \ref{Def:AdmissibleL}. Then the point-wise limit $\bm{\delta_{\rm L}}$ of the local derivations $\delta_{\LM_n}(A)=\ii[H_{\LM_n},A]$ defines a densely defined $*$-derivation $\delta_{\LM}$ on $\A$. More specifically, each $U_p$ for $p\in P$ belongs to the domain of $\delta_{\LM}$ and
    \begin{equation}
        \delta_{\LM}(U_p) \;=\; \ii  U_p \lambda_p, \quad \lambda_p \in \A_L,
    \end{equation}
    where, if  we use the isomorphism $\A_L\simeq C(\widehat L)$ and define the support of a character as 
    \begin{equation}\label{Eq:Support}
    {\rm supp}\, \chi \;: =\;\{ l\in \bm{\mathrm L}\mid \chi(l)\neq 1\},
    \end{equation}
    then
\begin{equation}\label{Eq:Lambda_P}
  \lambda_p(\chi)
\;=\;
\sum_{l\in{\rm supp}\, \chi_{[p]}}
\big(
\delta_1\big( \chi(l)\big)-\delta_1\big(\omega(p,l)\chi(l)\big)
\big),  
\end{equation}
with $\delta_1: \mathbb T \to \{0,1\}$ being the characteristic function of the subset $\{1\}$.
\end{proposition}

\begin{proof}
   Each ${\bm \delta}_{\LM_n}$ is, by construction, an inner $\ast$-derivation. We have
    \[
    [1-P_l,U_p] \;=\; U_p P_l - P_l U_p \;=\; \tfrac{1}{|G|}U_p\sum_{j=1}^{|G|} \big (U_l^j - \omega(p,l)^j U_l^j \big) .
    \]
    Passing to the $C(\widehat L)$ presentation of $\A_L$, we see that
    \[
    \tfrac{1}{|G|}\Big [\sum_{j=1}^{|G|} \big (U_l^j - \omega(p,l)^j U_l^j\big)\Big ](\chi)\;=\;\tfrac{1}{|G|} \sum_{j=1}^{|G|} \big (\chi(l)^j - \omega(p,l)^j \chi(l)^j \big),
    \]
    and, since both $\omega(p,l)$ and $\chi(l)$ are $|G|$-th roots of unity, the right side is just 
    \[
     \delta_1(\chi(l)) - \delta_1(\omega(p,l)\chi(l)).
    \]
    As a consequence
    \[
    \ii [H_{\LM_n},U_p] \;=\; \ii U_p \lambda_p^n, \quad \lambda_p^n(\chi) \;=\; \sum_{l\in \LM_n \cap \, {\rm supp}\, \chi_{[p]}}
\big(
\delta_1\big( \chi(l)\big)-\delta_1\big(\omega(p,l)\chi(l)\big)\big).
\]
Note that assumption $L \in \mathcal L$ assures us that each ${\rm supp}\, \chi_{[p]}$ is finite. Then the action stabilizes as we increase $n$, because $\LM_n$ will eventually cover the support of $\chi_{[p]}$, in which case $\ii [H_{\LM_n},U_p] = \ii U_p \lambda_p$. Hence, the limits of $\delta_{\LM_n}(U_p)$ exist inside $\A$ for each $p \in P$ and, since $\{U_p\}$ generates $\A$, all statements follow.
\end{proof}

\begin{remark}\label{Re:ConChi0}
    Note that $\lambda_p(\chi_0) = |{\rm supp}\, \chi_{[p]}|<\infty$.  \hfill $\blacktriangleleft$
\end{remark}

From the above calculation, we can actually see that $U_p$ are analytic vectors for $\delta_{\LM}$ and $\delta_{\LM}$ is conservative ({\it e.g.} $\delta_{\LM}$ passes the test of \cite[Proposition~3.2.22]{bratteli-robinson-97}). Then \cite[Theorem~3.2.50]{bratteli-robinson-97} assures us that $\delta_{\LM}$ generates a strongly continuous one-parameter group
of $*$-automorphisms 
$$
\mathbb R\ni t\to \alpha_{\LM}^t\in\mathrm{Aut}(\A), \quad \lim_{t \to 0} \frac{\alpha_{\LM}^t(A)-A}{t} \;=\; \delta_{\LM}(A), \quad \forall \, A \in {\rm Dom}(\delta_{\LM}),
$$
which we shall refer to as the dynamics associated with the generating set $\LM$ of the stabilizer
group $L$.

\subsection{Ground-states: General considerations} 

The algebraic characterization of the ground-states for dynamics $\alpha_{\LM}$ is given by the following definition.
\begin{definition}
    A state $\SM$ of $\A$ is a ground-state for $\alpha_{\LM}$ if 
    $$-\ii \SM\big(A^\ast \delta_{\LM}(A)\big)\;\geq \; 0, \quad \forall\, A \in {\rm Dom}(\delta_{\LM}).$$
   We denote the set of ground-states of $\alpha_{\LM}$ by $\mathfrak G_{\LM}$.
\end{definition}

Among the ground-states, there exists a special subclass of states introduced first in \cite{AffleckCMP1988}. See also \cite{DET1}.

\begin{definition}
    A ground-state $\SM$ of $\alpha_{\LM}$ is said to be \emph{frustration-free} if 
    $$
    \SM(P_l)\;=\;1, \quad \forall\, l\in \bm{\mathrm L}.$$
    The set of frustration-free ground-states will be denoted by $\mathfrak G^F_{\LM}$.
\end{definition}
There is a good chance that the frustration-free states coincide with the KMS$_\infty$ states introduced in \cite{CM}, as the states that are weak-$\ast$ limits of $\beta$-KMS states as $\beta$ tends to infinity. This has been confirmed in \cite{PoloProdan2026} for abelian Kitaev models on the plane.

\begin{definition}
Let $\alpha\colon \R\to {\rm Aut}(\A)$ be a dynamics, and let $\mathtt C\subseteq \mathtt S(\A)$ be a code. We say that $\alpha$ \emph{stabilizes} $\mathtt C$ if its set of frustration-free ground-states coincides with $\mathtt C.$
\end{definition}

Notice that the definition of the dynamics $\alpha_{\LM}$ involves a choice of generating set $\bm{\mathrm L}$ and  depends on that data. However, the following characterization shows that its frustration-free ground-state space is independent of this choice and, moreover, that the resulting dynamics stabilizes the code.

\begin{proposition}\label{prop: code FF}
If $L\in \mathcal L$, then $\mathfrak G^F_{\LM} = \mathtt C_{L}$. Therefore, $\delta_{\LM}$ stabilizes the code $\mathtt C_{L}$, independently of the choice of generators.
\end{proposition}
\begin{proof}
For $\SM \in \mathfrak G^F_{\LM}$, we have $\SM(1-P_l)=0$ for all $l \in \bm{\mathrm L}$ and, from Cauchy--Schwarz inequality,  it is also true that $\SM((1-P_l)A)=0$ for all $A \in \A$. Then $\SM(U_l) = \SM(P_l U_l)=\SM(P_l)=1$. Now, if $l\in L$, then $l=l_1\cdots l_n$ for some $l_i\in \bm{\mathrm L}$ and, moreover, $U_l=U_{l_1}\cdots U_{l_n}$. Lemma~\ref{Lem:SwithU} then assures us that $\SM(U_l)=1.$ Therefore, $\mathfrak G^F_{\LM} \subseteq \mathtt C_{L}$. Conversely, if $\SM \in \mathtt C_{L}$, then $\SM(U_l)=1$ for all $l \in \bm{\mathrm L}$, and Lemma~\ref{Lem:SwithU} assures us that $\SM(P_l) =1$.

Last statement is a consequence of Proposition \ref{prop: code generators}.
\end{proof}

\begin{remark}
While $\mathfrak G_{\LM}^F$ depends only on $L$ and is therefore independent of the choice of $\bm{\mathrm L}$, the full ground-state space $\mathfrak G_{\LM}$ generally is not, as we shall see in Section~\ref{sec: biased ising}. \hfill $\blacktriangleleft$
\end{remark}

 \begin{proposition}\label{prop: spectrum}
Assume $L \in \mathcal L$ is Lagrangian.  Let $(\pi,\mathcal H,\psi_0)$ be the GNS representation of the
frustration-free state $\SM_0$, and with abuse in the notation let $H_{\LM}$ be the positive
self-adjoint generator acting on $\mathcal{H}$ implementing the dynamics $\alpha_{\LM}$. Then $
\operatorname{Spec}(H_{\LM})\subseteq \mathbb N_0 .$
\end{proposition}

\begin{proof}
Let $
\mathcal D_0
=
\operatorname{span}
\{
\pi(U_p)\psi_0
\mid
p\in P
\}.$
Since $\psi_0$ is cyclic, $\mathcal D_0$ is dense in
$\mathcal H$. One has
$$
H_{\LM}\pi(U_p)\psi_0
\;=\;
\pi(U_p)\pi(\lambda_p)\psi_0 \;=\; \lambda_p(\chi_0) \pi(U_p)\psi_0.$$
Since $\lambda_p(\chi_0) \in \N_0$ by \ref{Re:ConChi0} for all $p\in P$, then every vector in $\mathcal D_0$ is a finite linear
combination of eigenvectors of $H_{\LM}$ with eigenvalues in
$\mathbb N_0$. For each $n\in\mathbb N_0$, let
\[
\mathcal H_n
\;=\;
\overline{
\operatorname{Span}
\big\{
\pi(U_p)\psi_0
\mid
\lambda_p(\chi_0)=n
\big\}
}.
\]
Then
\[
\mathcal H
\;=\;
\overline{\bigoplus_{n\in\mathbb N_0}\mathcal H_n},
\]
and the restriction of $H_{\LM}$ to $\mathcal H_n$ is multiplication by
$n$. Hence the spectral measure of $H_{\LM}$ is supported on
$\mathbb N_0$ and $
\operatorname{Spec}(H_{\LM})
\subseteq
\mathbb N_0.$
\end{proof}
Previous Proposition with \ref{prop: code FF} and \ref{coro: unique FF} verifies the following important result:
\begin{theorem}\label{theo: gapped}
     If $L\in \mathcal{L}$ is Lagrangian, then $\delta_{\LM}$ has a unique frustration-free ground-state $\SM_0$, which is moreover gapped in the sense
     $$-\ii \SM_0(A^*\delta_{\LM}(A))\;\geq\; \SM_0(A^*A)-|\SM_0(A)|^2,\qquad \forall A\in {\rm Dom}(\delta_{\LM}).$$
\end{theorem}

\begin{proposition}
The dynamics $\alpha_{\LM}$ is periodic with period $2\pi$. Hence it
defines a strongly continuous action of the circle group $ \mathbb T \cong \mathbb R/2\pi\mathbb Z$ on $\A$ by
$$
    z \cdot X\;:=\;\alpha_{\LM}^t(X),
    \qquad z\;=\;e^{it}.$$
Moreover, the map
\begin{equation}\label{eq: expectation fixed}
        E_{\LM}(X)
    \;:=\;
    \frac{1}{2\pi}\int_0^{2\pi}\alpha_{\LM}^t(X)\,dt,
    \qquad X\in \A,
\end{equation}
is a conditional expectation onto the fixed-point algebra
\[
\A^{\alpha_{\LM}}\;
    :=\;
    \big\{A\in \A\mid \alpha_{\LM}^t(A)=A
    \text{ for all } t\in\mathbb R\big\}.
\]
\end{proposition}
\begin{proof}
    This is a direct consequence of Proposition \ref{prop: spectrum} and that $\A$ is simple.
\end{proof}

\subsection{Dynamics by groupoid method}\label{sec: dynamics by group}

We are in the special situation where the dynamics $\alpha_{\LM}$ restricts to the identity on the subalgebra $C(\widehat L)$ corresponding to the unit space of $\G_L$. For a large class of groupoids which cover $\G_L$, such dynamics can be efficiently described in terms of groupoid 1-cocycle, as we explain next.

Let $\mathcal G$ be a general topological groupoid. A continuous \emph{$1$-cocycle}
with values in an abelian topological group $A$ is a continuous map
$c\colon\G \to A$ such that $c(\zeta_1\zeta_2)\;=\;c(\zeta_1)+c(\zeta_2)$ for all composable pairs $(\zeta_1,\zeta_2)$. They form an abelian group which is usually denoted by $Z^{1}(\G,A)$. If $c\in Z^{1}(\G,A)$ and $\widehat A$ is the Pontryagin dual of $A$, then 
$$
\widehat A \ni \tilde a \;\mapsto\; [\alpha_c^{\tilde a}(f)](\zeta)\;:=\;\langle \tilde a,c(\zeta)\rangle f(\zeta), \quad f\in C_c(\G),
$$
extends to a $C^\ast$-dynamical system $(C^\ast\G,\widehat A,\alpha_c)$ \cite[Proposition~5.1]{RenaultBook}. In particular, if $A=\R$, the map
 $$(\alpha_c^t f)(\zeta)\;:=\;e^{\ii t c(\zeta)} f(\zeta),\qquad t\in \R, \ f\in C_c(\mathcal{G})$$
defines a strongly continuous family of $*$-automorphisms $t\mapsto \alpha_c^t$ on $C^*\mathcal{G}$. Furthermore, $C_c(\G)$ consists of entire analytic elements for $\alpha_c$. The (pre-)generator of $\alpha_c$ is the $*$-derivation  $\bm{\delta}_c(f)(\zeta):=\ii c(\zeta)f(\zeta)$. For $c\in Z^1(\G,\T)$, we can also define an automorphism $\alpha_c$ of $C^*\G$ by setting
 $$(\alpha_cf)(\zeta)\;:=\; c(\zeta) f(\zeta),\qquad f\in C_c(\mathcal{G}).$$

 \begin{theorem}\label{theo: cocycle} Let $L$ be Lagrangian and denote by $\G_L$ the Weyl groupoid of the $C^*$-diagonal pair $(\A_\sigma,\A_L)$. Then there exists a cocycle $c\in Z^1(\G_L,\R)$ such that $\alpha_{\LM} = \alpha_c$.
\end{theorem}

\begin{proof}
Since the dynamics $\alpha_{\LM}$ restricts to the identity on $\A_L\simeq C(\G_L^0)$ and $\G_L$ is a  principal \'etale groupoid, by \cite[Corollary~3.2.6]{KomuraDocMath2025}, each automorphism
$\alpha_{\LM}^t$ is implemented by a unique $\T$-valued $1$-cocycle
$c^t\in Z^1(\G_L ,\T)$. Moreover, since $\alpha_{\LM}$ is a strongly continuous action of\/ $\R$,
the map $(t,[p],\chi)\mapsto c^t([p],\chi)$
is jointly continuous, and as such for each arrow $([p],\chi)\in\G_L$ the function
$t\mapsto c^t([p],\chi)$ defines a continuous group homomorphism
from $\R$ to $\T$.
It follows that there exists a unique real number $c([p],\chi)\in\R$
such that
$$
c^t([p],\chi)\;=\;e^{\ii t\,c([p],\chi)},
\qquad\text{for all }t\in\R.$$
By joint continuity, we obtain a continuous real-valued cocycle
$c\in Z^1(\G_L,\R)$ satisfying $\alpha_{\LM}^t=\alpha_c^t$ for all $t\in\R$.
\end{proof}

This theorem is crucial because it diagonalizes the dynamics and
encodes it through the simpler object $c.$
Moreover, even when $L$ is not Lagrangian, the dynamics can still be
described by a groupoid cocycle:
\begin{corollary}
    Let $L\subset P$ isotropic. If $\tilde{L}\in \mathcal L$ is any admissible Lagrangian extension of $L$, then there exists $c\in Z^1(\G_{\tilde L},\R)$ such that $\alpha_{\LM}=\alpha_c.$
\end{corollary}
\begin{proof}
    Since $\alpha_{\LM}$ is the identity restricted to $\A_{\tilde L},$ then the same argument in the proof of Theorem \ref{theo: cocycle} verifies the existence of such a cocycle.
\end{proof}
We can describe explicitly the cocycle $c_{\LM}$ on $C^\ast \G_L$ which describes the dynamics $\alpha_{\LM}$, even though we do not have a concrete form of the isomorphism $\A \simeq C^\ast \G_L$. Let us introduce a few conventions about $C^\ast \G_L$. For each $[p] \in P/L$ we define $V_{[p]} \in C_c(\G_L)$ as 
$$
V_{[p]}([q],\chi) \;=\; \delta_{[p]}([q]).$$
For $f \in C(\widehat L)$, we will also consider elements of the form 
$$
(\delta_1 f)([p],\chi)\;=\;\delta_1([p]) f(\chi),$$
which generate $C(\widehat L)$ inside $C^\ast \G_L$. Together, $f$'s and $V_p$'s generate the whole $C_c(\G_L)$ and 
$$
V_{[p]}^\ast f \, V_{[p]} \;=\;[ p] \triangleright f, \quad ([p] \triangleright f)(\chi)\; =\; f([p] \triangleright \chi).$$
Note that, if $A \in C_c(\G_L)$, then
\[
A \;=\; \sum_{[p] \in P/L} V_{[p]} A([p],\text{--}),
\]
where the sum involves only finitely many nonzero terms and each $A([p],\text{--})$ belongs to $C(\widehat L)$.

\begin{proposition}
    The 1-cocycle implementing the dynamics on $C^\ast \G_L$ associated with the Lagrangian $L$ is $c_{\LM}([p],\chi) : = \lambda_p(\chi)$, which admits the following equivalent expression:
    \begin{equation}\label{Eq:GoodCL}
    c_{\LM}([p],\chi) \;=\; |{\rm supp} \, (\chi_{[p]} \chi) \cap {\rm supp}\, \chi_{[p]}| - |{\rm supp}\, \chi \cap {\rm supp}\, \chi_{[p]}|.
    \end{equation}
   The same formula also works when $L$ is isotropic.
\end{proposition}

\begin{proof}
    We know that the isomorphism $\A \simeq C^\ast \G_L$ sends $\A_L$ into $C(\widehat L)$, hence $H_{\LM_n}$ is mapped into $h_n \in C(\widehat L)$, $h_n(\chi) = \sum_{l \in \LM_n} (1-\delta_1(\chi(l)))$. We can compute
    \[
    [h_n,V_{[p]}] \;=\; V_{[p]}(V_{[p]}^\ast h_n V_{[p]}-h_n) \;=\; V_{[p]}( [p] \triangleright h_n -h_n) \;=\; V_{[p]} \lambda_p^n.
    \]
    Then, for $A \in C_c(\G_L)$, we have
    \[
    (\delta_{\LM_n}(A)) \;=\; \ii \sum V_{[p]} A([p],\text{--})\lambda_p^n,
    \]
    and, if we define $c_{\LM_n}([p],\chi) : = \lambda_p^n(\chi)$, a direct check shows that
    \[
    (\delta_{\LM_n}(A))([p],\chi) \;=\; \ii A([p],\chi) c_{\LM_n}([p],\chi).
    \]
    By taking $\LM_n$ large enough so as to cover the supports of $\chi_{[p]}$ for all $[p]$ engaged in the above sum, we have 
     \[
    (\delta_{\LM_n}(A))([p],\chi) \;=\; \ii A([p],\chi) c_{\LM}([p],\chi),
    \]
    and the statement follows because $C_c(\G_L)$ is a core of $\delta_{\LM}$.

    For an isotropic subgroup $L_0\subseteq L$, the same formula holds by replacing $\bm{\mathrm L}$ with $\LM_0$, with the supports in \eqref{Eq:GoodCL} understood relative to $\LM_0$.
\end{proof}

\section{Computation of the ground-state space}

\subsection{The boundary groupoid}

Renault in \cite{RenaultBook} points us to the following closed subset
\begin{equation}
 Z_{\LM} \;:=\; 
\left\{
\chi\in\widehat L\mid
c_{\LM}([p],\chi)\geq0 ,\, \forall \, [p]\in P/L
\right\}\; \subset \;\G_L^0.
\end{equation}
It was called the boundary set of $c_{\LM}$ in \cite{LLN2019} and the geodesic set in \cite{Thomsen2021}. For orientation, we state the following fact:

\begin{lemma}\label{Lem:ChiGStates}
Let $\chi\in Z_{\LM}$. Then the induced state $\SM_\chi$ on $C^\ast \G_L$ is a ground-state for $\alpha_{\LM}$. Reciprocally, if $s$ is a ground-state for $\alpha_{\LM}$, then its restriction to $C(\widehat L)$ induces a positive measure on $\widehat L$ whose support is contained in $ Z_{\LM}$.
\end{lemma}

\begin{proof}
For $A \in C_c(\G_L)$, $\delta_{\LM}(A) = \ii c_{\LM} A$ and, recalling that $A^\ast(p,\chi)=\overline {A([p]^{-1},[p] \triangleright \chi)}$,
\[
 s_\chi(-\ii A^\ast \delta_{\LM}(A)) = \sum_{[p] \in P/L} c_{\LM}([p],\chi) |A([p],\chi)|^2.
 \]
This sum contains a finite number of terms because $A$ has compact support and each term is nonnegative for $\chi \in Z_{\LM}$. For the reciprocal statement, we have
$$
-\ii s((V_{[p]} f)^\ast \delta_{\LM}(V_{[p]} f)) \;= \;s(|f|^2 c_{\LM}([p],\text{--})) \;=\; \int_{\widehat L} |f(\chi)|^2c_{\LM}([p],\chi) d\mu_s(\chi).
$$
for all $[p] \in P/L$. If the support of $\mu_s$ contains a $\chi$ with $c_{\LM}([p],\chi)<0$, then we can always find $f \in C(\widehat L)$ with support in a small vicinity of $\chi$ such that integral is negative. The latter cannot happen if $s$ is a ground-state, hence ${\rm supp}\, \mu_s \subseteq Z_{\LM}$.
\end{proof}

$ Z_{\LM}$ will play a much bigger role. Indeed, we know that $ Z_{\LM}$ is not empty because it contains  $\chi_0$. Hence, we can consider the corresponding reduction of the Weyl groupoid $\G_L$,
\[
\G_{\LM}^b \;:=\;\G_L| Z_{\LM}
\;=\;
\bigl\{\xi \in \G_{\LM} \mid \mathfrak s(\xi), \mathfrak r(\xi)\in Z_{\LM}\bigr\} = \mathfrak s^{-1}(Z_{\LM}) \cap \mathfrak r^{-1}(Z_{\LM}),
\]
where recall $\mathfrak s, \mathfrak r\colon \G_L\to \widehat L$ are the source and range maps, respectively. We will refer to $\G_{\LM}^b$ as the boundary groupoid of $\alpha_{\LM}$, as in \cite{LLN2019}. Since $c_{\LM}$ is continuous and takes values in a discrete subset of
$\mathbb R$, its kernel ${\rm Ker}(c_{\LM})$ is a clopen subgroupoid of
$\mathcal G_L$, and hence is étale. Then \cite[Lemma~1.3]{LLN2019} assures us that $\G_{\LM}^b$ is \'etale and we are within the conditions of Theorem~1.4 of \cite{LLN2019}:

\begin{theorem}[\cite{LLN2019}]\label{Th:AffineIso}
There is an affine homeomorphism of the state space of \(C^\ast \G_{\LM}^b\) onto the
\(\alpha_{\LM}\)-ground-state space $\mathfrak G_{\LM}$. More precisely,  every
state $\psi\in \mathtt S(C^\ast\G_{\LM}^b)$ determines a unique ground-state
\(\mathtt s_\psi\in\mathfrak G_{\LM}\) satisfying
\begin{equation}
    \mathtt s_\psi(f)
\;=\;
\psi\bigl(f|_{\G_{\LM}^b}\bigr),
\qquad f\in C_c(\G_L).
\end{equation}
\end{theorem}
\subsection{Computation of the ground-states}\label{sec: computation GS}
Theorem \ref{Th:AffineIso} is crucial because it reduces
the computation of the ground-state space of the Lagrangian model $\delta_{\LM}$ to that of the state space of the
$C^*$-algebra $C^*\G_{\LM}^b$. Our goal now is to determine such a space, and we will do so by using the technical results from \cite{Clark2007}.

\medskip

Let us recall that $Z_{\LM}$ coincides with the unit space of $\G_{\LM}^b$. Consider the continuous map
$\pi\colon \G^b_{\LM}\to Z_{\LM}\times Z_{\LM}$ given by
$$
\pi([p],\chi)
\;=\;
\big(\mathfrak r([p],\chi),\mathfrak s([p],\chi)\big)
\;=\;
([p]\triangleright\chi,\chi).
$$
Its image defines the orbit equivalence relation on $Z_{\LM}$ induced by the action of $\G_{\LM}^b$ on $Z_{\LM}$. We denote the corresponding orbit space by
$\Gamma_{\LM}:=Z_{\LM}/\G^b_{\LM}$ and equip it with
the quotient topology. As we shall see in remark \ref{Re:LocAutEquiv}, the elements
of $\Gamma_{\LM}$ label the sectors of the model, parametrizing
 suitable equivalence classes of the irreducible
representations of $C^*\G^b_{\LM}$.  

\begin{proposition}\label{Prop:Orbit1} For each $\chi\in Z_{\LM}$, its equivalence class is the
groupoid orbit $\Oo_\chi$ consisting precisely of its
zero-energy translations:
$$
\Oo_\chi
\;=\;
\big\{
[p]\triangleright\chi
\mid \; [p]\in P/L,\;
c_{\LM}([p],\chi)=0
\big\}.
$$
\end{proposition}

\begin{proof}
Let $\chi'\in \Oo_\chi$. By definition, there exists $[p]\in P/L$ such that $\chi'=[p]\triangleright \chi$. The cocycle identity together with the fact that $\chi'\in Z_{\LM}$ yield
$$0\;\leq\; c_{\LM}([p]^{-1},\chi')\;=\;-c_{\LM}([p],\chi)\;\leq \;0$$
where in the last inequality we used that $\chi\in Z_{\LM}$. Thus $c_{\LM}([p],\chi)=0.$
 Conversely, let $[p]\in P/L$ satisfy
$c_{\LM}([p],\chi)=0$, and set
$\chi'=[p]\triangleright\chi$. For every $[q]\in P/L$,
the cocycle identity gives
$$
c_{\LM}([q],\chi')
\;=\;
c_{\LM}([qp],\chi)-c_{\LM}([p],\chi)
\;=\;
c_{\LM}([qp],\chi)
\;\geq\; 0,
$$
since $\chi\in Z_{\LM}$. Thus $\chi'\in Z_{\LM}$,
and hence $\chi'\in\Oo_\chi$.
\end{proof}

 We denote the orbits by $\Oo_\gamma$, with $\gamma\in\Gamma_{\LM}$, and the spectrum of $C^\ast \G^b_{\LM}$ by
$\widehat{C^* \G^b_{\LM}}$. The latter consists of the set of unitary equivalence classes of irreducible representations of $C^* \G^b_{\LM}$, equipped with the hull-kernel topology (see \cite[Section 5.4]{MurphyBook}). 

\begin{proposition} There is a natural map $\iota \colon \Gamma_{\LM}\to \widehat{C^* \G^b_{\LM}}$ given by $\iota(\gamma)=[\iota_\gamma]$, where $\iota_\gamma$ is the left regular representation
$$
(\iota_\gamma(f)\xi)(\chi)
\;=\;
\sum_{\substack{[p]\in P/L\\c_{\LM}([p],\chi)=0}}
f\big([p]^{-1},[p]\triangleright\chi\big)\,
\xi\big([p]\triangleright\chi\big),\qquad \xi\in \ell^2(\Oo_\gamma),\;\chi\in \Oo_\gamma.
$$
\end{proposition}

\begin{proof} Since any AF groupoid is principal \cite{GTS}, then $\G_L$ is principal by Lemma \ref{lem: AF} and its reduction $\G^b_{\LM}=\G_L|_{Z_{\LM}}$ is also principal.
Consequently, each regular representation $\iota_\gamma$
is irreducible \cite[Lemma~2.4]{MuhlyWilliams1990},
so $\iota$ is well defined. 
\end{proof}

\begin{lemma}\label{Lem: spectrum}
    The map $\iota\colon \Gamma_{\LM}\to \widehat{C^* \G^b_{\LM}}$ is an injective continuous open map. Moreover, it is a homeomorphism if the sector space $\Gamma_{\LM}$ is $T_0.$
\end{lemma}
\begin{proof}
   Notice that $\iota$ coincides with the map $\psi$ defined in \cite[Section 3]{Clark2007} for the groupoid $\G^b_{\LM}$.  Moreover,  since $\G_{\LM}^b$ is principal,  \cite[Proposition 3.6]{Clark2007} implies that $\iota$ is a continuous open injection. The last statement is a consequence of \cite[Proposition 5.1]{Clark2007}.
\end{proof}
\begin{conjecture}\label{conj: T0}
Prove or disprove that for every admissible Lagrangian model $\delta_{\LM}$,  the sector space
$\Gamma_{\LM}$ is $T_0$.
\end{conjecture}

If this conjecture holds, then $\Gamma_{\LM}$
parametrizes all pure ground-states of the model up to
unitary equivalence of their GNS representations. In Section~\ref{Sec:SpaceGS}, we establish a sufficient
condition on $Z_{\LM}$ ensuring that $\Gamma_{\LM}$ is $T_0$.
In Section~\ref{Sec:Applications}, we verify this condition
for several physical models and explicitly determine their
full ground-state spaces.

\medskip 

We now describe explicitly the ground-states arising from
the sector space $\Gamma_{\LM}$. Consider
$$
\Hh_{\LM}
\;:=\;
\bigsqcup_{\gamma\in\Gamma_{\LM}}\ell^2(\Oo_\gamma),
\qquad
\pr(\Hh_{\LM})
\;:=\;
\bigsqcup_{\gamma\in\Gamma_{\LM}}
\pr\big(\ell^2(\Oo_\gamma)\big),
$$
where $\pr(H)$ denotes the projective space of a Hilbert
space $H$, and by a slight abuse of notation, $\pr(\Hh_{\LM})$ denotes
the fiberwise projectivization of $\Hh_{\LM}$. For each $\gamma\in\Gamma_{\LM}$ and each unit vector
$\xi\in\ell^2(\Oo_\gamma)$, the irreducible representation
$\iota_\gamma$ determines a pure state on $C^*\G^b_{\LM}$ via
$$
\psi_{\gamma,\xi}(A)
\;:=\;
\langle\xi,\iota_\gamma(A)\xi\rangle,
\qquad A\in C^*\G_{\LM}^b.
$$
By Theorem~\ref{Th:AffineIso}, this state gives rise to a pure
ground-state $\SM_{\gamma,\xi}\equiv \SM_{\psi_{\gamma,\xi}}$ of $\delta_{\LM}$.
For $f\in C_c(\G_L)$, it is explicitly given by
\begin{align}\label{eq: GS2}
\SM_{\gamma,\xi}(f)
&=\;
\left\langle
\xi,\iota_\gamma\big(f|_{\G_{\LM}^b}\big)\xi
\right\rangle\\
&=\;
\sum_{\chi\in\Oo_\gamma}
\sum_{\substack{[p]\in P/L\\
c_{\LM}([p],\chi)=0}}
\overline{\xi(\chi)}\,
f\big([p]^{-1},[p]\triangleright\chi\big)\,
\xi\big([p]\triangleright\chi\big).
\end{align}
 Moreover, for unit vectors $\xi\in\ell^2(\Oo_\gamma)$
and $\eta\in\ell^2(\Oo_{\gamma'})$,
$$
\SM_{\gamma,\xi}\;=\;\SM_{\gamma',\eta}
\quad\Longleftrightarrow\quad
\gamma\;=\;\gamma'
\;\text{and}\;
\eta\;=\;z\xi
\;\;\text{for some }\;z\in\mathbb T.
$$
Consequently, the assignment
$$
\pr(\Hh_{\LM})\longrightarrow
\operatorname{Ext}(\mathfrak G_{\LM}),
\qquad
(\gamma,[\xi])\longmapsto\SM_{\gamma,\xi},
$$
with $\xi$ chosen of unit norm, is well defined and injective. Here $\operatorname{Ext}(\mathfrak G_{\LM})$ are the set of extrema or pure ground-states. 
By Lemma \ref{Lem: spectrum}, if $\Gamma_{\LM}$ is $T_0$ this map is also surjective and determines all the pure ground-states of $\delta_{\LM}$. Now we are in a position to state one of the main results of this work.

\begin{theorem}\label{theo: GS}
For each admissible Lagrangian model $\delta_{\LM}$,
the assignment $(\gamma, [\xi])\mapsto \SM_{\gamma,\xi}$ is an injective  map from $\pr(\Hh_{\LM})$ to ${\rm Ext}(\mathfrak G_{\LM})$. If $\Gamma_{\LM}$ is $T_0$ it is surjective and hence determines a bijection  $
\pr(\Hh_{\LM})
\cong
\operatorname{Ext}(\mathfrak G_{\LM}).
$
\end{theorem}
Theorem~\ref{theo: GS} provides a large family
of pure ground-states for the models $\delta_{\LM}$,
exhausting all pure ground-states whenever $\Gamma_{\LM}$
is $T_0$. We now describe how this family gives rise to
more general ground-states. Namely, each positive trace-class operator $\rho_\gamma$ on
$\ell^2(\Oo_\gamma)$ with $\operatorname{Tr}(\rho_\gamma)=1$
determines a ground-state by
\begin{equation}\label{eq: decomposition gs}
    \SM_{\gamma,\rho_\gamma}(f)
\;:=\;
\operatorname{Tr}\!\left(
\rho_\gamma\,
\iota_\gamma\big(f|_{\G_{\LM}^b}\big)
\right),
\qquad f\in C_c(\G_L).
\end{equation}
Finite or countable convex combinations of these states
are again ground-states:
\begin{equation}\label{eq: decompositon convex}
    \SM
\;=\;
\sum_{\gamma\in\Gamma_{\LM}}
t_\gamma\SM_{\gamma,\rho_\gamma},
\qquad
t_\gamma\geq0,\qquad
\sum_{\gamma\in\Gamma_{\LM}}t_\gamma=1,
\end{equation}
where at most countably many coefficients are nonzero
and the series converges in norm. Thus, the sector space provides an explicit family of
pure ground-states and their finite or countable mixtures.
When $\Gamma_{\LM}$ is $T_0$, the full ground-state space
is the weak$^*$-closed convex hull of this family of pure
ground-states.

\subsection{Ground-state classification for finite-support configurations}\label{Sec:SpaceGS} Throughout this section, we assume that the boundary set
$Z_{\LM}$ consists entirely of characters with finite support.
Under this assumption, we describe the structure of
$C^*\G_{\LM}^b$ and obtain a complete classification of
the ground-states of $\delta_{\LM}$.
In Section~\ref{Sec:Applications}, we present explicit
models satisfying this condition.

% Consider the standard action of the ground groupoid on its unit space $ Z_{\LM}$. Then $ Z_{\LM}$ can be partitioned by the orbits of this action, and we will only consider the case when the number of orbits are finite. As we argued in the introduction, this is the situation most useful in practice.

We recall that the action of $P/L$ on the unit space $\widehat L$ of $\G_L$ is supplied by the embedding 
$$
[p] \mapsto \chi_{[p]} \;=\; \omega(p,\text{--})$$
of $P/L$ into $\widehat L$. Since the $\chi_{[p]}$ have finite support, this embedding lands in the subgroup $\Xi_{\LM}$ of $\widehat L$ of characters with finite support, \ie

$$\Xi_{\LM}\;=\;\big\{ \chi\in \widehat L\mid 
\{l\in\LM:\chi(l)\neq 1\}
\text{ is finite}\big\}.$$
Clearly $\Xi_{\LM}$ is a countable dense subgroup of $\widehat L$ which is invariant under the action of $P/L$. Moreover, the cocycle $c_{\LM}$ for each $\chi\in \Xi_{\LM}$ admits the description
\begin{equation}\label{eq: cocycle finite}
    c_{\LM}([p],\chi)\;=\;|{\rm supp}( \chi_{[p]} \chi)|-|{\rm supp}\, \chi|
\end{equation}
where $|\cdot|$ stands for the cardinal of a set. We have the exact sequence of abelian groups
$$
    1\longrightarrow P/L\longrightarrow \Xi_{\LM} \longrightarrow \Xi_{\LM}/(P/L) \longrightarrow 1$$
where the cosets $\gamma=[\chi]$ in the abelian group $\widetilde\Gamma_{\LM}=\Xi_{\LM} /(P/L)$ encode the orbits in $\widehat L$ under the $P/L$-action of characters $\chi$ with finite support.

\begin{definition}
The elements of the abelian group $\widetilde\Gamma_{\LM}$ are called the
\emph{sectors} of the model $\delta_{\LM}$ and we refer to $\widetilde\Gamma_{\LM}$ as the group of sectors.
\end{definition}

\begin{remark}\label{Re:LocAutEquiv}
Let us explain the physical motivation for this terminology. Suppose
that $\chi,\chi'\in\Xi_{\LM}$ belong to the same sector, \ie, they
determine the same element of $\widetilde \Gamma_{\LM}$. Then there exists
$[p]\in P/L$ such that $\chi=[p]\triangleright\chi'$. If
$\SM_\chi$ and $\SM_{\chi'}$ denote the unique corresponding pure states on
$\A$, then
$$
\SM_\chi(X)\;=\;\SM_{\chi'}(U_p^* X U_p),
\qquad X\in\A,$$
where $U_p$ is a local unitary in $\A$. Thus, states in the same sector differ only by a local automorphism
and are therefore physically indistinguishable at infinity, in the
sense that they coincide on every observable localized outside the
finite support of $p$. In
Section~\ref{Sec:Applications}, we will see that these sectors recover
the superselection sectors of abelian quantum double and color code
models.  \hfill $\blacktriangleleft$
\end{remark}

\medskip

Since $Z_{\LM}\subseteq\Xi_{\LM}$, the inclusion induces
a natural injection
$\Gamma_{\LM}\to\widetilde\Gamma_{\LM}$. The following Proposition shows that this map is also
surjective.

\begin{proposition}
Every class in $\widetilde\Gamma_{\LM}$ has a representative
in $Z_{\LM}$. Consequently, the natural map
$\Gamma_{\LM}\to\widetilde\Gamma_{\LM}$ is a bijection.
\end{proposition}
\begin{proof}
Let $\gamma\in\widetilde\Gamma_{\LM}$ and denote by $q_L\colon \Xi_{\LM}\to \widetilde \Gamma_{\LM}$ the quotient map. Since
$|{\rm supp}\,\chi|$ takes values in $\mathbb N_0$ on
$\Xi_{\LM}$, the well-ordering principle implies that there exists a lift $\chi_\gamma\in \Xi_{\LM}$ of $\gamma$ such that
$$
|{\rm supp}\,\chi_\gamma|
\;=\;
\min\bigl\{
|{\rm supp}\,\chi|
\mid [\chi]\in q_L^{-1}(\gamma)
\bigr\}.
$$
For every $[p]\in P/L$, we have
$\chi_{[p]}\chi_\gamma\in q_L^{-1}(\gamma)$. Therefore, using the previous with equation \eqref{eq: cocycle finite} one gets
$$
c_{\LM}([p],\chi_\gamma)
\;=\;
|{\rm supp}\,(\chi_{[p]}\chi_\gamma)|
-
|{\rm supp}\,\chi_\gamma|
\geq 0.
$$
Thus $\chi_\gamma\in Z_{\LM}$, proving that every class
in $\widetilde\Gamma_{\LM}$ has a representative in
$Z_{\LM}$. The natural injection
$\Gamma_{\LM}\to\widetilde\Gamma_{\LM}$ is therefore
surjective, and hence a bijection.
\end{proof}
% \begin{proof}
% Since $|{\rm supp}\, \chi|$ takes values in $\mathbb N_0$ over
% $\Xi_{\LM}$, the well-ordering principle implies that there exists $\chi_\gamma \in \gamma$ such that
% $$
% |{\rm supp}\, \chi_\gamma|
% \;=\;
% \min\bigl\{|{\rm supp}\, \chi|, \ \chi\in \gamma \bigr\}.$$
% On the other hand,  by definition $\chi_{[p]} \chi_\gamma\in q_L^{-1}(\gamma)$ for all $[p]\in P/L$. 
% \begin{align*}
%     c_{\LM}([p],\chi_\gamma)\;=\;|{\rm supp}\, \chi_{[p]} \chi_\gamma|-|{\rm supp}\, \chi_\gamma|\;\geq\;0
% \end{align*}
% for all $[p]\in P/L$. Thus, $\chi_\gamma\in Z_{\LM}$ which verifies that $\chi_\gamma\in \mathscr O_\gamma.$

% The final statement is a direct consequence of Lemma \ref{Lem:ChiGStates}.
% \end{proof}

By a slight abuse of notation, we identify
$\widetilde\Gamma_{\LM}$ with $\Gamma_{\LM}$ as sets
and use $\Gamma_{\LM}$ for both, retaining the quotient
topology induced by $Z_{\LM}\to\Gamma_{\LM}$.  
Now, consider the map $\chi \mapsto |{\rm supp}\, \chi|\in \mathbb N_0$ on $ Z_{\LM}$, which we call $T$.

\begin{proposition}\label{prop: T0}
If $Z_{\LM}\subseteq\Xi_{\LM}$, then $\Gamma_{\LM}$
is $T_0$.
\end{proposition}

\begin{proof}
Let $\gamma,\gamma'\in\Gamma_{\LM}$ be distinct.
Since the cocycle vanishes on $\G_{\LM}^b$, equation \eqref{eq: cocycle finite} yields that the function $T$ is constant on $\Oo_\gamma$ for any $\gamma\in \Gamma_{\LM}$. Denote its values on $\Oo_\gamma$ and $\Oo_{\gamma'}$
by $n$ and $m$, respectively. Interchanging the orbits
if necessary, assume that $n\geq m$. Choose $\chi\in\Oo_\gamma$ and consider the open
cylinder set
$$ U\;
:=\;
\bigl\{
\chi'\in Z_{\LM}
\mid
\chi'(l)=\chi(l),\;\forall l\in{\rm supp}\,\chi
\bigr\}.
$$
We claim that $U\cap\Oo_{\gamma'}=\varnothing$.
Indeed, any $\eta\in U\cap \Oo_{\gamma'}$  would satisfy
${\rm supp}\,\chi\subseteq{\rm supp}\,\eta$, so
$n\leq m$. Thus $n=m$, and the two supports coincide.
Since $\eta$ and $\chi$ agree on this support and are
both trivial elsewhere on $\LM$, they coincide,
contradicting $\gamma\neq\gamma'$.

Because $\G_{\LM}^b$ is étale, the saturation $[U]:=\mathfrak r\bigl(\mathfrak s^{-1}(U)\bigr)
$
is open in $Z_{\LM}$. It contains $\Oo_\gamma$ and
is disjoint from $\Oo_{\gamma'}$. Its image in
$\Gamma_{\LM}$ is therefore an open set containing
$\gamma$ but not $\gamma'$, proving the $T_0$ property.
\end{proof}

The previous Proposition together with Theorem \ref{theo: GS} implies the following paramount classification result.

\begin{theorem}\label{theo: finite ground-states}
Let $\delta_{\LM}$ be an admissible Lagrangian model
such that $Z_{\LM}\subseteq\Xi_{\LM}$. Then the assignment$
(\gamma,[\xi])\mapsto\SM_{\gamma,\xi}$
defines a bijection
$$
\bigsqcup_{\gamma\in\Gamma_{\LM}}
\pr\bigl(\ell^2(\Oo_\gamma)\bigr)
\;\cong\;
\operatorname{Ext}(\mathfrak G_{\LM}).
$$
Moreover, every ground-state admits a decomposition as in \eqref{eq: decompositon convex}.
\end{theorem}

Another important consequence of Proposition~\ref{prop: T0} is that
the $C^*$-algebra $C^*\G^b_{\LM}$ is postliminal
\cite[Theorem~5.3]{Clark2007}. This means that the image of every irreducible representation contains all compact operators on its representation space. If, in addition, $T$ is uniformly bounded on $Z_{\LM}$, then $C^*\G^b_{\LM}$ is solvable in the sense of \cite{Dynin}.
We conclude this section by establishing this stronger structural
property, assuming henceforth that $T$ is uniformly bounded on $Z_{\LM}$.

\begin{lemma}\label{lemm: open invariant}
Let  $m$ be the maximum value of $T$ in $Z_{\LM}$. Then the sub-level sets
$$
X_n\;:=\;\big\{\chi\in Z_{\LM}\mid  \ T(\chi) \geq m-n\big\}, \quad n=0,\ldots,m,$$
are open in $ Z_{\LM}$ and invariant under $\G_{\LM}^b$ action.
\end{lemma}

\begin{proof}
Let $\chi$ be from $X_n$. Then $\chi(l)\neq 1$ for at least $m-n$ distinct elements $l_i$ of $\LM$. Choose a cylinder open neighborhood $\mathcal C$ of $\chi$ in $\widehat L$, made up of characters that take same values of $\chi$ at the $l_i$. Then every $\chi'\in \mathcal C \cap Z_{\LM}$ has $ T(\chi') \geq m-n$, hence $\chi$ has an open neighborhood entirely contained by $X_n$. Finally, since $T$ is constant on each orbit, $X_n$ is a union
of orbits and is therefore invariant.
\end{proof}

\begin{proposition}
    $X_n \setminus X_{n-1}$ is a discrete subset of $ Z_{\LM}$ which is closed in $X_n$.
\end{proposition}

\begin{proof}
   Let $\chi \in X_n \setminus X_{n-1}$ and consider a cylinder open neighborhood $\mathcal C$ of $\chi$ in $\widehat L$, made up of characters that coincide with $\chi$ on its support. Then $\mathcal C \cap Z_{\LM}$ can not contain any $\chi' \in X_n\setminus X_{n-1}$ distinct from $\chi$, because that will imply $T(\chi') >m-n$.
\end{proof}

\begin{remark}
    Since $X_n\setminus X_{n-1}$ is closed in $X_n$, we can be sure that the orbits contained in $X_n \setminus X_{n-1}$ do not have accumulation points in $X_n$. However, this does not exclude possible accumulation points in $X_k\setminus X_n$, $k>n$, which is often the case in solved examples. \hfill $\blacktriangleleft$
\end{remark}
\begin{proposition}
    $C^\ast \G_{\LM}^b$ is a solvable $C^\ast$-algebra. Specifically, the sets $X_n$ provide a filtration of $ Z_{\LM}$ by open invariant subsets,
    $$X_0\;\subseteq \;X_1\;\subseteq\; \cdots \;\subseteq\; X_m\;=\; Z_{\LM}$$
    such that $X_n\setminus X_{n-1}$ are discrete and invariant subset (possibly empty) of $ Z_{\LM}$. 
    This in turn supplies a filtration of  $C^\ast\G_{\LM}^b$ by ideals
    \[
    0 \longrightarrow C^*(\G_{\LM}^b|X_0) \stackrel{\psi_1}{\longrightarrow} C^*(\G_{\LM}^b|X_1) \stackrel{\psi_2}{\longrightarrow} \cdots \stackrel{\psi_m}{\longrightarrow} C^*(\G_{\LM}^b|X_m) \;=\; C^*\G_{\LM}^b
    \]
    such that
$${\rm Coker}(\psi_n)\;\simeq\;\bigoplus_{\gamma \in T^{-1}(m-n)} \mathcal{K}\big(\ell^2(\mathscr{O}_\gamma)\big)$$
where $\mathcal{K}(\cdot)$ stands for the algebra of compact operators.
\end{proposition}
\begin{proof}
 The existence of the cofiltration and filtration is a direct consequence of Lemma \ref{lemm: open invariant}. To finish the proof we just need to verify final statement. Let us first notice that
    $${\rm Coker}(\psi_n)\;=\;C^\ast (\G_{\LM}^b|(X_n\setminus X_{n-1}))\;\simeq\;\bigoplus_{\gamma \in T^{-1}(m-n)}C^*(\G_{\LM}^b|{\mathscr O_\gamma}),$$
   where we used that the set  $X_n\setminus X_{n-1}=\bigsqcup_{\gamma \in T^{-1}(m-n)}\mathscr O_\gamma $ is discrete and countable. To complete the proof, it remains to observe that $\G_{\LM}^b|\mathscr O_\gamma$ is principal and $\mathscr O_\gamma$ is discrete, hence $C^\ast (\G_{\LM}^b |\mathscr O_\gamma) \simeq \mathcal K\bigl(\ell^2(\mathscr O_\gamma)\bigr)$.
\end{proof}

\section{Applications}\label{Sec:Applications}
In this section, we present several examples illustrating how our main theorem can be used to explicitly compute the ground-states of explicit models. 

\subsection{Abelian quantum doubles}\label{Sec:Kitaev}

These models \cite{Kit1} (see also \cite{PoloProdan2026}) form a distinguished class of homological stabilizer models \cite{Anderson2013}. Although our general theory and the techniques developed in this section also apply to higher-dimensional versions of the models \cite{ADL2025}, we restrict our attention to the abelian Kitaev's models deployed on the plane. The main goal here is to exemplify how the general machinery works in a familiar setting, and how it reproduces the space of ground-states computed in \cite{CNN}.

We consider the uniform CW-decomposition $\Sigma$ of the plane implemented by the ordinary square lattice, as well as its dual $\widetilde \Sigma$,  both embedded in the same plane. We recall that the $0$-cells in $\widetilde \Sigma_0$ lie at the barycenters of the $2$-cells in $\Sigma_2$. Two 1-cells $e \in \Sigma_1$ and $\tilde e \in \widetilde \Sigma_1$ are conjugated if they intersect. The 1-cells of $\Sigma$ are assigned with arbitrary orientations once and for all. This automatically fixes the orientations of the 1-cells of $\widetilde \Sigma$. We choose the dual orientations such that, if $e \in \Sigma_1$ is rotated counter-clockwise by $\frac{\pi}{2}$, which we write as $r\cdot e$, we obtain its conjugate. Equivalently, if we rotate $\tilde e$ with the reversed rotation, $\bar r \cdot \tilde e$, we obtain its conjugated edge. The tilde symbol will also be used for the conjugation maps between $\Sigma$ and $\widetilde \Sigma$: For $\sigma\in \Sigma$, $\tilde \sigma$ is its dual in $\widetilde \Sigma$ and, for $\tilde \sigma \in \widetilde \Sigma$, $\tilde{\tilde \sigma}$ is its conjugate in $\Sigma$. Lastly, for a domain $\Omega$ of the plane, $\partial \Omega$ will denote its boundary and, for a closed non self-intersecting loop $\gamma$ in the plane, $\partial^{-1}\gamma$ will denote the domain bounded by $\gamma$, including $\gamma$ itself.

\begin{figure}
    \centering
    \includegraphics[width=0.99\linewidth]{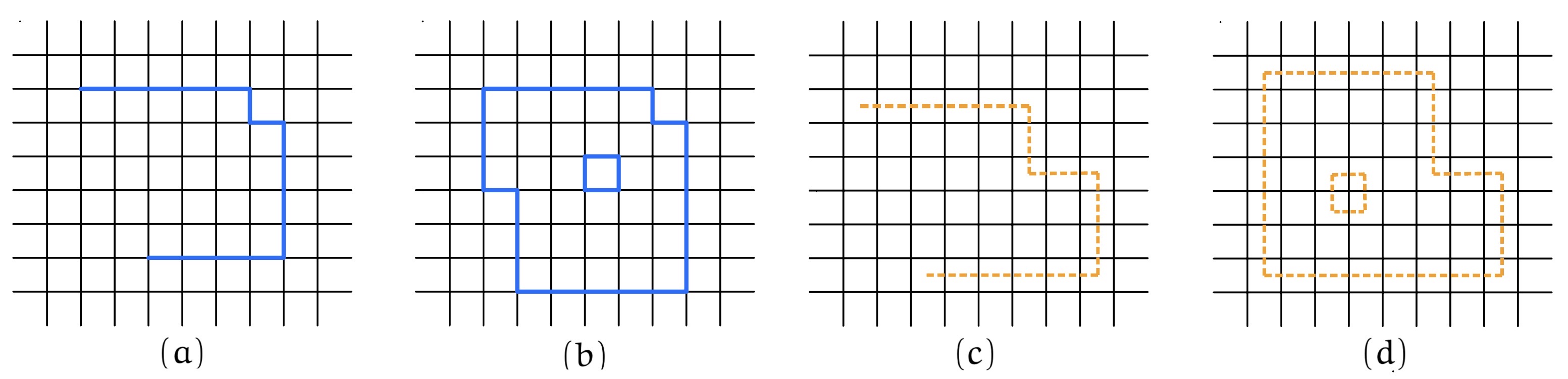}
    \caption{CW-decomposition of the plane together with examples of open and closed ribbons on the direct and dual complexes. The ribbons on the dual complex will be consistently drawn with dashed lines.}
    \label{Fig:Ribbons}
\end{figure}

We take the group of Pauli labels to be
$$
P_1 \;=\; C_c(\Sigma_1,G) \oplus C_c(\widetilde \Sigma_1,\widehat G).
$$
A ribbon is a finite oriented edge path in either the direct or the dual complex, with no self-intersections, except that
its initial and terminal vertices may coincide in the closed case. Examples of open and closed ribbons for the direct and dual complexes are shown in Fig.~\ref{Fig:Ribbons}. The start/end points of an open ribbon $\rho$ will be indicated by $\rho_\mp$, respectively. Let $g\in G$ and $\varphi\in\widehat G$, and let $\rho$ and
$\tilde\rho$ be direct and dual ribbons, respectively.
One defines $F_\rho^g\in C_c(\Sigma_1,G)$ and
$F_{\tilde\rho}^\varphi\in C_c(\widetilde\Sigma_1,\widehat G)$ via the rules
\begin{equation}\label{Eq:FS}
F_\rho^g(e) \;=\; \left\{
\begin{aligned}
    & g^{w(\rho,e)}, \ {\rm if} \ e \in \rho\\
    & 1, \ {\rm otherwise}
\end{aligned} \right.,
\qquad
F_{\tilde \rho}^\varphi(\tilde e) \;=\; \left\{
\begin{aligned}
    & \varphi^{w(\tilde \rho,\tilde e)}, \ {\rm if} \ \tilde e \in \tilde \rho,\\
    & 1, \ {\rm otherwise}
\end{aligned} \right.
\end{equation}
where $w=+1$ if the orientations of the ribbons and edges match and $w=-1$ otherwise. These elements generate $P_1$, and their representation in the algebra $C^\ast(P_1,\gamma_1)\simeq {\rm M}_{|G|}^{\otimes \Sigma_1}$, satisfy the same relations as the ribbon operators introduced by Kitaev \cite{Kit1}. Indeed, denoting their canonical lifts to $\mathcal P_1$
by the same symbols, we have
$$
F_{\rho_1}^g F_{\rho_2}^{g'} \;=\; F_{\rho_2}^{g'} F_{\rho_1}^g, \quad F_{\tilde \rho_1}^{\varphi} F_{\tilde \rho_2}^{\varphi'} \;=\; F_{\tilde \rho_2}^{\varphi'} F_{\tilde \rho_1}^\varphi,$$
and
\begin{equation}\label{Eq:CommRel1}
    F_{\rho_1}^g F_{\tilde \rho_2}^\varphi \;= \;\omega_1(F_{\tilde \rho_2}^\varphi,F_{\rho_1}^g)\; F_{\tilde \rho_2}^\varphi F_{\rho_1}^g\; =\; \prod_{  e,\;e \cap (\rho_1 \cap \tilde \rho_2) \neq \emptyset} \varphi(g)^{w(\rho_1,e)w(\tilde \rho_2,r \cdot e)} \; F_{\tilde \rho_2}^\varphi F_{\rho_1}^g.
\end{equation}

For each vertex $v\in\Sigma_0$, let $\tilde\rho_v$ be the
shortest counterclockwise-oriented closed dual ribbon
surrounding $v$, and  similarly, for each dual vertex $\tilde v\in\widetilde\Sigma_0$,
let $\rho_{\tilde v}$ be the shortest counterclockwise-oriented
closed direct ribbon surrounding $\tilde v$ (see Figs.~\ref{Fig:Ribbons}(b,d)). Set
$$
F_v^\varphi \;:=\; F_{\tilde\rho_v}^\varphi,\quad F_{\tilde v}^g \;:=\; F_{\rho_{\tilde v}}^g
\qquad \varphi\in\widehat G,\; g\in G.
$$
The representation of these elements in $\A$ are the so-called vertex and face operators \cite[Section 2.1]{PoloProdan2026}. Thus, we shall refer to them as the face and vertex elements, respectively.
\begin{proposition}\label{Prop:IdentityW}
The following identity holds
    $$
    \sum_{e,\; e \cap (\rho_1  \cap \tilde \rho_2) \neq \emptyset} w(\rho_1,e)w(\tilde \rho_2,r \cdot e) \;=\;  I(\rho_1,\tilde \rho_2)
    $$
    where $I(\rho_1,\tilde \rho_2)$ is the signed intersection index of the two curves.
\end{proposition}

\begin{proof}
  Note that each term of the sum is independent of the orientation of the edge $e$. Hence, we can assume that all edges of $\rho_1$ are oriented in the same direction as $\rho_1$. Then $w(\rho_1,e)=1$ and the remaining factor $w(\tilde \rho_2, r\cdot e)$ measures if the orientations of $\rho_1$ and $\tilde \rho_2$ at their transversal intersections match the orientation of the plane.
\end{proof}

Other useful identities, seen also inside $\mathcal P_1$, are
\[
F_{-\rho}^g \;=\; F_{\rho}^{ g^{-1}}\; =\; (F_\rho^g)^{-1}, \quad F_{-\tilde \rho}^\varphi \;=\; F_{\tilde \rho}^{\varphi^{-1}}\; =\; (F_{\tilde \rho}^\varphi)^{-1}
\]
\[
F_\rho^{g_1} F_\rho^{g_2} \;=\; F_\rho^{g_1g_2}, \quad F_{\tilde \rho}^{\varphi_1}F_{\tilde \rho}^{\varphi_2}\; =\; F_{\tilde \rho}^{\varphi_1 \varphi_2}
\]
\[
F_{\rho_1}^g F_{\rho_2}^g \;= \;F_{\rho_1 \rho_2}^g, \quad F_{\tilde \rho_1}^\varphi F_{\tilde \rho_2}^\varphi \;=\; F_{\tilde \rho_1 \tilde \rho_2}^\varphi
\]
If $\rho_1$ and $\rho_2$ as well as $\tilde \rho_1$ and $\tilde \rho_2$ can be concatenated.

\begin{remark}
    According to the first set of identities, we are over accounting for the ribbon elements if we sample all possible ribbon orientations. For open ribbons, there is no canonical way to fix the orientation, so we allow double counting. However, closed ribbons will be considered with the counter-clockwise orientation. \hfill $\blacktriangleleft$
\end{remark}

\begin{remark}\label{Re:RibProd}
    A convention that we will always implement is that a product of open ribbon elements will always be reduced using the concatenation rule. For example, if $\rho_i$ are ribbons with the same terminal vertex and $g_i \in G$ are such that $\prod g_i =1$ for $i=\overline{1,n}$, then 
    \[
    F_{\rho_1}^{g_1} \cdots F_{\rho_n}^{g_n}\;=\; F_{\rho'_1}^{g_1} \cdots F_{\rho'_{n-1}}^{g_{n-1}}, \quad \rho'_i \;=\; \rho_i \cup (-\rho_n).
    \]
   Here concatenation $\rho_i \cup (-\rho_n)$ is understood with cancellation
   of backtracking and decomposition into simple paths
   and closed ribbons when necessary. As such, we can always assume that, $g_1 \cdots g_k \neq 1$ in a product $\prod F_{\rho_i}^{g_i}$ whenever the open ribbons $\rho_i$ have overlapping end points or overlapping start points. Similar conclusions apply for the ribbon elements induced from the dual complex. \hfill $\blacktriangleleft$
\end{remark} 

\begin{proposition}
    The group generated by closed ribbon elements, direct or dual, is commutative in $\mathcal P_1$ and, as such,  they generate an isotropic subgroup $L$ of $P_1$.  
\end{proposition}
\begin{proof}
    This is a direct consequence of \ref{Prop:IdentityW}.
\end{proof}

By definition, the vertex and face elements belong to $L$.
In fact, they generate $L$.
\begin{proposition}
    The vertex and face elements supply a generating set $\bm{\mathrm L}$ of $L.$
\end{proposition}
\begin{proof}
For counterclockwise-oriented closed ribbons, cancellation
along interior edges gives
\begin{equation}\label{Eq:ProdId}
F_\rho^g \;=\;
\prod_{\tilde v\in\operatorname{int}(\partial^{-1}\rho)}
F_{\tilde v}^g,
\qquad
F_{\tilde\rho}^\varphi \;=\;
\prod_{v\in\operatorname{int}(\partial^{-1}\tilde\rho)}
F_v^\varphi.
\end{equation}
where $\operatorname{int}(\partial^{-1}\rho)$ denotes the interior
of the region bounded by $\rho$.  Since closed ribbon elements generate $L$, the claim follows.
\end{proof}

\begin{theorem}
    $L$ is a Lagrangian subgroup of $P_1$. 
\end{theorem}
\begin{proof}

    Suppose $L$ is not a Lagrangian subgroup. Then there exists a non-trivial element $p \in L^\perp \setminus L$ and, since the ribbon elements generate $P_1$, then $p$ is a finite product of ribbon elements and at least one ribbon $\rho_1$ must be open, which we assume it is a direct ribbon. The arguments are identical if it was a dual ribbon. 
    Let $v$ be the terminal
vertex of $\rho_1$, and consider all direct open ribbons
$\rho_i$ having an endpoint at $v$. Reversing their
orientations and inverting their labels if necessary,
we may assume that they all terminate at $v$.  By remark~\ref{Re:RibProd}, we can assume that $\prod g_i \neq 1$, since otherwise one can replace
these ribbons by ribbons joining their other endpoints, thereby reducing the number
of open ribbons. We have from \eqref{Eq:CommRel1} and \ref{Prop:IdentityW} that
    \[
    \omega_1(F_{ v}^\varphi,p) \;=\; \omega_1\big(F_{ v}^\varphi,\prod F_{\rho_i}^{g_i}\big)\;=\; \varphi\big ( \prod g_i\big ).
    \]
    so the conclusion is that $p$ pairs non-trivially with elements of $L$, hence a contradiction.
\end{proof}

There is one copy of $\widehat G$ per vertex of the direct complex and one copy of $G$ per vertex of the dual complex, with no relations between distinct copies. As such, the Lagrangian subgroup, its Pontryagin dual and $\Xi_{\LM}$ can be conveniently presented as 
\[
L \simeq  C_c(\Sigma_2,G) \oplus C_c(\widetilde \Sigma_2,\widehat G), \quad \widehat L \simeq  \prod_{\Sigma_2} \widehat G \times \prod_{\widetilde \Sigma_2} G, \quad \Xi_{\LM}   \simeq  C_c(\Sigma_2,\widehat G) \oplus C_c(\widetilde \Sigma_2, G).
\] 
where we identified each vertex $v\in\Sigma_0$ with its dual face
$\tilde f\in\widetilde\Sigma_2$, and each dual vertex
$\tilde v\in\widetilde\Sigma_0$ with its corresponding face
$f\in\Sigma_2$. Thus,
$
F_v^\varphi = F_{\partial\tilde f}^\varphi$ and $F_{\tilde v}^g = F_{\partial f}^g.
$
There are two reasons that make these presentations optimal. First, the analysis is mostly about manipulations of the supports of characters, with the latter defined in \eqref{Eq:Support} as the set of generators in $\LM$ that are evaluated by the character to non-trivial values. The above presentation transforms this definition into the ordinary support of a function. Secondly, the generators of $L$ and the action of $P_1/L$ on $\widehat L$ take the simplest forms possible. Indeed, $F_{\partial f}^g$ and $F_{\partial \tilde f}^\varphi$ become the functions $\delta_f^g$ and $\delta_{\tilde f}^\varphi$, which assign $g$ to $f$ and $\varphi$ to $\tilde f$, respectively, and the neutral element to all other faces. Given $\hat \phi \phi \in \widehat L$, its evaluation on our chosen generating set of $L$ is
\[
\hat \phi \phi(\delta_f^g) \;=\; [\hat \phi(f)](g), \qquad \hat \phi \phi(\delta_{\tilde f}^\varphi) \;=\; \varphi(\phi(\tilde f)).
\]
Same rules apply when we see the generators $F_{\partial f}^g$ and $F_{\partial \tilde f}^\varphi$ as function over $\widehat L$, under the identification $C^\ast L \simeq C(\widehat L)$. The rules can be extended to other closed ribbons by \eqref{Eq:ProdId}.

Furthermore, if $p=(p_1,\hat p_2)\in P_1$, then $\chi_{[p]} = \hat \phi \phi \in \widehat L$ can be computed from the definition \eqref{Eq:ChiP} of $\chi_{[p]}$ and the above rules, namely,
\[
[\hat \phi(f)](g) \;=\; \chi_{[p]}(F_{\partial f}^g)\; =\; \omega_1\big ((p_1,\hat p_2),(F_{\partial f}^g,1)\big ) 
\] 
and
\[
\varphi(\phi(\tilde f))\; =\; \chi_{[p]}(F_{\partial \tilde f}^\varphi) \;= \;\omega_1\big ((p_1,\hat p_2),(1,F_{\partial \tilde f}^\varphi)\big ).
\]
These lead to
\begin{equation}\label{Eq:HP}
\begin{aligned} 
& \hat \phi(f) \;=\; \hat \phi_{\hat p_2}(f) \;=\; \prod_{e \in \partial f} \hat p_2(r\cdot e)^{w(\partial f,e)}, \\
& \phi(\tilde f) \;=\; \phi_{p_1}(\tilde f) \;= \;\prod_{\hat e \in \partial \tilde f} p_1(\bar r \cdot \hat e)^{-w(\partial \tilde f, \hat e)}
\end{aligned}
\end{equation}
These identities confirm that each $\chi_{[p]}$ has finite support, hence $L \in \mathcal L$ is admissible.

\begin{proposition}\label{Prop:ChiF}
If $\rho$ and $\tilde \rho$ are open ribbon element, then
    \begin{equation}\label{Eq:ChiF}
    \chi_{[F_{\tilde \rho}^\varphi]}(f) \;=\; \left \{
    \begin{aligned}
        & \varphi, \;\; {\rm if} \ f = \widetilde{\tilde \rho_+} \\
        & \varphi^{-1},\;\; \ {\rm if} \ f= \widetilde{\tilde \rho_-}\\
        & 1,\;\; \ \rm{\rm otherwise}
    \end{aligned} \right . \ ,
    \qquad
    \chi_{[F_\rho^g]}(\tilde f) \;=\; \left \{
    \begin{aligned}
        & g,\;\; \ {\rm if} \ \tilde f = \widetilde{\rho_+} \\
        &  g^{-1} ,\;\;\ {\rm if} \ \tilde f= \widetilde{\rho_-}\\
        & 1,\;\; \ \rm{\rm otherwise}
    \end{aligned} \right . \ .
    \end{equation}
\end{proposition}

\begin{proof}
    After replacing $\hat p_2$ and $p_1$ in \eqref{Eq:HP} with the expressions from \eqref{Eq:FS}, we see the edges in \eqref{Eq:HP} appear in combinations like $w(\rho,e)w(\partial \tilde f, r \cdot e)$ and $w(\hat \rho,\hat e)w(\partial f, \bar r \cdot \hat e)$, the result is fully independent of the orientations of the edges. It is also independent of the shape of the ribbon, because closed ribbon elements commute. Then, for a face that does not overlap with ends of the ribbon, we can change the shape of the ribbon to avoid such face entirely and, as such, for those cases the result should be 1, as stated. For the other cases, it is sufficient to work out particular examples, since the evaluation engages only the end edges of the ribbon. Given such edge, we can always rotate the page to bring it to a standard position and, since the result is independent of the edge orientations, we are done, provided the relevant particular cases check.
\end{proof}

\begin{figure}
    \centering
    \includegraphics[width=0.9\linewidth]{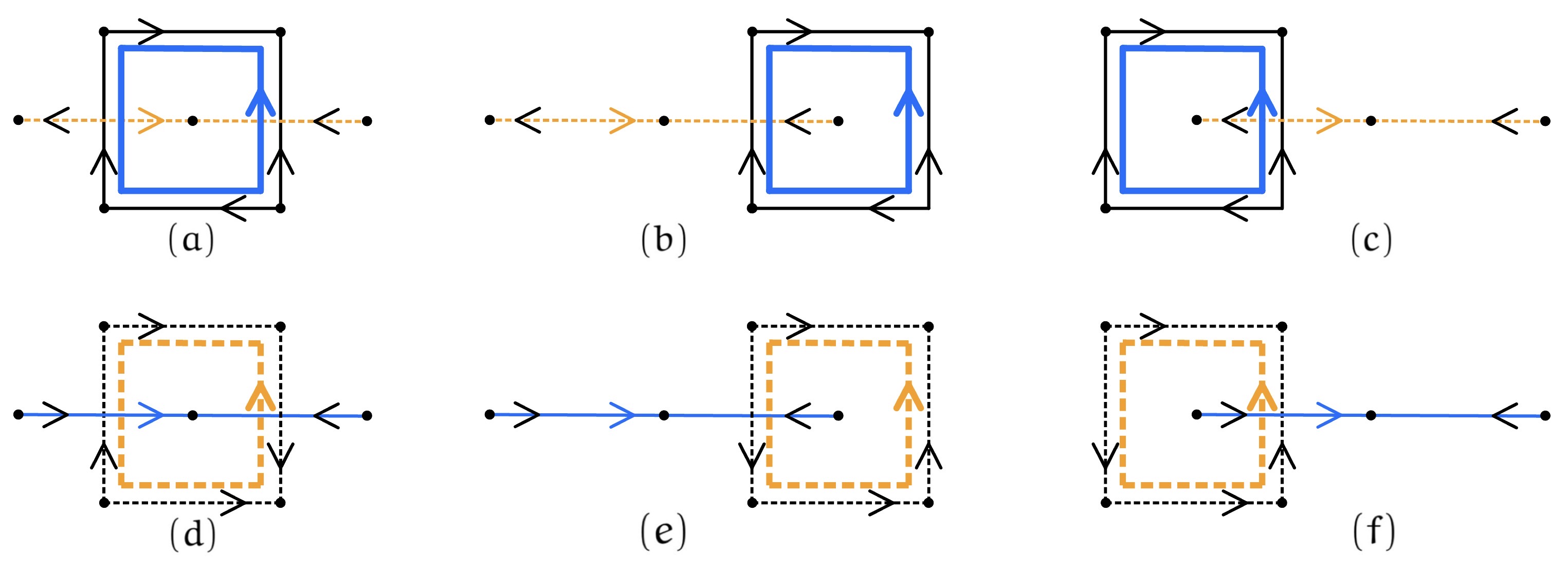}
    \caption{Six cases of ribbon-face configurations covering the following situation: (a,d) the face intersects the ribbon but avoids the ends; (b,e) The face overlaps with the starting vertex of the ribbon; (c,f) The face overlaps with the ending vertex of the ribbon. The ribbons and their orientations are shown in color, while the edges and their orientations are shown in black. The orientations were chosen arbitrarily but consistent with the rules stated in the text.}
    \label{Fig:RibbonOnFace}
\end{figure}

\begin{example}
We work out the cases shown in Fig.~\ref{Fig:RibbonOnFace}.
First, we substitute the ribbon elements from \eqref{Eq:FS}
into \eqref{Eq:HP}. For convenience, the oriented edges
$r\cdot e$ and $\bar r\cdot\hat e$ are already shown in
the diagrams, allowing a direct evaluation of the
$w$-factors. In case (a), the dual ribbon crosses the boundary of $f$
twice, with opposite contributions, giving
$\chi_{[F_{\tilde\rho}^\varphi]}(f)
=\varphi\varphi^{-1}=1$.
Similarly, in case (d),
$\chi_{[F_\rho^g]}(\tilde f)=gg^{-1}=1$.
Each remaining case has a single intersection point.
Cases (b) and (e) correspond to initial endpoints, giving
$\chi_{[F_{\tilde\rho}^\varphi]}(f)=\varphi^{-1}$
and $\chi_{[F_\rho^g]}(\tilde f)=g^{-1}$, respectively.
Cases (c) and (f) correspond to terminal endpoints, giving
$\chi_{[F_{\tilde\rho}^\varphi]}(f)=\varphi$
and $\chi_{[F_\rho^g]}(\tilde f)=g$, respectively.
These outcomes agree with \eqref{Eq:ChiF}.
\end{example}
\begin{proposition}
The following is a convenient presentation of the quotient group:
    \begin{equation}\label{Eq:PL}
    P_1/L \;\simeq\; \big\{(\hat P,P) \in C_c( \Sigma_2,\widehat G) \times C_c(\widetilde \Sigma_2,G)\; |\; \prod \hat P( f) =1,\ \prod P(\tilde f)=1\big\}. 
    \end{equation}
    Furthermore, if $[p]=(\hat P,P) \in P_1/L$ and $\chi = \hat \phi \phi  \in \widehat L$, then 
    \[
    [p] \triangleright \chi \;=\; (\hat P \hat \phi)(P \phi).
    \]
\end{proposition}

\begin{proof}
    We know that $[p] \mapsto \chi_{[p]}$ is an embedding of $P_1/L$ into $\widehat L$. Using \ref{Prop:ChiF} and the fact that the ribbon elements generate $P_1$, we see that this embedding is into the right side of \eqref{Eq:PL}. Reciprocally, any $\hat P$ and $P$ satisfying the constraints from the statement can be written as finite products of characters as in \eqref{Eq:ChiF}. 
\end{proof}
\begin{proposition}[\cite{PoloProdan2026}]
   The group of sectors is given by $\Gamma_{\LM}=\Xi_{\LM}/(P_1/L) \simeq G \times \widehat G.$
\end{proposition}

We now focus on the computation of $ Z_{\LM}$ and verify that it is contained in $\Xi_{\LM}$. If $\chi = \hat \phi \phi \in \widehat L$ and $[p]=(\hat P,P) \in P_1/L$, then \eqref{Eq:GoodCL} gives
\[
\begin{aligned}
c_{\LM}([p],\chi)  \;=\;&   |{\rm supp}\, \hat P \cap {\rm supp} (\hat P \hat \phi)| -  |{\rm supp}\, \hat P \cap {\rm supp}\, \hat \phi| \\
& + |{\rm supp}\, P \cap {\rm supp}(P \phi)| -  |{\rm supp}\, P \cap {\rm supp}\, \phi|
\end{aligned}
\]
\begin{proposition}
Let $\phi_ f^\varphi\in C_c(\Sigma_2,\widehat G)$ assign value $\varphi$ to face $f$ and value 1 to all the other faces, and similarly $\phi_{\tilde f}^g\in C_c(\widetilde \Sigma_2,G)$ assign  $g$ to $\tilde f$ and 1 otherwise. Then
    \[
    Z_{\LM} \;=\;\big\{  \phi_{ f}^\varphi \phi_{\tilde f}^g \; |\; (f,\tilde f)\in \Sigma_2 \times \widetilde \Sigma_2, \ (g,\varphi) \in G \times \widehat G\big\}\;\subset\; \Xi_{\LM}.
    \]

\end{proposition}
\begin{proof}
Each $\chi=\phi_{ f}^\varphi \phi_{\tilde f}^g $ has support
containing at most one face and at most one dual face. On the other hand, every non-trivial component $(\hat P,P)\in P_1/L$ has support containing at least
two faces, since
$\prod_f\hat P(f)=1$ and $\prod_{\tilde f}P(\tilde f)=1$.
Multiplication by $\chi$ can cancel at most one contribution
in each component, so
$$
\big|\operatorname{supp}(\hat P P\chi)
\cap\operatorname{supp}(\hat P P)\big|
\;\geq\;
\big|\operatorname{supp}\chi
\cap\operatorname{supp}(\hat P P)\big|.
$$
Thus, $c_{\LM}((\hat P,P),\chi)\geq0$, and hence $\chi\in Z_{\LM}$. We now show that these are the only characters in $Z_{\LM}$.
Suppose $\chi=\hat\phi\phi$ and that
$\hat\phi(f_i)\neq1$ at two distinct faces $f_1,f_2$.
Define $\hat P\in C_c(\Sigma_2,\widehat G)$ by
$$
\hat P(f_1)\;=\;\hat\phi(f_1)^{-1},
\qquad
\hat P(f_2)\;=\;\hat\phi(f_1),
\qquad
\hat P(f)\;=\;1 \qquad f\notin\{f_1,f_2\}.
$$
Then $\hat P$ satisfies the constraint in \eqref{Eq:PL},
and $\operatorname{supp}\hat P=\{f_1,f_2\}$.
Since $(\hat P\hat\phi)(f_1)=1$, we have
$$
\big|\operatorname{supp}\hat P
\cap\operatorname{supp}(\hat P\hat\phi)\big|
\;\leq\;1
\;<\;2
\;=\;
\big|\operatorname{supp}\hat P
\cap\operatorname{supp}\hat\phi\big|.
$$
Hence $c_{\LM}((\hat P,1),\chi)<0$, so
$\chi\notin Z_{\LM}$. The same argument applies if $\phi$ takes non-trivial
values at two or more dual faces, completing the proof.
\end{proof}

The preceding computation shows that
$Z_{\LM}\subseteq\Xi_{\LM}$. Hence, the results of Section~\ref{Sec:SpaceGS} yield the
following classification of the ground-states of the abelian
quantum double model, recovering the classification obtained
in \cite{CNN} and refining it through an explicit
parametrization of the pure ground-states within each sector.
\begin{theorem}
For the abelian quantum double model, the boundary groupoid
$\G^b_{\LM}$ has $|G|^2$ orbits
$$
\mathscr O_{(g,\varphi)}
\;=\;
\big\{\phi_f^\varphi\phi_{\tilde f}^{g}\mid 
(f,\tilde f)\in\Sigma_2\times\widetilde\Sigma_2\big\},
\qquad (g,\varphi)\in G\times\widehat G.$$
The pure ground-states are parametrized by
\[
\operatorname{Ext}(\mathfrak G_{\LM})
\;\simeq\;
\bigsqcup_{(g,\varphi)\in G\times\widehat G}
\mathbb P\bigl(\ell^2(\mathscr O_{(g,\varphi)})\bigr),
\]
where
\[
\mathscr O_{(g,\varphi)}\;\simeq\;
\begin{cases}
\{\chi_0\}, & g=1,\ \varphi=1,\\
\widetilde\Sigma_2, & g\neq1,\ \varphi=1,\\
\Sigma_2, & g=1,\ \varphi\neq1,\\
\Sigma_2\times\widetilde\Sigma_2, & g\neq1,\ \varphi\neq1.
\end{cases}
\]
The trivial sector consists of the single frustration-free state $s_{\chi_0}$.
\end{theorem}

\subsection{Color code model} In this section we compute the set of ground-states and the sectors of the color code model \cite{Bombin2012,CaoJiaTan2026,Kubica2015}.

\begin{figure}
    \centering
    \includegraphics[width=0.8\linewidth]{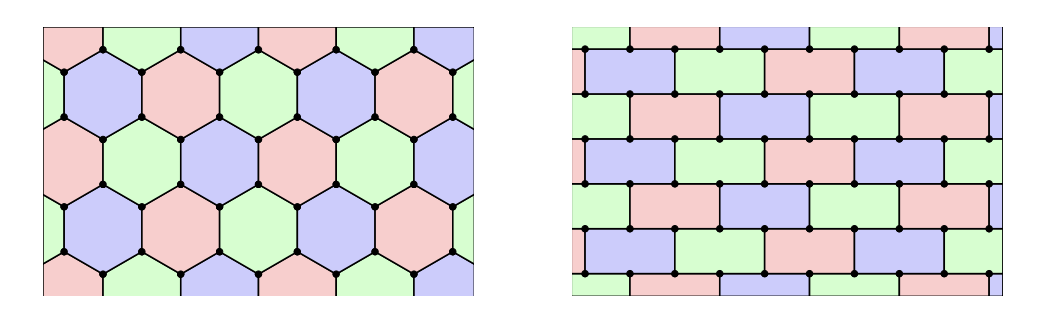}
    \caption{Figure taken from \cite[Figure 1]{CaoJiaTan2026}. Examples of trivalent cellulations  of the plane.}
    \label{fig:placeholder}
\end{figure}

\medskip 
Let $\Sigma$ be a trivalent cellulation of the plane as in
Figure~\ref{fig:placeholder}, equipped with the indicated
proper $3$-coloring
${\rm col}\colon \Sigma_2\to\{r,g,b\}$.
Thus, the three faces incident to each vertex have distinct
colors, and every face has six boundary edges.
The dual cellulation $\widetilde\Sigma$ is a triangulation whose vertices inherit a proper $3$-coloring from the faces of $\Sigma$:
if $\widetilde v_f$ is dual to $f$, then $\operatorname{col}(\widetilde v_f):=\operatorname{col}(f)$. We consider $G=\Z_2$ and the Pauli module,
$$
P_0\;=\; C_c(\Sigma_0,\Z_2)\oplus C_c(\widetilde{\Sigma}_2,\widehat{\Z_2})$$
endowed with the standard cocycle $\gamma_0$.  The observable algebra coincides with the spin $\frac{1}{2}$-algebra
$\A= C^*(P_0, \gamma_0)\simeq {\rm M}_2^{\otimes \Sigma_0}$. Consequently, the qubits and their associated Pauli operators are placed at the vertices of the lattice.

\medskip 

For each $f\in\Sigma_2$ and each
$\tilde  v\in\widetilde{\Sigma}_0$, define 
$l_f\in C_0(\Sigma,\mathbb Z_2)$ and $
l_{\tilde v}\in
C_2(\widetilde{\Sigma},\widehat{\mathbb Z}_2)$
by 
$$
l_f(v)\;=\;
\begin{cases}
-1,& v\in\partial f,\\
1,& \text{otherwise},
\end{cases}
\qquad
l_{\tilde v}(\tilde f)\;=\;
\begin{cases}
\varepsilon_1,& \tilde v\in\partial\tilde f,\\
\varepsilon_0,& \text{otherwise},
\end{cases}
$$
where $\varepsilon_1$ and $\varepsilon_0$ denote the nontrivial and
trivial characters of $\mathbb Z_2$, respectively, and $\partial f$ stands for the vertices in the boundary of the face $f$. The color-code
stabilizer subgroup is
\begin{equation}\label{eq: generators cc}
L
\;:=\;
\left\langle
(l_f,1),(1,l_{\tilde v})
\mid
f\in\Sigma_2,\;
\tilde v\in\widetilde{\Sigma}_0
\right\rangle
\;\subseteq\; P_0.
\end{equation}
\begin{proposition}\label{prop: code isotropic}
    $L$ is isotropic.
\end{proposition}
\begin{proof}
    It is enough to verify the condition on the generators. Indeed, let us first notice
$$
\omega_0\big((l_f,1),(l_g,1)\big)\;=\;1,
\qquad
\omega_0\big((1,l_{\widetilde v}),(1,l_{\widetilde w})\big)\;=\;1.$$
Moreover, for a face $f\in\Sigma_2$ and a dual vertex
$\widetilde v\in\widetilde\Sigma_0$ which is dual to the face $g\in \Sigma_2$ via $\beta$ (hence $g = \tilde{\tilde v}$) one has
$$\omega_0\big((l_f,1),(1,l_{\widetilde v})\big)\;=\;\prod_{x\in\partial f\cap\partial g}\varepsilon_1(-1)\;=\;(-1)^{|\partial f\cap\partial g|}.$$
If $f\neq g$, then the two faces are either disjoint or share an edge, and hence $|\partial f\cap\partial g|$ is either $0$ or $2$. If $f=g$,
then $|\partial f|$ is six. Therefore, for all the cases
$
\omega_0
\big((l_f,1),(1,l_{\widetilde v})\big)=1,
$ and $L$ is isotropic.
\end{proof}

From now on, we denote by $\LM$ the set of generators of $L$
described in \eqref{eq: generators cc}, and we identify
$\widehat{\mathbb Z}_2$ with $\mathbb Z_2=\{1,-1\}$. For each cell $c$, let $\delta_c$ denote the configuration
equal to $-1$ at $c$ and $1$ elsewhere.
We  define the incidence map $d_\Sigma\colon C_c(\Sigma_2,\mathbb Z_2)\to C_c(\Sigma_0,\mathbb Z_2)$ determined by the relation
$$
d_\Sigma(\delta_f)\;:=\;\prod_{v\in\partial f}\delta_v,
$$

and similarly $
d_{\widetilde\Sigma}\colon C_c(\widetilde\Sigma_2,\widehat{\mathbb Z}_2)
\to
C_c(\widetilde\Sigma_0,\widehat {\mathbb Z}_2).$   They have as a adjoints the maps
$$d_{\Sigma}^*(\delta_v)\;=\;\prod_{v\in \partial f}\delta_f, \qquad  d_{\widetilde \Sigma}^*(\delta_{\tilde v})\;=\;\prod_{\tilde v\in \partial \tilde f}\delta_{\tilde f}$$
\begin{proposition}\label{prop: image}
    It holds true that
    $L={\rm Im}(d_\Sigma)\oplus {\rm Im}(d^*_{\widetilde \Sigma})$.
\end{proposition}
\begin{proof}
For every face $f\in\Sigma_2$, one has
$$
d_\Sigma(\delta_f)
\;=\;
\prod_{v\in\partial f}\delta_v
\;=\;
l_f.$$
Similarly, for every dual vertex $\widetilde v\in\widetilde\Sigma_0$, one gets $
d_{\widetilde\Sigma}^{*}(\delta_{\widetilde v})
=
l_{\widetilde v}.$ 
Since the elements $\delta_f$ generate
\(C_2(\Sigma,\mathbb Z_2)\), it follows that $
\left\langle l_f\mid f\in\Sigma_2\right\rangle
=
\operatorname{Im}(d_\Sigma).$ Likewise,   $
\left\langle
l_{\widetilde v}
\mid
\widetilde v\in\widetilde\Sigma_0
\right\rangle
=
\operatorname{Im}(d_{\widetilde\Sigma}^{*}).$
The conclusion follows from the definition of \(L\).
\end{proof}
\begin{proposition}
    $L$ is a Lagrangian subgroup of $P_0$.
\end{proposition}
\begin{proof}
   We verified that $L$ is isotropic in Proposition \ref{prop: code isotropic}, so it remains
to show that $L^\perp=L$.  Let $l=(l_1,l_2)\in L^\bot$ be non-trivial, then $$\omega_0(l,l_f)\;=\;1\;=\;\omega_0(l,l_{\tilde v})\qquad \forall f, \tilde v$$
Assume that the lattice support of $l=(l_1,l_2)$ in $\Sigma$ coincides with finite region $\Lambda$ of $\Sigma.$ The above orthogonality relations imply that
$l_1$ is non-trivial in the boundary of every $f\in \Lambda_2$ in an even number of vertices,
and similarly $l_2$ is non-trivial in the star of every $\tilde v\in \widetilde \Lambda_0$ in an even
number of faces. Hence $l_1\in {\rm Ker}(d_{\Sigma}^*)$ and $l_2\in {\rm Ker}(d_{\widetilde \Sigma}).$ To conclude, the standard description of color-code cycles
on the shrunk lattices, together with the contractibility
of the plane, implies exactness of the color-code incidence
complex at its middle term, yielding
$${\rm Im}(d_\Sigma)\;=\;{\rm Ker}(d^*_{\Sigma}),\qquad {\rm Im}(d_{\widetilde \Sigma}^*)\;=\; {\rm Ker}(d_{\widetilde \Sigma})$$
and hence $l\in L$ by Proposition \ref{prop: image}.
\end{proof}

Now we can describe the color-code dynamics. Consider the face and vertex operators 
\begin{equation}
    U_f\;=\;U_{(l_f,1)},\qquad U_{\tilde v}\;=\;U_{(1,l_{\tilde v})}
\end{equation}
which are local elements in $\A$. They provide the local interactions for the color code model given by
$$P_f\;=\;\frac{{\bf 1}+U_f}{2},\qquad P_{\tilde v}\;=\;\frac{{\bf 1}+U_{\tilde v}}{2}$$
For a finite region $\Lambda$ of $\Sigma$ one can consider the net of local code Hamiltonian
\begin{equation}
    H_{\LM}^\Lambda\;=\;\sum_{f\in \Lambda_2} \big({\bf 1}-P_f\big)+\sum_{\tilde v\in\widetilde \Lambda_0}\big({\bf 1}-P_{\tilde v}\big)
\end{equation}
which induce the dynamics $\bm{\delta_{\LM}}$ on $\A.$ We begin  by recalling following characterization recently established in \cite[Theorem 2.5]{CaoJiaTan2026},  which also follows from our general framework  by Corollary \ref{coro: unique FF}.
\begin{proposition}
    The color code dynamics $\delta_{\LM}$ has a unique frustration-free gapped ground-state.
\end{proposition}
Now we proceed to compute the full set of ground-states for $\delta_{\LM}$. We start by describing the group of local interactions $P_0/L$. Recall that $P_0$ is generated by the $\delta_v\in C_c(\Sigma_0,\Z_2)$ and $\delta_{\tilde f}\in C_c(\widetilde \Sigma_2,\Z_2)$ for $v\in \Sigma_0$ and $\tilde f\in \widetilde \Sigma_2.$  
 \begin{proposition}\label{prop: Lcc}
The group $\widehat L$ can be identified with the Cantor space
     $$\widehat L\;\simeq \; \Z_2^{\Sigma_2}\times \Z_2^{\widetilde\Sigma_0}$$
     and the action of $P_0/L$ on $\widehat L$ is given by translations by the subgroup
$$
L'
\;:=\;
\operatorname{Im}(d_\Sigma^*)
\oplus
\operatorname{Im}(d_{\widetilde\Sigma})
\;\subseteq\;
\mathbb Z_2^{\Sigma_2}
\times
\mathbb Z_2^{\widetilde\Sigma_0}.$$
 \end{proposition}
 \begin{proof}
     The first part is a direct consequence of that there are no relation between the elements $l_f$ and $l_{\tilde v}$ and moreover, they satisfy $l_f^2=1 ={l_{\tilde v}}^2$.
     Thus, $\widehat L$ is topologically generated by the dual elements determined by the relations
     $$\hat{l}_f(l_{f'})\;=\; (-1)^{\delta_{f,f'}},\qquad \hat{l}_{\tilde v}(l_{\tilde v'})\;=\; (-1)^{\delta_{\tilde v,\tilde v'}}$$
     On the other hand, recall that the action of $P_0/L$ is given by translations of the subgroup of characters
     $\chi_{[p]}(l)= \omega_0(p,l)$.      Since $P_0$ is generated by the elements $\delta_v$,
 and $\delta_{\tilde f}$, it is enough to compute the
corresponding characters. For every face $f\in\Sigma_2$ and dual vertex
$\tilde v\in\widetilde\Sigma_0$, one has
$$\chi_{[\delta_v]}(l_f)
\;=\;1
,\qquad 
\chi_{[\delta_v]}(l_{\tilde w})\;=\;\begin{cases}
-1,&\tilde{w}\in\partial \tilde f\;\;\text{and}\;\;\beta_0(v,\tilde f)=1,\\
1,&\text{otherwise}
\end{cases}.
$$
Similarly,
$$
\chi_{[\delta_{\widetilde f}]}(l_{\tilde v})\;
=\;
1,\qquad 
\chi_{[\delta_{\widetilde f}]}(l_g)\;=\;\begin{cases}
-1,&v\in\partial  g\;\;\text{and}\;\;\beta_0(v,\tilde f)=1,\\
1,&\text{otherwise}\end{cases}.
$$
Therefore, 
the subgroup of translations is precisely $
\operatorname{Im}(d_\Sigma^{*})
\oplus
\operatorname{Im}(d_{\widetilde\Sigma}),$
where a primal vertex flips the three dual vertices on the
boundary of its dual face, and a dual face flips the three
primal faces incident to its dual vertex.
 \end{proof}

Let us notice that the set of generators $\LM$ of $L$ 
provides an admissible presentation of $L$ since they are uniformly localized in the cellulation $\Sigma$ and $\widetilde \Sigma$.  Moreover, in this case, the group $\Xi_{\LM}$ of finitely supported functions agrees with
$$
\Xi_{\LM}
\;=\;
C_c(\Sigma_2,\mathbb Z_2)
\oplus
C_c(\widetilde\Sigma_0,\widehat {\mathbb Z_2})
\;\subseteq
\;\widehat L
$$

The idea now is to compute the group of sectors $\Gamma_{\LM}=\Xi_{\LM}/(P_0/L)$. Indeed,  for $
x\in C_2(\Sigma,\mathbb Z_2),$
define its color parities by
$$
q_c(x)
\;:=\;
\prod_{\substack{f\in\Sigma_2\\
\operatorname{col}(f)=c}}
x(f),
\qquad
c\in\{r,g,b\}.$$
This product is well-defined since $x$ is finitely supported. Similarly, using the color of the primal face dual to
$\widetilde v\in\widetilde\Sigma_0$, define
$$
\widetilde q_c(y)
\;:=\;
\prod_{\substack{\widetilde v\in\widetilde\Sigma_0\\
\operatorname{col}(\widetilde v)=c}}
y(\widetilde v),\qquad
y\in C_0(\widetilde\Sigma,\widehat{\mathbb Z_2}).$$
Now we consider the charge map $\varphi \colon \Xi_{\LM}\to\mathbb Z_2^4$
given by
$$
\varphi(x,y)
\;=\;
\Big(
q_r(x)q_g(x),\,
q_r(x)q_b(x),\,
\widetilde q_r(y)\widetilde q_g(y),\,
\widetilde q_r(y)\widetilde q_b(y)
\Big).
$$

\begin{proposition}\label{prop: sectors cc}
The group of sectors of the planar color code  satisfies
$
\Gamma_{\LM}\simeq \Z_2^4$.
\end{proposition}
\begin{proof}
Since every vertex is incident to precisely one red, one green and one
blue face, one has
$$
d_\Sigma^*(\delta_v)
\;=\;
\delta_{f_r(v)}\,
\delta_{f_g(v)}\,
\delta_{f_b(v)}.$$
Thus, translation by $d_\Sigma^*(\delta_v)$ changes the three color
parities simultaneously
$
(q_r,q_g,q_b)
\mapsto
(-q_r,-q_g,-q_b).$
Therefore, the two quantities $
q_r\,q_g,$ $q_r\,q_b$
are invariant under $\operatorname{Im}(d_\Sigma^*)$. Likewise, every dual face has precisely one vertex of each color on its
boundary. Hence $
\widetilde q_r\,\widetilde q_g$ and $
\widetilde q_r\,\widetilde q_b$ are also invariant under ${\rm Im}(d_{\widetilde \Sigma})$. By Proposition \ref{prop: Lcc}, it follows that $\varphi$ is invariant under $P_0/L$ and hence constant on every
$P_0/L$-orbit. Thus $\varphi$ drops to a well-defined surjective group homomorphism $\tilde \varphi\colon \Gamma_{\LM}\to \Z_2^4$.  To finish the proof is enough to verify that $\tilde\varphi $ is injective. Indeed, suppose that $
\varphi(x,y)=\varphi(x',y').$ The color-parity vector of $x\,x'$ is either $(1,1,1)$
or $(-1,-1,-1)$. In the latter case, multiplying by
$d_\Sigma^*(\delta_v)$ for any vertex $v$ changes all
three parities to 1. The remaining excitations
can then be paired by color and annihilated using
finite products of string elements
\cite[Section 3.2]{CaoJiaTan2026}. Since
$d_\Sigma^*(\delta_v)\in\operatorname{Im}(d_\Sigma^*)$,
we conclude that $
x\,x'\in\operatorname{Im}(d_\Sigma^*).$ The same argument on the dual lattice, using
$d_{\widetilde\Sigma}(\delta_{\tilde f})$ for a dual
face $\tilde f$, gives $
y\,y'\in\operatorname{Im}(d_{\widetilde\Sigma}).$
Therefore, $(x',y')(x,y)^{-1}\in P_0/L,$ so $(x,y)$ and $(x',y')$ lie in the same orbit and $[(x,y)]=[(x',y')] $. This proves the injectivity and completes the proof. 
\end{proof}

For each  $f\in\Sigma_2$ and 
$\widetilde v\in\widetilde\Sigma_0$, consider the characters
$\widehat l_f,\widehat l_{\widetilde v}\in\widehat L$ defined in the proof of Proposition \ref{prop: Lcc}.  We define
$$
\chi_{f,\widetilde v}
\;:=\;
\widehat l_f\widehat l_{\widetilde v}
,\quad \chi_{0,\widetilde v}\;=\; \widehat l_{\widetilde v},\quad \chi_{f,0}\;=\;\widehat l_f,\quad \chi_{0,0}\;=\;\chi_{0}
$$
These characters provide a description of the boundary set $Z_{\LM}.$
\begin{proposition}\label{prop: cc Z}
    It holds true that
    $$Z_{{\LM}}\;=\;\big\{\chi_{f,\widetilde v},\,\chi_{0,\widetilde v},\, \chi_{f,0},\,\chi_0\mid f\in \Sigma_2,\; \widetilde v\in \widetilde \Sigma_0\big\}\;\subset \; \Xi_{\LM}$$
\end{proposition}
\begin{proof}
    This is a consequence of the existence of string elements for the color code model \cite[Section 3.2]{CaoJiaTan2026}. Indeed, let $\chi\in Z_{\LM}$ be non-trivial in two generators $l_{f_1}$ and $l_{f_2}$, \ie 
    $$\chi(l_{f_1})\;=\;-1\;=\;\chi(l_{f_2})$$
 If $f_1$ and $f_2$ have different colors, choose a face
$f_3$ of the remaining color. The configuration
$\widehat l_{f_1}\widehat l_{f_2}\widehat l_{f_3}\in \Xi_{\LM}$
has trivial parity under $\varphi$, so Proposition~\ref{prop: sectors cc}
provides $p\in P_0$ such that
$\chi_{[p]}=\widehat l_{f_1}\widehat l_{f_2}\widehat l_{f_3}$.
Thus, $c_{\LM}([p],\chi)$ equals $-1$ or $-3$,
depending on whether $\chi(l_{f_3})=1$ or $-1$,
respectively, giving a contradiction.  If $f_1$ and $f_2$ have the same color, there exists a string element $[p]\in P_0/L$ such that $\chi_{[p]}(l_{f_1})=-1=\chi_{[p]}(l_{f_2})$ and trivial elsewhere \cite[Section 3.2]{CaoJiaTan2026}. Hence
$c_{\LM}([p],\chi)= -2$, a contradiction. The previous argument shows that $\chi$ can be non-trivial on at
most one face generator $l_f$. By applying the same argument to the dual
lattice, $\chi$ can also be non-trivial on at most one vertex generator
$l_{\widetilde v}$. Therefore,
 $$Z_{{\LM}}\;\subset\;\big\{\chi_{f,\widetilde v},\,\chi_{0,\widetilde v},\, \chi_{f,0},\,\chi_0\mid f\in \Sigma_2,\; \widetilde v\in \widetilde \Sigma_0\big\}$$
Conversely, any non-trivial translation induced
by $[p]$ has support of size at least two in each affected component. Therefore, a configuration containing at most one excitation of each
type cannot decrease its energy under such a translation and hence
belongs to $Z_{\LM}$.
\end{proof}

The preceding proposition ensures that the results of
Section~\ref{Sec:SpaceGS} apply to the color code dynamics,
yielding an explicit description of its full ground-state
space. We now construct representative ground-states
in each sector. Fix faces $f_r,f_g,f_b\in\Sigma_2$ and  vertices
$\widetilde v_r,\widetilde v_g,\widetilde v_b
\in\widetilde\Sigma_0$, one of each color. Set
$
f_0=\widetilde v_0=\varnothing,$ such that
$\widehat l_{f_0}
=
\widehat l_{\widetilde v_0}
=
1.$ Then the sixteen sectors admit the representatives
$$
\left\{
\chi_{f_c,\widetilde v_d}
\;\middle|\;
c,d\in\{0,r,g,b\}
\right\}.$$
Let $\SM_{c,d}\equiv \SM_{\chi_{f_c,\widetilde v_d}}$ denote the unique pure extension to $\A$. We now present the main result of this section, which,
to the best of our knowledge, provides the first complete
classification of the ground-states of the planar color
code model.
\begin{theorem}\label{thm: color-code-ground-states}
The pure ground-states of the  color-code dynamics
$\delta_{\LM}$ form exactly sixteen unitary equivalence
classes, represented by
$\{\SM_{c,d}\mid c,d\in\{0,r,g,b\}\}$.
In particular, the state $\SM_0\equiv \SM_{0,0}$ is the unique frustration-free
gapped ground-state of the model.
\end{theorem}
\subsection{A biased one-dimensional Ising model}\label{sec: biased ising}
Let $\Sigma$ be the standard CW-decomposition of $\mathbb R$,
with vertices $\Sigma_0=\mathbb Z$ and edges $\Sigma_1$ joining
consecutive integers. We orient each edge of $\Sigma_1$ from $v$ to $v+1$. For each edge $e\in\Sigma_1$ we denote by $e_-$ and $e_+$ its initial and final vertex, respectively. The relevant group for the Ising model is 
\[
P_0 \;=\; C_c(\Sigma_0,\mathbb Z_2) \oplus C_c(\widetilde \Sigma_1, \widehat{\mathbb Z}_2).
\]
With the standard extension $\gamma_0$, the algebra of  observables becomes $\A \simeq {\rm M}_2^{\otimes \Sigma_0}.$
Pauli matrices $\sigma^a$, $a=x,y,z$ with the identity supply a convenient basis for ${\rm M}_2$, and we will write $\sigma_v^a$ for the Pauli matrix located at vertex $v \in \Sigma_0=\mathbb Z$ in ${\rm M}_2^{\otimes \Sigma_0}$.

 Let the Lagrangian subgroup be $L=C_c(\Sigma_0,\mathbb Z_2)$, whose corresponding sub-$C^\ast$-algebra is the standard diagonal of ${\rm M}_2^{\otimes \Sigma_0}$ and $P_0/L = C_c(\widetilde \Sigma_1, \widehat{\mathbb Z}_2)$. However, we will choose the following generating set of $L$:
\[
l_0(v) \;=\; \left \{
\begin{aligned}
    & -1,\; \ {\rm if} \ v=0\\
    & +1,\; \ {\rm otherwise}
\end{aligned} \right . , \qquad
  l_e(v) \;=\; \left \{
  \begin{aligned}
      & -1,\; \ {\rm if} \ v \in \{e_+,e_-\}\\
      & +1,\; \ {\rm otherwise}
  \end{aligned} \right .
  \]
with one $l_e$ for each $e\in \Sigma_1$. Inside ${\rm M}_2^{\otimes \Sigma_0}$, these are represented by the unitary operators
 \[
U_{l_0}\;=\;\sigma_0^z,
\qquad
U_{l_e}\;=\;\sigma_{e_-}^z\sigma_{e_+}^z,
\]
  which indeed generate the standard diagonal $\A_L=C^\ast(\sigma_v^z\mid  v \in \Sigma_0)$ of ${\rm M}_2^{\otimes \Sigma_0}$. The dynamics associated to this generating set is the formal Hamiltonian 
\begin{equation}\label{Eq:IsingH}
H_L \;=\;(1-P_{l_0}) + \sum_{e\in \Sigma_1} (1-P_{l_e})\; =\; \tfrac{1}{2}(1-\sigma_0^z) + \tfrac{1}{2}\sum_{v\in \mathbb Z} (1-\sigma_v^z \sigma_{v+1}^z)
\end{equation}
which is the classic Ising Hamiltonian biased at site $0$. Since $L$ is an admissible Lagrangian subgroup, one obtains that it has a unique gapped frustration-free ground-state by Corollary \ref{coro: unique FF}.

We now consider the following convenient generating set for $P_0/L = C_c(\widetilde \Sigma_1,\widehat{\mathbb Z}_2)$. Let $\tilde \rho$ be a finite continuous ribbon made up of edges in $\widetilde \Sigma_1$ and $\varepsilon_1$ be the non-trivial character of $\mathbb Z_2$, and define the elements of $P_0/L$
\begin{equation}\label{Eq:FString}
F_{\tilde \rho}(\tilde e) \;=\;\left \{ 
\begin{aligned}
    & \varepsilon_1,\; \ {\rm if} \ \tilde e \in \tilde \rho\\
    & \varepsilon_0,\; \ {\rm otherwise}
\end{aligned} \right .
\end{equation}
which in the algebra of physical observables are represented by $\prod_{\tilde e \in \tilde \rho} \sigma_{\tilde{\tilde e}}^x.$ Then $F_{\tilde \rho}$ generate $C_c(\widetilde \Sigma_1,\widehat{\mathbb Z}_2)$. The ribbon will be oriented as the real axis is, and the end (start) vertices of $\tilde \rho$ will be denoted by $\tilde \rho_+ \in \widetilde \Sigma_0$ ($\tilde \rho_- \in \widetilde \Sigma_0$).  There is one $l_j$ generator per each 1-cell of $\Sigma$, hence we have the following convenient presentations:
\begin{equation}\label{Eq:LLHat}
L \;\simeq\; C_c(\{v_0\} \cup \Sigma_1,\mathbb Z_2), \quad \widehat L \;\simeq\; \mathbb Z_2 \times \prod_{\Sigma_1} \mathbb Z_2, \quad \Xi_{\LM} \;\simeq\; C_c(\{v_0\}\cup\Sigma_1,\mathbb Z_2).
\end{equation}
where $v_0=0 \in \Sigma_0$ is the site carrying $l_0$. There is a major advantage in these presentations. Indeed, the support of a character was defined in \eqref{Eq:Support} as the set of generators that are evaluated by the character to non-trivial values. In \eqref{Eq:LLHat}, the support of a character coincides with the support of the actual function, and this is again a crucial aspect of our analysis.

\begin{proposition}\label{Prop:ChiString} The elements $F_{\tilde \rho}$ generate $P_0/L$ and, in the presentations \eqref{Eq:LLHat}, we have
 \[
    \chi_{[F_{\tilde \rho}]}(e) \;=\; \left \{
    \begin{aligned}
        & -1,\; \ {\rm if} \ e \in \{\widetilde{\tilde \rho_+},\widetilde{\tilde \rho_-}\} \\
        & +1,\; \ \rm{\rm otherwise}
    \end{aligned} \right. ,\qquad \chi_{[F_{\tilde \rho}]}(v_0) \;=\; \left \{
    \begin{aligned}
        & -1,\; \ {\rm if} \ v_0 \in \tilde{\tilde{\rho}}\\
        & +1,\; \ \rm{\rm otherwise}
    \end{aligned} \right. 
    \]
    where $\widetilde{\tilde\rho_\pm}$ are the edges
dual to the endpoints of $\tilde\rho$, and
$\tilde{\tilde\rho}$ denotes the set of vertices
dual to the edges of $\tilde\rho$.
\end{proposition}

\begin{proof}
    This is a direct consequence of the representations inside the algebra $\A.$
\end{proof}

\begin{proposition}\label{Prop:IsingPL} It holds that
    $\Gamma_{\LM} = \Xi_{\LM}/(P_0/L) \simeq \mathbb Z_2 \times \mathbb Z_2$.
\end{proposition}

\begin{proof}
Let $\chi_-\in\Xi_{\LM}$ be the character assigning $-1$
to $v_0=0$ and $+1$ to all edges. We introduce the subsets of edges $\Sigma_1^<=\{e \in \Sigma_1\mid e_+ \leq 0\}$ and $\Sigma_1^> = \{e \in \Sigma_1\mid e_- \geq 0\}$, as well as two
configurations $\chi_<,\chi_>\in\Xi_{\LM}$ such that
\[
{\rm supp}\,\chi_<
\;\subseteq\;
\Sigma_1^<,
\qquad
|{\rm supp}\,\chi_<|\equiv1\pmod2,
\]
\[
{\rm supp}\,\chi_>
\;\subseteq\;
\Sigma_1^>,
\qquad
|{\rm supp}\,\chi_>|\equiv1\pmod2.
\]
In particular, $\chi_<(v_0)=\chi_>(v_0)=1$. We claim that $\chi_0,\chi_-,\chi_<,\chi_>$ belong to
distinct orbits of $\G_L$, where recall that $\chi_0$ is the trivial character. For this, we must examine the effects of applying the ribbons \eqref{Eq:FString}, which reduce to multiplications by the characters introduced in \ref{Prop:ChiString}. Applying on $\chi \in \Xi_\LM$ a ribbon whose endpoint
edges lie on the same side of the origin preserves the
parity of $|{\rm supp} \, \chi \cap \Sigma_1^<|$ and of $|{\rm supp}\, \chi \cap \Sigma_1^>|$, as well as the value
at $v_0$. A ribbon whose endpoint edges lie on opposite
sides changes both these parities and flips the value at $v_0$.
Thus, a product of ribbon elements taking $\chi_<$ to
$\chi_>$ would contain an odd number of ribbons crossing
the origin and would therefore change the value at $v_0$
to $-1$, a contradiction.
Moreover, application of any ribbon preserves the parity of
the total edge support, separating $\chi_<$ and $\chi_>$
from $\chi_0$ and $\chi_-$. Finally, an element of $P_0/L$
preserving both side parities contains an even number
of ribbons crossing the origin and therefore cannot take
$\chi_0$ to $\chi_-$.

To show that these four orbits exhaust $\Xi_{\LM}$,
let $\chi\in\Xi_{\LM}$. By forming pairs of edges in ${\rm supp}\, \chi \cap \Sigma_1^<$ and in ${\rm supp}\,\chi \cap \Sigma_1^>$, and by applying the ribbons joining each such pair, we
reduce ${\rm supp}\, \chi$ to at most one edge on each side of the origin.
If one edge remains on each side, a ribbon joining them
removes both and flips the value at $v_0$, leaving either
$\chi_0$ or $\chi_-$. If exactly one edge remains and
the value at $v_0$ is $-1$, a ribbon crossing the origin
moves that edge to the opposite side and changes this
value to $+1$. Finally, any two odd finite edge
configurations on the same side, with value $+1$ at
$v_0$, differ by a product of ribbons contained on that
side. Hence the remaining possibilities belong to
the orbits of $\chi_<$ or $\chi_>$.

Lastly, $\chi_<\chi_>$ belongs to the orbit of $\chi_-$,
and $\chi^2=\chi_0$ for every character. Consequently,
the assignments
$$
[\chi_0]\mapsto(+1,+1),\quad
[\chi_<]\mapsto(-1,+1),\quad
[\chi_>]\mapsto(+1,-1),\quad
[\chi_-]\mapsto(-1,-1)$$
define an isomorphism
$\Gamma_{\LM}\simeq\mathbb Z_2\times\mathbb Z_2$.
\end{proof}

\begin{proposition} Let $\chi_e \in \widehat L$ assign value $-1$ to edge $e\in \Sigma_1$ and value $+1$ to any other edge, as well as $\chi_e(v_0) =1$. Then
    \[
    Z_{\LM} \;=\; \{\chi_0\} \sqcup \{\chi_-\} \sqcup \{\chi_e\mid  e \in \Sigma_1\}.
    \]
    Furthermore, the orbits of the boundary groupoid consist of
    
  \[
\begin{aligned}
\mathscr O_{(+1,+1)} \;&=\; \{\chi_0\},
&\qquad \mathscr O_{(-1,-1)} \;&=\; \{\chi_-\},\\
\mathscr O_{(-1,+1)} \;&=\; \{\chi_e\mid e_+\leq0\},
&\qquad \mathscr O_{(+1,-1)}\; &=\; \{\chi_e\mid e_-\geq0\}.
\end{aligned}
\]
    Consequently,
    \[
   {\rm Ext}( \mathfrak G_{\LM})\;\equiv \;\{\mathtt s_{\chi_0}, \mathtt s_{\chi_-} \}\sqcup \mathbb P\big(\ell^2(\Sigma_1^<)\big )\; \sqcup \; \mathbb P\big(\ell^2(\Sigma_1^>)\big ).
    \]
\end{proposition}

\begin{proof}
    Let $\varphi$ be a character with support made up of at most one point. Since all non-trivial $\chi_{[p]}$ have supports made up of two or more points,  then, if $A \Delta B$ is the symmetric difference of sets $A,B$, we have
    \[
    \begin{aligned}
    c_{\LM}([p],\varphi) \;&  =\; |{\rm supp}\, \chi_{[p]} \varphi \cap {\rm supp}\, \chi_{[p]}| - |{\rm supp}\, \varphi \cap {\rm supp}\, \chi_{[p]}|\\
    & = \;|({\rm supp}\, \chi_{[p]}\; \Delta \; {\rm supp}\, \varphi) \cap {\rm supp}\, \chi_{[p]}| - |{\rm supp}\, \varphi \cap {\rm supp}\, \chi_{[p]}|\;\geq\; 0,
    \end{aligned}
    \]
    hence $\varphi \in Z_{\LM}$. On the other hand, if we consider a character with support made up of two or more points, we can always reduce their number by applying  a (finite) ribbon element. Since $ Z_{\LM}$ only contains compactly supported characters, the remaining statements are direct consequences of \ref{Prop:IsingPL} and of our general theory.
\end{proof}

\begin{figure}
    \centering
    \includegraphics[width=0.56\linewidth]{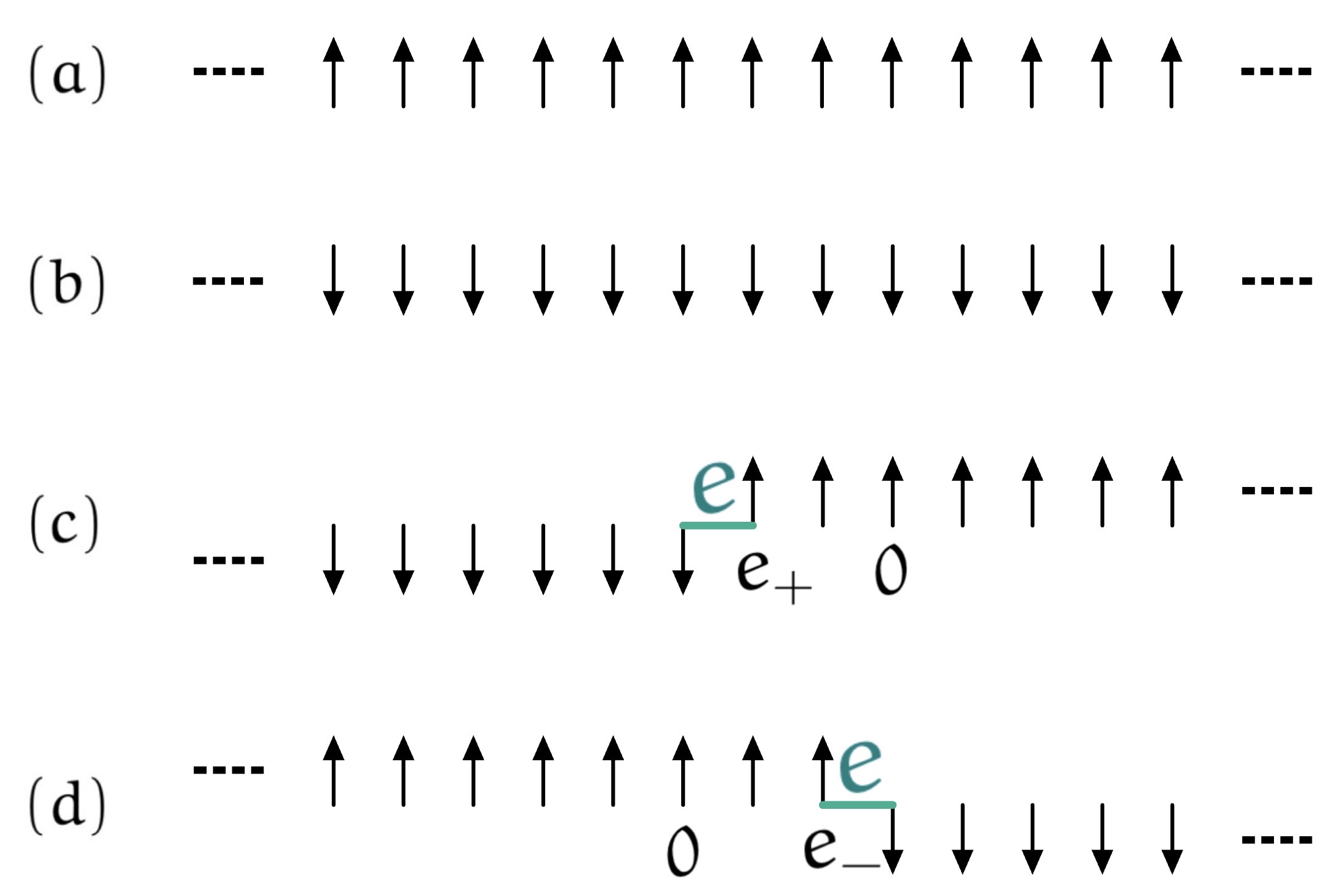} 
    \caption{Physical representation of the ground-states of the biased Ising model induced from (a) $\chi_0$, (b) $\chi_-$, (c) $\chi_e$ for $e \in \Sigma_1^<$, and (d) $\chi_e$ for $e \in \Sigma_1^>$. Here, the arrows are attached to the vertices from $\Sigma_0$ and an up (down) arrow (or spin) represents the element $+1$ ($-1$) of $\mathbb Z_2$.}
    \label{Fig:Ising}
\end{figure}

\begin{remark}
The states $\mathtt s_{\chi_0}$ and $\mathtt s_{\chi_-}$
are the all-up and all-down product states, respectively.
The states $\mathtt s_{\chi_e}$ describe a single domain
wall: a kink from $-1$ to $+1$ when $e\in\Sigma_1^<$,
and an antikink from $+1$ to $-1$ when $e\in\Sigma_1^>$.
The corresponding projective Hilbert spaces also include
pure ground states given by superpositions of
domain-wall positions within each sector.
Only the all-up state is frustration-free.   \hfill $\blacktriangleleft$
\end{remark}

There is a standard presentation of the Lagrangian $L= C_c(\Sigma_0,\mathbb Z_2)$ and of its dual $\widehat L = \prod_{\Sigma_0} \widehat{\mathbb Z}_2$. Here, the characters inducing the ground-states are represented as follows: $\chi_0(v) =1$ and $\chi_-(v)=-1$ for all $v \in \Sigma_0$, and for $e \in \Sigma_1^<$,
\[
\chi_e(v)\;=\; \left \{
\begin{aligned}
   & -1,\;\ {\rm if} \ v < e_+, \\
   & +1,\; \ {\rm otherwise},
\end{aligned}
\right . 
\]
while for $e \in \Sigma_1^>$,
\[
\chi_e(v)\;=\; \left \{
\begin{aligned}
   & -1 \ {\rm if} \ v > e_- ,\\
   & +1 \ {\rm otherwise}.
\end{aligned}
\right . 
\]
They are pictorially illustrated in Fig.~\ref{Fig:Ising} in terms of spin configurations.

\begin{remark}
    If instead we chose the standard generators of the Lagrangian $L=C_c(\Sigma_0,\mathbb Z_2)$, given by $l_j(v) =-1$ if $v=j$ and $l_j(v) =1$ otherwise, then $\Xi_{\LM} = C_c(\Sigma_0,\Z_2)$ and the ribbon elements take a completely different form:
     \[
    \chi_{[F_{\tilde \rho}]}(v)\; =\; \left \{
    \begin{aligned}
        & -1,\; \ {\rm if} \ v \in \tilde{\tilde \rho} \\
        & +1,\; \ \rm{\rm otherwise}
    \end{aligned} \right . \ .
    \]
    As a consequence, $\Gamma_{\LM} = \{1\}$. Furthermore, the Hamiltonian becomes $H_L = \frac{1}{2}\sum_{j \in \Sigma_0} (1-\sigma_j^z)$ for which $ Z_{\LM} = \{\chi_0\}$. This shows that space of the ground-states that are not frustration-free is highly dependent on the chosen generating set.   \hfill $\blacktriangleleft$
\end{remark}

\subsection{The unbiased one-dimensional Ising model}

This is an example which involves a non-Lagrangian subgroup of $P_0$ from the previous example. The unbiased Ising model is defined by the formal Hamiltonian
\begin{equation}\label{Eq:IsingH2}
H_J\;=\; \sum_{e\in \Sigma_1} (1-P_{l_e}) \;=\; \tfrac{1}{2}\sum_{v\in \mathbb Z} (1-\sigma_v^z \sigma_{v+1}^z)
\end{equation}
corresponding to the partial generating set $\mathtt J=\{l_e\}_{e\in \Sigma_1}$, which give rise to a strict isotropic subgroup $J \subset L$. As we emphasized in our general theory, its dynamics can be analyzed within the Lagrangian framework for $L$, by simply changing the groupoid 1-cocycle from the previous section to
\[
 c_{\mathtt J}([p],\chi) \;=\; |\overline{\rm supp}\; \chi_{[p]}\chi \cap \overline{\rm supp}\; \chi_{[p]}| - |\overline{\rm supp}\; \chi \cap \overline{\rm supp}\; \chi_{[p]}|
\]
for any $\chi \in \widehat L$, where $
\overline{\rm supp}\, \chi := {\rm supp}\, \chi \setminus \{v_0\}.$

\begin{proposition}
    Let $\bm{\delta}_{\mathtt J}$ be the dynamics in $\A$ generated by the Ising model \eqref{Eq:IsingH2}. Then it has two pure frustration-free ground states given $\SM_{\chi_0}$ and $\SM_{\chi_-}$, where $\chi_-$ was defined in the previous example. 
\end{proposition}
\begin{proof}
Since $J\subset L$ and $L$ is Lagrangian, we have
$L\subseteq J^\perp$. Conversely, let $p\in J^\perp$.
For each vertex $v$, choose another vertex $w$ outside the
support of $p$. Let $l_v,l_w\in L$ be the configurations
equal to $-1$ at $v,w$, respectively, and $1$ elsewhere.
One knows $l_vl_w\in J$ and hence
$$
1\;=\;\omega_0(p,l_vl_w)
\;=\;\omega_0(p,l_v)\omega_0(p,l_w)
\;=\;\omega_0(p,l_v).$$
Because the $l_v$ generate $L$, we obtain
$p\in L^\perp=L$. Hence $J^\perp=L$ and one gets
$$J^\bot/J\;=\;L/J\;\simeq\; C_c(\{v_0\}\cup \Sigma_1,\Z_2)/C_c( \Sigma_1,\Z_2)\;\simeq\;\Z_2$$
Since any $2$-cocyle on $\Z_2$ is cohomologically trivial, the logical algebra satisfies 
$$
\A^{\rm log}_{J}
\;\simeq \;C^*\mathbb Z_2
\;\simeq\; \C\oplus \C ,
$$
By Theorem~\ref{teo: GSFF}, every frustration-free ground
state is thus a convex combination of the all-up and
all-down product states $\SM_{\chi_0}$ and $\SM_{\chi_-}$.
These are pure, proving the claim.
\end{proof}

Clearly, $J\subset L$ is and admissible isotropic. A similar computation to the biased case provides the complete set of ground states for the unbiased Ising model:
\begin{theorem}\label{thm: Ising-ground-states}
Consider the unbiased Ising dynamics $\bm{\delta}_{\mathtt J}$.
Let $\chi_e^\pm\in\widehat L$ take value $-1$ at
$e\in\Sigma_1$, value $+1$ at every other edge, and
$\chi_e^\pm(v_0)=\pm1$. Then
\[
Z_{\mathtt J}
\;=\;
\{\chi_0,\chi_-\}
\sqcup
\{\chi_e^\pm\mid e\in\Sigma_1\}.
\]
The boundary groupoid has four orbits:
\[
\begin{aligned}
\mathscr O_{(+1,+1)}\;&=\;\{\chi_0\},
&\qquad \mathscr O_{(-1,+1)}
\;=\;\{\chi_e^+\mid e_+\leq0\}\cup\{\chi_e^-\mid e_+>0\},\\
\mathscr O_{(-1,-1)}\;&=\;\{\chi_-\}
&\qquad \mathscr O_{(+1,-1)}
\;=\;\{\chi_e^+\mid e_-\geq0\}\cup\{\chi_e^-\mid e_-<0\}.
\end{aligned}
\]
Consequently, the pure ground states are parametrized by
\[
\operatorname{Ext}(\mathfrak G_{\mathtt J})
\;\simeq\;
\{\SM_{\chi_0},\SM_{\chi_-}\}
\sqcup
\mathbb P\bigl(\ell^2(\Sigma_1)\bigr)
\sqcup
\mathbb P\bigl(\ell^2(\Sigma_1)\bigr).
\]
\end{theorem}

The two non-trivial domain-wall sectors describe kinks
and antikinks interpolating between the all-up and
all-down ferromagnetic states. Their product-state
representatives are depicted in
Fig.~\ref{Fig:Ising}(c,d), now without restrictions on
the domain-wall positions, and correspond to the
soliton states described in \cite{Araki1985}.

\appendix
\section{Technical lemmas}
In this section, we prove technical Lemmas used in Section \ref{sec: pauli}. We recall that $\sigma$ is a $2$-cocyle on $P$ with a non-degenerated  bicharacter $\omega$.

\medskip

We assume that $\Sigma$ is infinite. By the fundamental theorem of finite abelian groups, one has a decomposition
$$
G \;\cong\; \bigoplus_{p\mid |G|} G_p,
$$
where $G_p$ denotes the $p$-primary component of $G$, consisting of all elements whose order is a power of $p$. Consequently, the primary decomposition theorem for abelian groups verifies that
$$
P\;\simeq \;\bigoplus_{\mathbb N}G
\;\cong\;
\bigoplus_{p\mid |G|} P_p,
\qquad
P_p\;:=\;\bigoplus_{\mathbb N} G_p.
$$
Clearly each $P_p$ is infinite. The bicharacter $\omega$ restricts to a nondegenerate bicharacter on each $P_p$, and distinct primary components are orthogonal. Indeed,
$$\omega(x,y)^{{\rm ord}(x)}\;=\;\omega({{\rm ord}(x)} x,y)\;=\;1\;=\;\omega(x,y)^{{\rm ord}(y)}$$
For $x\in P_p$ and $y\in P_q$.  Since these orders are relatively prime for $p\neq q$, one obtains $\omega(x,y)=1$.
\begin{lemma}\label{lem: infinite}
If $L\subseteq P$ is Lagrangian,  then $L_p:=L\cap P_p$ is infinite for every $p\mid{ |G|}$.
\end{lemma}

\begin{proof}
Fix $p\mid |G|$. Suppose that $L_p$ is finite. Since the restriction of
$\omega$ to $P_p$ is nondegenerate, the quotient $P_p/(L_p^\perp\cap P_p)$
embeds into $\widehat{L}_p$ via $\omega$, hence is finite. Therefore,
$L_p^\perp\cap P_p$ must be infinite. In particular, we may choose $
x\in (L_p^\perp\cap P_p)\setminus L_p.$ If $q\neq p$, elements of $P_p$ pair trivially with elements of $P_q$.
Since
\[
L\;=\;\bigoplus_{q\mid |G|}L_q,
\]
the element $x$ is orthogonal to all of $L$. Hence $x\in L^\perp=L$.
But $x\in P_p$, so $x\in L\cap P_p=L_p$, a contradiction. Thus $L_p$
is infinite.
\end{proof}

\begin{lemma}\label{lem: finite subgroup}
    For each $m\in \N$ there exist a finite subgroup $F_m$ of $ L$ such that
    $|F|\;=\; |G|^m$.
\end{lemma}
\begin{proof}
Write
$$
|G|\;=\;\prod_{p\mid |G|}p^{a_p}.$$
By  lemma \ref{lem: infinite}, for every $p\mid |G|$, the group $L_p$ is infinite. Therefore, it contains finite subgroups of arbitrarily large order. Moreover, every finite abelian $p$-group of order at least $p^r$ contains a subgroup of order $p^r$. Hence, for each $m$, we may choose a
finite subgroup $
F_{p,m}\subseteq L_p$
such that $
|F_{p,m}|=p^{a_p m}.$
Indeed, if $L_p$ contains a finite subgroup of order at least $p^{a_p m}$,
then, being a finite abelian $p$-group, it contains a subgroup of order
exactly $p^{a_p m}$. Now define
\[
F_m\;:=\;\bigoplus_{p\mid |G|}F_{p,m}\;\subseteq \;L.
\]
 $F_m$ is finite and clearly
$$
|F_m|
\;=\;
\prod_{p\mid |G|}|F_{p,m}|
\;=\;
\prod_{p\mid |G|}p^{a_p m}
\;=\;
|G|^m.
$$
This proves the claim.
\end{proof}


\begin{thebibliography}{99}
\bibitem{ADL2025}
J. Acu\~na Flores, G. De Nittis, J. Lorca Espiro,
\textit{Higher abelian quantum double models},
arXiv:2509.12864 (2025).

\bibitem{AffleckCMP1988} 
I. Affleck, T. Kennedy, E. H. Lieb, H. Tasaki, 
\textit{Valence bond ground-states in isotropic quantum antiferromagnets}, 
Commun. Math. Phys. {\bf 115} (1988), 477-528.

\bibitem{AlickiJPA2007} 
R. Alicki, M. Fannes, M. Horodecki,
\textit{A statistical mechanics view on Kitaev's proposal for quantum memories},
J. Phys. A: Math. Theor. \textbf{40} (2007), 6451--6465.

\bibitem{And1}
J. Anderson,
\textit{Extensions, restrictions, and representations of states on $C^*$-algebras},
Trans. Amer. Math. Soc. \textbf{249} (1979), 303--329.

\bibitem{Anderson2013}
J. T. Anderson,
\textit{Homological stabilizer codes},
Ann. Phys. \textbf{330} (2013), 1--22.

\bibitem{Araki1985}
H. Araki, T. Matsui, \textit{ground-states of the XY -model}, 
Commun. Math. Phys. {\bf 101} (1985), 213–245.

\bibitem{AB2}
R. J. Archbold, J. W. Bunce,
\textit{Extensions of states of $C^*$-algebras, II},
Proc. Roy. Soc. Edinburgh Sect. A \textbf{92} (1982), 113--122.

\bibitem{Ball2023}
S. Ball, A. Centelles, F. Huber, 
\textit{Quantum error-correcting codes and their geometries}, Ann. Inst. H. Poincar\'e D Comb. Phys. Interact. {\bf 10} (2023), 337–405.


\bibitem{BHNV1}
A. Bols, M. Hamdan, P. Naaijkens, S. Vadnerkar,
\textit{The category of anyon sectors for non-abelian quantum double models},
Commun. Math. Phys. \textbf{407} (2026), Article 2.


\bibitem{bratteli-robinson-97}
O. Bratteli, D. W. Robinson,
\textit{Operator Algebras and Quantum Statistical Mechanics 1 \& 2},
Springer, Berlin, 1997.

\bibitem{Bombin2012}
H. Bombin, G. Duclos-Cianci, D. Poulin,
\textit{Universal topological phase of two-dimensional stabilizer codes},
New J. Phys. \textbf{14} (2012), 073048.

\bibitem{BrownBook}
K. S. Brown,
\textit{Cohomology of groups},
Springer-Verlag, New York, 1982.

\bibitem{BuerschaperJMP2013} 
O. Buerschaper, J. M. Mombelli, M. Christandl, M. Aguado, 
\textit{A hierarchy of topological tensor network states}, 
J. Math. Phys. {\bf 54} (2013), 012201.


\bibitem{CRSS1997}
A.~R.~Calderbank, E.~M.~Rains, P.~W.~Shor and N.~J.~A.~Sloane,
\emph{Quantum error correction and orthogonal geometry},
Phys. Rev. Lett. \textbf{78} (1997), 405--408.
\bibitem{CaoJiaTan2026}
S. Cao, Z. Jia, S. Tan,
\textit{Topological quantum color code model on infinite lattice},
arXiv:2601.12409 (2026).

\bibitem{CNN}
M. Cha, P. Naaijkens, B. Nachtergaele,
\textit{The complete set of infinite volume ground-states for Kitaev's abelian quantum double models},
Commun. Math. Phys. \textbf{357} (2018), 125--157.

\bibitem{Penneys}
C. Y. Chuah, B. Hungar, K. Kawagoe, D. Penneys, M. Tomba, D. Wallick, S. Wei,
\textit{Boundary algebras of the Kitaev quantum double model},
J. Math. Phys. \textbf{65} (2024), 012101.

\bibitem{Clark2007}
L. O. Clark,
\textit{Classifying the types of principal groupoid $C^*$-algebras},
J. Operator Theory \textbf{57} (2007), no. 2, 251--266.

\bibitem{ClarkPAMS2025}
L. O. Clark, A. an Huef, R. P. Lima, and C. F. Sehnem, 
\textit{Equivalence of definitions of AF groupoid}, 
Proc. Amer. Math. Soc. \textbf{153} (2025), 1313–1324.

\bibitem{CM}
A. Connes, M. Marcolli, 
\textit{From physics to number theory via noncommutative geometry}, 
in: \textit{Frontiers in Number Theory, Physics, and Geometry. I}, 
pp. 269–347. Springer, Berlin (2006).
\bibitem{Dennis2002}
E.~Dennis, A.~Kitaev, A.~Landahl and J.~Preskill,
\emph{Topological quantum memory},
J. Math. Phys. \textbf{43} (2002), 4452--4505.

\bibitem{Dynin}
A. Dynin, 
\textit{Inversion problem for singular integral operators: $C^\ast$-approach},
Proc. Natl. Acad. Sci. USA {\bf 75} (1978), 4668-4670.

\bibitem{FarsiMJM2019}
C. Farsi, A. Kumjian, D. Pask, and A. Sims, 
\textit{Ample groupoids: equivalence, homology, and Matui’s HK conjecture}, 
M\"unster J. Math. \textbf{12} (2019), 411–451.

\bibitem{Naaij2}
L. Fiedler, P. Naaijkens,
\textit{Haag duality for Kitaev's quantum double model for abelian groups},
Rev. Math. Phys. \textbf{27} (2015), 1550021.

\bibitem{GTS}
T. Giordano, I. F. Putnam, C. F. Skau,
\textit{Affable equivalence relations and orbit structure of Cantor dynamical systems},
Ergodic Theory Dynam. Systems \textbf{24} (2004), 441--475.

\bibitem{Gottesman1997}
D.~Gottesman,
\emph{Stabilizer Codes and Quantum Error Correction},
Ph.D. thesis, California Institute of Technology (1997).
arXiv:quant-ph/9705052.
\bibitem{GottesmanFT1998}
D.~Gottesman,
\emph{Theory of fault-tolerant quantum computation},
Phys. Rev. A \textbf{57} (1998), 127--137.



\bibitem{Haah2013}
J.~Haah,
\textit{Commuting Pauli Hamiltonians as maps between free modules},
Commun. Math. Phys. \textbf{324}, no.~2, 351--399 (2013).
\bibitem{Hamdan2024}
M. Hamdan,
\textit{Infinite volume ground-states in the non-abelian quantum double model: on anyon excitations in the plane},
PhD thesis, Cardiff University, 2024.

\bibitem{Kit0}
A. Y. Kitaev,
\textit{Quantum error correction with imperfect gates},
in: O. Hirota, A. S. Holevo, C. M. Caves (eds.),
\textit{Quantum communication, computing, and measurement},
Springer, Boston, MA, 1997, 181--188.

\bibitem{Kit00}
A. Y. Kitaev,
\textit{Quantum computations: Algorithms and error correction},
Russian Math. Surveys {\bf 52} (1997), 1191-1249.

\bibitem{Kit1}
A. Y. Kitaev,
\textit{Fault-tolerant quantum computation by anyons},
Ann. Phys. \textbf{303} (2003), 2--30.

\bibitem{Kit2}
A. Y. Kitaev,
\textit{Anyons in an exactly solved model and beyond},
Ann. Phys. \textbf{321} (2006), 2--111.

\bibitem{Koma1998}
T. Koma, B. Nachtergaele,
\textit{The complete set of ground-states of the ferromagnetic XXZ chains},
Adv. Theor. Math. Phys. {\bf 2} (1998), 533–558.

\bibitem{KomuraDocMath2025}
F. Komura,
\textit{$*$-homomorphisms between groupoid $C^*$-algebras},
Doc. Math. \textbf{31} (2026), 197--221.

\bibitem{Krieger}
W. Krieger,
\textit{On a dimension for a class of homeomorphism groups},
Math. Ann. \textbf{252} (1979/80), 87--95.

\bibitem{KR1}
A. Kumjian, J. Renault,
\textit{KMS states on $C^*$-algebras associated to expansive maps},
Proc. Amer. Math. Soc. \textbf{134} (2006), 2067--2078.

\bibitem{Kum}
A. Kumjian,
\textit{On $C^*$-diagonals},
Can. J. Math. \textbf{38} (1986), 969--1008.

\bibitem{Kubica2015}
A. Kubica, M. E. Beverland,
\textit{Universal transversal gates with color codes: A simplified approach},
Phys. Rev. A \textbf{91} (2015), 032330.

\bibitem{LLN2019}
M. Laca, N. S. Larsen, S. Neshveyev,
\textit{ground-states of groupoid $C^*$-algebras, phase transitions and
arithmetic subalgebras for Hecke algebras},
J. Geom. Phys. \textbf{136} (2019), 268--283.

\bibitem{LevinPRB2005}
M. A. Levin, X.-G. Wen,
\textit{String-net condensation: A physical mechanism for topological phases},
Phys. Rev. B \textbf{71} (2005), 045110.

\bibitem{Matsui1996}
T. Matsui,
\textit{On ground-states of the one-dimensional ferromagnetic XXZ chain},
Lett. Math. Phys. {\bf 37} (1996), 397–403.

\bibitem{Mat}
H. Matui,
\textit{Homology and topological full groups of étale groupoids on totally disconnected spaces},
Proc. Lond. Math. Soc. \textbf{104} (2012), 27--56.

\bibitem{MuhlyWilliams1990}
P. S. Muhly and D. P. Williams,
\textit{Continuous trace groupoid $C^*$-algebras},
Math. Scand. \textbf{66} (1990), 231--241.

\bibitem{MurphyBook} G. J. Murphy, {\sl $C^\ast$-algebras and operator theory}, (Academic Press, Boston, 1990).

\bibitem{Naaij1}
P. Naaijkens,
\textit{Localized endomorphisms in Kitaev's toric code on the plane},
Rev. Math. Phys. \textbf{23} (2011), 347--373.

\bibitem{Nesh1}
S. Neshveyev,
\textit{KMS states on the $C^*$-algebras of non-principal groupoids},
J. Operator Theory \textbf{70} (2013), 513--530.

\bibitem{OgataI}
Y. Ogata,
\textit{A Class of asymmetric gapped Hamiltonians
on quantum spin chains and its characterization I},
Commun. Math. Phys. {\bf 348} (2016), 847–895. 

\bibitem{OgataII}
Y. Ogata,
\textit{A Class of asymmetric gapped Hamiltonians
on quantum spin chains and its characterization II},
Commun. Math. Phys. {\bf 348} (2016), 897–957.

\bibitem{Ogata2022} 
Y. Ogata,
\textit{Classification of gapped ground-state phases in quantum spin systems},
Proc. Int. Cong. Math. {\bf 5} (2022), 4142–4161.

\bibitem{Perez}
D. P\'erez-García,
\textit{The quest for a classification of 2D quantum phases of matter},
Proc. Int. Cong. Math. {\bf 5} (2026), 534-550.

\bibitem{DET1}
D. Polo Ojito, E. Prodan, T. Stoiber,
\textit{On frustration-free quantum spin models},
arXiv:2507.03201, 2025.

\bibitem{DE1}
D. Polo Ojito, E. Prodan,
\textit{On a $C^*$-diagonal generated by the toric code},
Advances in Operator Theory {\bf 11} (2026), 48.

\bibitem{PoloProdan2026}
D. Polo Ojito, E. Prodan,
\textit{The full set of KMS--states for abelian Kitaev models},
arXiv:2603.27436 (2026).

% \bibitem{ProdanPRB2009}
% E. Prodan, F. D. M. Haldane,
% \textit{Mapping the braiding properties of the Moore-Read state},
% Phys. Rev. B {\bf 80} (2009), 115121.



\bibitem{Ren1}
J. Renault,
\textit{Cartan subalgebras in $C^*$-algebras},
Irish Math. Soc. Bull. \textbf{61} (2008), 29--63.

\bibitem{RenaultBook}
J. Renault,
\textit{A groupoid approach to $C^\ast$-algebras},
Springer, Berlin, 1980.

\bibitem{RoeBook}
J. Roe,
\textit{Lectures on coarse geometry},
University Lecture Series \textbf{31},
American Mathematical Society, Providence, RI, 2003.

\bibitem{RubaYang2024}
B.~Ruba and B.~Yang,
\textit{Homological invariants of Pauli stabilizer codes},
Commun. Math. Phys. \textbf{405}, Article 126 (2024).

\bibitem{RubaYang2025}
B.~Ruba and B.~Yang,
\textit{Witt groups and bulk-boundary correspondence for stabilizer states},
arXiv:2509.10418v2 (2025).

\bibitem{Terha2015}
B. M. Terhal, 
\textit{Quantum error correction for quantum memories}, 
Rev. Mod. Phys. {\bf 87} (2015), 307.

\bibitem{Thomsen2014} 
K. Thomsen, 
\textit{KMS weights on graph $C^\ast$-algebras II. Factor types and ground-states},
arXiv:1412.6762 (2014)

\bibitem{Thomsen2020}
K. Thomsen, 
\textit{Phase transition in the CAR algebra},
Adv. Math. {\bf 372} (2020), 107312.


\bibitem{Thomsen2021}
K. Thomsen,
\textit{ground-states for generalized gauge actions on UHF algebras},
Commun. Math. Phys. \textbf{386} (2021), 57--85.

% \bibitem{NestQIC2011}
% M. Van den Nest,
% \textit{Efficient classical simulations of quantum Fourier
% transforms and normalizer circuits over Abelian groups},
% arXiv:1201.4867 (2012).

\bibitem{WangBook}
Z. Wang,
\textit{Topological quantum computation},
CBMS Regional Conference Series in Mathematics \textbf{112}, 2010.

\bibitem{WilczekBook}
F. Wilczek,
\textit{Fractional statistics and anyon superconductivity},
World Scientific Publishing Co., Inc., Teaneck, NJ 1990.

\bibitem{WilliamsBook}
D. P. Williams,
\textit{A toolkit for groupoid $C^*$-algebras},
American Mathematical Society, Providence, RI, 2019.
\end{thebibliography}
    \end{document}